\documentclass[twocolumn]{aastex62}

\newcommand{\arcs}{$^{\prime\prime}$} % Arcseconds
\newcommand{\beq}{\begin{equation}\begin{aligned}}
\newcommand{\eeq}{\end{aligned}\end{equation}}

\newcommand{\msun}{M$_\odot$}

\usepackage{scalerel,amssymb}
\usepackage{amsmath}
\usepackage{enumitem}
\usepackage[]{url}
\usepackage{lineno}

\accepted{\today}
\submitjournal{ApJ}

\shorttitle{Star Clusters of Nearby Dwarf Galaxies}
\shortauthors{Carlsten et al.}

\begin{document}

\title{ELVES II: GCs and Nuclear Star Clusters of Dwarf Galaxies; The Importance of Environment}

\correspondingauthor{Scott G. Carlsten}
\email{scottgc@princeton.edu}

\author[0000-0002-5382-2898]{Scott G. Carlsten}
\affil{Department of Astrophysical Sciences, 4 Ivy Lane, Princeton University, Princeton, NJ 08544}

\author[0000-0002-5612-3427]{Jenny E. Greene}
\affil{Department of Astrophysical Sciences, 4 Ivy Lane, Princeton University, Princeton, NJ 08544}

\author[0000-0002-1691-8217]{Rachael L. Beaton}
\altaffiliation{Hubble Fellow}
\affiliation{Department of Astrophysical Sciences, 4 Ivy Lane, Princeton University, Princeton, NJ 08544}
\affiliation{The Observatories of the Carnegie Institution for Science, 813 Santa Barbara St., Pasadena, CA~91101\\}

\author[0000-0003-4970-2874]{Johnny P. Greco}
\altaffiliation{NSF Astronomy \& Astrophysics Postdoctoral Fellow}
\affiliation{Center for Cosmology and AstroParticle Physics (CCAPP), The Ohio State University, Columbus, OH 43210, USA}

\begin{abstract}
We present the properties of the globular clusters (GCs) and nuclear star clusters (NSCs) of low-mass ($10^{5.5}<M_\star<10^{8.5}$ \msun) early-type satellites of Milky Way-like and small group hosts in the Local Volume (LV) using deep, ground-based data from the ongoing Exploration of Local VolumE Satellites (ELVES) Survey. This sample of 177 dwarfs significantly increases the statistics for studying the star clusters of dwarfs in low-density environments, offering an important comparison to samples from nearby galaxy clusters. The LV dwarfs exhibit significantly lower nucleation fractions at fixed galaxy mass than dwarfs in nearby clusters. The mass of NSCs of LV dwarfs show a similar scaling of $M_{\star,\mathrm{NSC}}\propto M_{\star,\mathrm{gal}}^{0.4}$ as that found in clusters but offset to lower NSC masses. To deal with foreground/background contamination in the GC analysis, we employ both a statistical subtraction and Bayesian approach to infer the average GC system properties from all dwarfs simultaneously. We find that the GC occupation fraction and average abundance are both increasing functions of galaxy stellar mass, and the LV dwarfs show significantly lower average GC abundance at fixed galaxy mass than a comparable sample of Virgo dwarfs analyzed in the same way, demonstrating that GC prevalence also shows an important secondary dependence on the dwarf's environment. This result strengthens the connection between GCs and NSCs in low-mass galaxies. We discuss these observations in the context of modern theories of GC and NSC formation, finding that the environmental dependencies can be well-explained by these models.

\end{abstract}

%\linenumbers

\section{Introduction}
Globular clusters (GCs) are generally old (age $\sim 10$ Gyr), massive ($M_\star\sim10^5$ \msun) stellar clusters that are ubiquitously found in nearby galaxies, with essentially all galaxies with $M_\star>10^{9}$ \msun~hosting a system of GCs. GCs are generally thought to form in a similar manner as young massive clusters (YMCs) observed in nearby merging and/or star-bursting galaxies \citep[e.g.][]{holtzman1992,whitmore1999} but instead at $z\sim2-3$ \citep{ashman1992, elmegreen1997, kravtsov2005_gcs, elmegreen2010, kruijssen2015, pfeffer2018}. The high densities and pressures in the ISM of galaxies at  $z\sim2-3$ are conducive to the formation of significant populations of massive clusters \citep{shapiro2010, kruijssen2012, reina-campos2017}. Understanding the physics of massive cluster formation at high redshift and the physical processes that connect the high redshift clusters to GCs we observe today are important open areas of research. Due to their age and prevalence, GCs have long been noted as important `fingerprints' to decipher the process of galaxy assembly \citep[for reviews of GCs in the context of galaxy formation, see][]{brodie2006, kruijssen2014, forbes2018, beasley2020}.

Dwarf galaxies ($M_\star\lesssim10^{9}$ \msun) offer an interesting window into the physics that sculpt the $z=0$ properties of GCs. Their relatively simple formation pathways (compared to the Milky Way, for instance) allows one to better isolate different physical mechanisms relevant to GCs. Knowledge of the GC systems of dwarf galaxies can then, in turn, inform our understanding of the build-up of the GC systems of more massive galaxies, like the Milky Way (MW), which consist of a large fraction of accreted, \textit{ex situ} GCs \citep[e.g.][]{leaman2013, kruijssen2019}. The GC systems of large samples of nearby dwarf galaxies have been studied in numerous past works, primarily using \emph{HST} observations \citep[e.g.][]{miller1998, lotz2004, sharina2005, miller2007, georgiev2009a, georgiev2010}. These works have studied how the properties of the GC systems depend on various properties of the host galaxy, including mass/luminosity, color, and morphological type.

Another, less explored, property of the host galaxy that appears to have an effect on the GC system is the larger scale environment of the host. There is moderate evidence that galaxies in denser environments have richer GC systems at fixed stellar mass. \citet{peng2008} found that dwarf galaxies in Virgo had higher GC specific frequencies when closer to the center of Virgo. While \citet{liu2019} did not find any difference in GC specific frequency between Fornax and Virgo dwarfs (even though the two clusters are almost an order of magnitude different in mass), they too found that dwarfs in denser regions of the clusters have somewhat higher specific frequencies. However, \citet{miller2007} did not find a trend in specific frequency with clustercentric distance amongst lower mass dwarf galaxies in Virgo. These studies have been limited to just the environments of Fornax and Virgo, and a study encompassing a much wider dynamic range in environment is needed. Investigating this is one of the key goals of the current work.

There have been some further early indications that the environment of a dwarf galaxy does indeed have an effect on its GC properties from recent work on `ultra-diffuse galaxies' (UDGs). UDGs are large $r_e\gtrsim1.5$ kpc dwarf galaxies ($10^7\lesssim M_\star \lesssim 10^9$ \msun) that have seen significant interest in recent years, at least partly because of unexpectedly rich GC systems \citep{pvd2016_df44, peng2016, pvd_2017_gcs}\footnote{However, see also \citet{saifollahi2020} for different results regarding a Coma UDG.}. These early results were for UDGs in the Coma cluster, and results for UDGs in other, less dense environments suggest that these ultra-rich GC systems might be unique to the Coma environment and are generally not found in lower mass clusters (like Virgo or Fornax) or groups \citep{prole2019, lim2020, somalwar2020}, although with a few exceptions \citep[e.g.][]{muller2021}. As UDGs are outliers in the mass-size relation for dwarf galaxies, it is unclear how this environmental dependence translates for the whole dwarf galaxy population.

In addition to GCs, another dense, bound stellar component that many galaxies, including dwarf galaxies, possess are nuclear star clusters (NSCs). NSCs are central stellar clusters with sizes comparable to GCs ($\sim$ few pc) but with a broad range in stellar mass ranging from roughly that of a GC to several orders of magnitude more \citep[$M_\star\sim10^5-10^8$\msun;~][]{walcher2005, cote2006, turner2012, georgiev2014}. For a recent review on the properties of NSCs see \citet{neumayer2020}. NSCs are found in a wide variety of galaxies, spanning different masses, morphological types, and environment \citep{binggeli1987, carollo1997, carollo1998, lisker2007, georgiev2009b, rsj2019}. For early-type galaxies, studies of Virgo and Fornax cluster galaxies and nearby groups show NSC occurrence is an increasing function of galaxy stellar mass up to $\sim 10^{9}$ \msun~where it reaches $\sim90$\%, beyond which it starts to decline \citep{turner2012, denbrok2014, ordenes-briceno2018_nuc_masses, rsj2019, habas2020}. 

Models for the formation of NSCs generally fall into two main classes. First, many works have shown they could be the result of dynamical friction-induced inspiral and subsequent merging of GCs \citep[e.g.][]{tremaine1975, capuzzo1993, oh2000, lotz2001, antonini2013, gnedin2014}. The second main proposed formation mechanism is that of \textit{in situ} formation of the NSC. In this model, dissipational gas transfer into the center of galaxies drives star formation and a buildup of a dense stellar nucleus. The gas can be sent to the center either via wet-mergers or other secular processes \citep[e.g.][]{mihos1994, milo2004}. It is almost certain that both mechanisms are at work \citep[e.g.][]{hartmann2011, leigh2012}. However, there is growing consensus that both models are not equally important at all host masses, but the relative importance of each pathway changes as a function of host mass. In particular, GC-inspiral is likely the dominant mechanism for dwarf ($M_\star < 10^{9}$ \msun) galaxies while the \textit{in situ} star formation is dominant at higher masses \citep{turner2012, rsj2019, neumayer2020}. 

The works referenced above have elucidated how various properties of an NSC depend on the properties of its host, including stellar mass and type. However, as with GCs, there is some evidence that the properties of NSCs (particularly their frequency) depend also on the larger scale environment in which the galaxy resides. Starting with \citet{ferguson1989}, there has been growing evidence that the nucleation fraction of dwarf early-type galaxies depends on the local density. The spatial distribution of nucleated dE's in Fornax and Virgo are more centrally concentrated than non-nucleated dE's \citep{lisker2007, ordenes-briceno_spatial}. Additionally, dEs in higher mass clusters, like Coma, are nucleated at a higher frequency than in Virgo or Fornax \citep{denbrok2014, rsj2019, zanatta2021}. \citet{rsj2019} tentatively showed that the nucleation fraction is even lower in the Local Group (LG); however, the statistics afforded by just the LG are not enough to make a strong conclusion. The cause of this environmental dependence is not clear.

With a few notable exceptions, the above-cited observational works that present the properties of GCs and NSCs in dwarf galaxies do so for dwarfs in nearby clusters. The primary goal of the current work is to systematically present the GC and NSC properties of dwarfs in a lower density environment, particularly in halos of nearby roughly Milky Way (MW)-sized galaxies in the Local Volume (LV; $D\lesssim12$ Mpc). As discussed more below, we focus only on early-type dwarfs, directly comparing to early-type dwarfs in nearby clusters. We find that adding the axis of environment opens a valuable new perspective on the formation of GCs and NSCs. 

Due to the low density of these environments, achieving the statistics necessary to meaningfully compare with samples of cluster dwarfs has been a major hurdle. It is not until recently that enough of the LV has been surveyed to build up a significant sample of dwarf satellites of MW-like galaxies with confirmed distances and well-quantified completeness. In particular, we use the dwarf catalogs from the ongoing Exploration of Local VolumE Satellites (ELVES) Survey. In Section \ref{sec:data}, we describe this sample, in Section \ref{sec:methods} we describe our analysis of the NSCs and GCs of these dwarfs, in Sections \ref{sec:results_nsc} and \ref{sec:results_gcs} we present our results regarding NSCs and GCs, respectively, in Section \ref{sec:disc} we discuss these results in the context of modern theories of how GCs and NSCs form, and we conclude in Section \ref{sec:concl}.

\section{Data}
\label{sec:data}

\subsection{ELVES Overview}
\label{sec:survey}
For our primary observational sample, we use results from the ongoing Exploration of Local VolumE Satellites (ELVES) survey. The overall goal of the survey is to survey the classical satellites down to luminosities of $M_V\sim-9$ and within $300$ projected kpc of all massive, $M_K < -22.4$ mag, hosts within 12 Mpc with full or nearly full distance information for the satellites. There are 29 hosts that meet this qualification, and further details on host selection and the list of hosts will be presented in a future paper. Satellite candidates are detected using deep, wide-field imaging combined with the detection algorithm specialized for finding low-surface brightness, diffuse dwarf galaxies of \citet{carlsten2020a} and \citet{greco2018}. Through extensive tests with injected artificial galaxies, we ensure that ELVES will be complete down to $M_V\sim-9$ and surface brightness of $\mu_{0,V}\lesssim 26.5$ mag arcsec$^{-2}$ \citep[see][for more detailed information on completeness]{carlsten2020b}.

The difficulty in this endeavor is in determining the distance to candidate satellites to confirm their association with a host. Since these groups are quite sparse (compared to Virgo or Fornax, for instance), contamination is often quite high, and candidates selected on their low-surface brightness, diffuse morphology alone can often consist of a majority \citep[$>80\%$, e.g.][]{sbf_m101, carlsten2020a} of background contaminants. In studying the star clusters of dwarfs, it is crucial we only consider satellites that have distance confirmation. Background, interloping galaxies likely would not have detectable nuclei or GCs and could artificially bring down the average inferred abundance of these clusters.

The candidate satellites are confirmed with a variety of distance measurements including archival redshifts and TRGB distances, but the majority of distances are measured via surface brightness fluctuations (SBF). SBF \citep[e.g.][]{tonry1988, sbf_calib, sbf_m101, greco2021} can produce relatively precise distance errors ($\lesssim15\%$) using modest ground based data. This is enough to confirm the association of candidate satellites or not in most cases.

The survey uses a mixture of archival CFHT/MegaCam imaging \citep{carlsten2020a, carlsten2020b, carlsten2020c} and the DESI Legacy Imaging Surveys \citep{decals}\footnote{\url{https://www.legacysurvey.org/}} which includes both the Beijing-Arizona Sky Survey \citep[BASS][]{bass1, bass2} and the DECam Legacy Survey (henceforth these surveys are collectively referred to as DECaLS). We find that we can readily detect dwarfs down to $M_V\sim-9$ even using the relatively shallow DECaLS imaging. While the CFHT/MegaCam data are high enough quality to apply SBF to even $M_V\sim-9$ mag dwarfs, the depth and PSF size of DECaLS are not adequate for SBF of dwarfs throughout the mass range of the survey. Thus, we are using deeper Gemini (program IDs: FT-2020A-060 and US-2020B-037) or Subaru/HSC data, where available, to measure the dwarf distances via SBF.

As discussed in \citet{carlsten2020b}, for some of the faintest and/or lowest surface brightness dwarfs, the SBF distances were inconclusive, leaving some candidates as `unconfirmed' or `possible' satellites. In this work, we only consider the confirmed satellites, e.g., those measured to be at the distance of their host. We are actively pursuing even deeper data for SBF or \emph{HST} data for TRGB measurements for the remaining candidates and expect that some will turn out to be real satellites. Thus, the satellite sample used in the current work is not actually complete to $M_V\sim-9$ mag. However, because the unconfirmed candidate satellites are generally the lowest luminosity ($M_V\gtrsim-10$ mag) candidates, their exclusion will not have a significant effect on the current work as such faint satellites are unlikely to host either NSCs or GCs.

A handful of hosts have been previously surveyed in the literature to at least the level of ELVES (or deeper), with distance confirmation for the satellites. We take the properties of the MW and M31 satellites from \citet{mcconnachie2012}. For the other four (M81, CenA, M94, M101) previously surveyed LV hosts, we only use the distance confirmation from the previous surveys. We derive all photometry ourselves (either from DECaLS or other archival data). The relevant references are: \citet{muller2019} and \citet{crnojevic2019} (CenA), \citet{chiboucas2013} (M81), \citet{smercina2018} (M94), and \citet{bennet2019} (M101). Note that the satellite samples of M81 and CenA are subsamples (including roughly 80\%) of their entire, known satellite systems for which we could find high-quality multi-band imaging.

The imaging data that we use in the current work to analyze NSCs and GCs is roughly about one-third CFHT/Megacam, one-half DECaLS, and the remainder Subaru/HSC. We prioritize the deeper CFHT and Subaru data, where available. For each dwarf we always have at least two imaging bands, either $g$ and $r$ or $g$ and $i$. The dwarfs are listed, along with their photometry (see Section \ref{sec:photo} below), in a table in Appendix \ref{app:photo_gcs}. In this table, we also list the data source for each dwarf (e.g. CFHT/Megacam, DECaLS, or other) and filter combination used. There is the concern that we are dealing with several different filter systems (DECam, HSC, BASS, MegaCam). All are Sloan-like filters but will have some differences. We make no attempt to bring the different measurements onto the same filter system. Instead, in a companion paper focused on the structural measurements of these galaxies (Carlsten et al, submitted),  we show that the filter systems do not differ by enough to alter any conclusions. The different filters differ generally by $\lesssim0.1$ mag.

We use slightly different samples of dwarfs for the NSC analysis and GC analysis. For the NSC analysis, we include satellites of 27 LV hosts (including the MW and M31), meaning that nearly all massive LV hosts have been surveyed to some extent. The list of hosts can be found in Appendix \ref{app:photo_gcs}. This is a sample of 177 early-type (see \S\ref{sec:morph}) satellites. However, due to the proximity of the MW and M31 satellites, we cannot analyze their GCs in the same way as the other LV hosts, and they are not included in the GC analysis. However, we do use the known GC properties of MW/M31 satellites as a check on our results. Additionally, due to their low galactic latitude (and high density of foreground stars), we do not analyze the satellites of CenA and NGC5236 for GCs. Thus, we only include the satellites of 23 LV hosts in the GC analysis, leading to a sample size of 140 early-type satellites.

We note that this is a larger host sample than used in the companion paper (Carlsten et al., submitted) which focuses on the structural scaling relations for ELVES dwarfs. In that work, we restricted the sample to hosts which have distance constraints on a majority of their candidate satellites due to the importance for that work to use a satellite sample that is as close to complete to $M_V\sim-9$ and $\mu_{0,V}\lesssim 26.5$ mag arcsec$^{-2}$ as possible. In this work, we instead maximize the sample statistics by including the confirmed satellites for a handful of other hosts who lack distance confirmation for the majority of their candidate satellites.  Along this vein, we also include a handful of confirmed satellites that are a little outside of the fiducial 300 projected kpc radial coverage of the ELVES Survey. The conclusions we find are unchanged if we do not include these extra satellites, although the statistics are lowered.

\subsection{Virgo Comparison Sample}
\label{sec:aux_data}

As mentioned in the introduction, many contemporary samples of dwarfs in denser, cluster environments exist. Of particular relevance, given their similar completeness levels to ELVES, are the Next Generation Virgo Survey (NGVS) \citep{ferrarese2012, rsj2019}, the Next Generation Fornax Survey (NGFS) \citep{munoz2015, eigenthaler2018, ordenes-briceno2018_nuc_masses}, and the Fornax Deep Survey (FDS) \citep{venhola2018, prole2019}. All three surveys have published analyses of the NSC properties of their dwarf samples. However, at the time of writing, only the FDS has published an analysis of the GCs for their entire dwarf sample \citep{prole2019}. This is clearly a valuable comparison for the LV sample, but given the vagaries involved in determining stellar masses and characterizing the GC systems of dwarfs (cf. Section \ref{sec:gc_preliminaries} below), it is important to have a cluster reference sample analyzed in as close to the same way as the LV sample as possible. Due to the archival availability of the data, we use the NGVS dwarf sample as the primary comparison sample.  We use the NGVS galaxy catalog from \citet{ferrarese2020} that considers only the core 4 deg$^2$ region of Virgo (roughly out to $r\lesssim R_\mathrm{vir}/5$). We acquire the raw CFHT/Megacam data from the Canadian Astronomy Data Center\footnote{\url{http://www.cadc-ccda.hia-iha.nrc-cnrc.gc.ca/}}, and reduce the data in the same way as the other CFHT/Megacam data used here \citep[see ][for details on the sky subtraction, etc.]{sbf_calib,carlsten2020a}. Since our focus is primarily on low-mass dwarf galaxies, we only reduce data for the NGVS dwarfs with $M_\star<10^9$\msun, as given  by the list of stellar masses from \citet{rsj2019}. This sample of NGVS dwarfs is analyzed in the same way as the main ELVES sample. In Appendix \ref{app:mass_distributions}, we show that the Virgo and LV dwarf samples have quite similar stellar mass distributions.

While the NGVS Virgo dwarf sample is our primary point of comparison, particularly for the GC analysis, we do make additional comparisons to other galaxy cluster samples, including from Fornax and Coma. These samples are described more in the Results sections where the comparisons are made.

%We also derive our own photometry for the NGFS dwarf sample from \citep{eigenthaler2018} that includes roughly the inner $r\lesssim R_\mathrm{vir}/4$ area of Fornax using Dark Energy Survey data (specifically the DECaLS reduction). These data are not deep enough for a GC analysis, however, we are still able to derive stellar masses in a consistent fashion (see Section \ref{sec:photo} below) to the LV sample which we use when comparing nucleation fractions at fixed stellar mass.

\subsection{Definition of `Environment'}
\label{sec:env}

Throughout this work we investigate the large-scale environmental dependence of the star cluster properties of dwarf galaxies, so it is important to clarify what is meant by `environment'. In our analysis, we use `environment' to denote the mass of the parent halo in which the dwarf satellites reside as subhalos. As we describe in more detail in Carlsten et al. (submitted), the LV hosts are significantly lower in halo mass than either cluster. Various estimates put the LV hosts in the halo mass range of $10^{12}\lesssim M_\mathrm{halo}/M_\odot \lesssim 10^{13}$. On the other hand, the Fornax cluster has a dynamical mass of $M_{200}\sim7\times10^{13}$~\msun~\citep{kourkchi2017}, the Virgo cluster has an estimated mass closer to $5-6\times10^{14}$~\msun~\citep{ferrarese2012, kourkchi2017, kashibadze2020}, and Coma is even more massive at $\sim 10^{15}$\msun~\citep{lokas2003}. 

The environmental difference may be exacerbated by the fact that the cluster samples are from the very central $r_\mathrm{proj}\lesssim R_\mathrm{vir}/4$ (see \S\ref{sec:aux_data} above) regions while the LV host satellite samples are generally complete to at least $r_\mathrm{proj}\gtrsim R_\mathrm{vir}/2$.

\subsection{Galaxy Type}
\label{sec:morph}

In this work, we only consider the GC and NSC properties of early-type dwarf galaxies in the ELVES sample. This is partly an observational necessity as late-type dwarfs contain a plethora of young star clusters, star-forming clumps, and HII regions that would greatly confuse any attempt to identify NSCs or count GCs. Late-type dwarfs require the resolution of \emph{HST} to be able to robustly identify these star clusters \citep[see e.g.][]{seth2004, georgiev2009b, georgiev2009a, georgiev2010}. Additionally, considering only the early-type dwarfs makes the comparison with the cluster samples (which are almost exclusively early-type) more direct. 

To split the dwarfs into early-type and late-type groups, we use a visual inspection of the dwarf morphology. Given the generally quite deep imaging data available (deep enough to apply SBF) and proximity of these dwarfs, we believe this is the most robust way to split dwarfs with the data available\footnote{These two classes also separate clearly in color-magnitude space. A simple cut in this space splits the dwarfs into two classes with the same result as the visual inspection about 90\% of the time.}. Dwarfs with smooth, featureless morphology are classed as early-types while dwarfs with clear star-forming regions, blue clumps, dust-lanes, or any other kinks in their surface brightness profile are classed as late-type. 

Measurements of H$\alpha$ and/or H\textsc{I} are available for about a third of the ELVES sample. We find that these measurements closely corroborate the visual morphology classification. The vast majority of early-type dwarfs with spectra do not have significant H$\alpha$ emission (equivalent widths $<2$\AA) while $\sim85$\% of the late-type dwarfs with spectra do have H$\alpha$ above this level (with more likely having H$\alpha$ emission that is simply below the sensitivity of the archival spectra, which is generally from SDSS). Based on this, we assert that the visual classification into late/early-type is essentially a physical distinction into star-forming and quenched dwarfs, respectively.

Throughout the paper, we focus only on dwarfs in the stellar mass range of $5.5 < \log(M_\star/{\rm M}_\odot) < 8.5$, and there are 177 such early-type dwarfs across all the LV hosts. For comparison, the NGVS sample includes 295 dwarfs.

\section{Methods}
\label{sec:methods}
 
In this section, we outline the steps we take in analyzing the NSCs and GCs of the LV sample. Briefly put, we first model and then subtract out the underlying light profile of the dwarf. In the image residuals, we look for a possible NSC. If one is found, we perform a second fit to the galaxy and NSC simultaneously using a multi-component model. Then, in the residuals of either the single- or multi-component fit, we look for candidate GCs. Due to the distance of the dwarf galaxies, GCs will not be spatially resolved. We take two different approaches to distinguish between actual GC candidates and contaminating sources (background galaxies or foreground stars). The first is a simple, statistical background subtraction, and the second is a more sophisticated likelihood-based approach. This second approach allows us to explore the radial distribution of GCs and fit for the fraction of dwarfs that possess GCs as a function of stellar mass. In the following sections, we give more details for each of these major steps.

\begin{figure*}
\includegraphics[width=\textwidth]{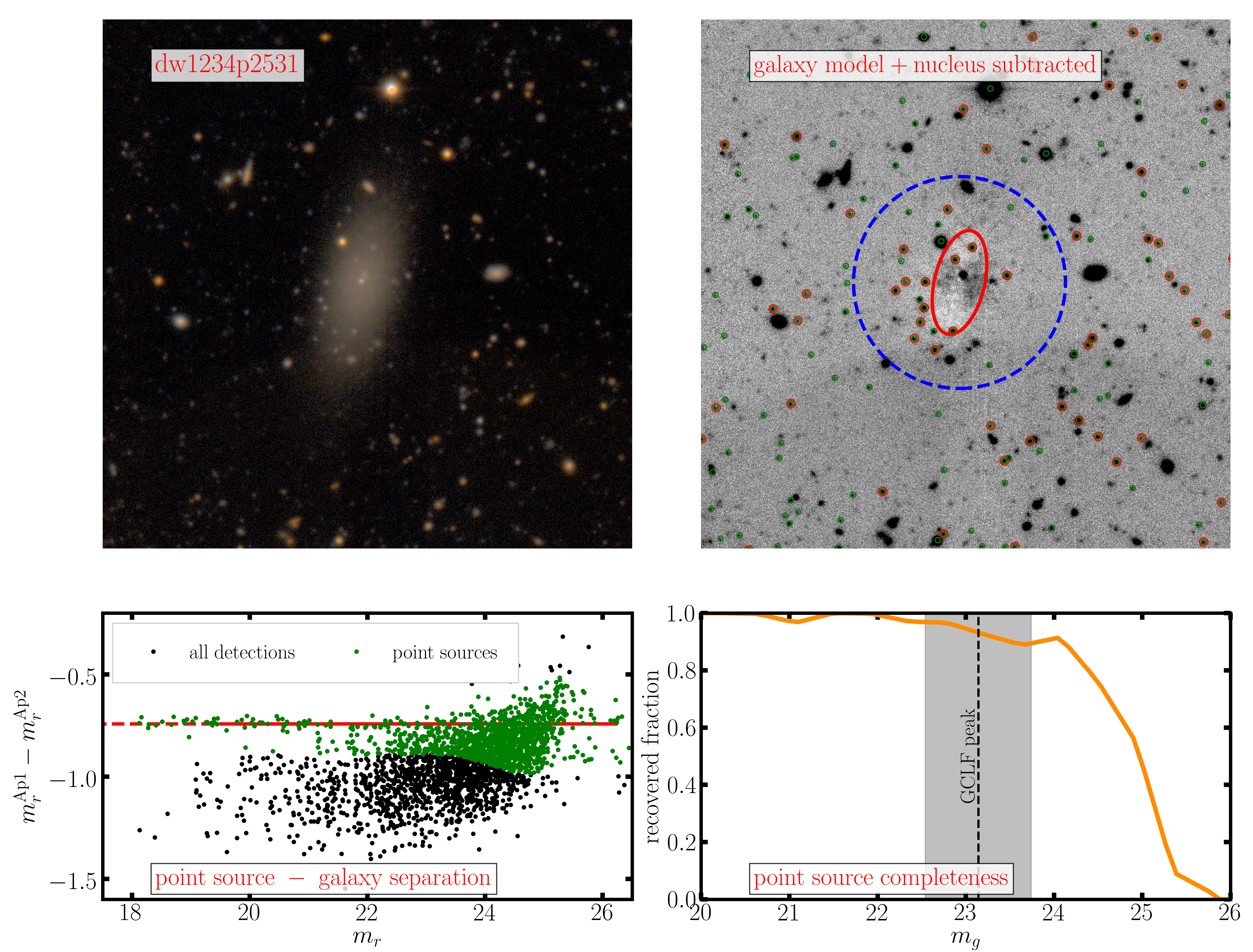}
\caption{Example of the NSC and GC analysis process. The top left shows a color image of a nucleated dwarf at $D=11.9$ Mpc; the image is $3^\prime$ to a side. The top right shows the residuals after the galaxy light and NSC have been modelled and subtracted. The red ellipse shows the ellipticity and effective radius of the dwarf. The dashed blue circle is the annulus within which we count GC candidates for the simple background subtraction method. The small green circles denote all the sources that pass the point source selection criterion (bottom left panel). The small red circles denote the sources that also pass the color cuts (see Section \ref{sec:color_cuts}). Bottom left shows the difference in aperture magnitudes (6 and 12 pixels) that we use to select point sources. The red line is the locus of point sources for this cutout. Bottom right shows the estimated completeness of these data as a function of $g$-band magnitude. Indicated on the plot is the expected GCLF peak at the distance of the dwarf. The data used generally extend multiple magnitudes below the GCLF peak.}
\label{fig:method}
\end{figure*}

\subsection{Galaxy and Nucleus Photometry}
\label{sec:photo}

We start the analysis with image cutouts of each dwarf. The cutouts are generally $\sim6^\prime$ to a side. This corresponds to $\gtrsim 20r_e$ for the LV dwarf sample. To model the underlying light profiles of the dwarfs, we rely on parametric 2D S\'{e}rsic profile fits. Since we are focusing only on the early-type dwarfs, S\'{e}rsics are generally quite good models for the surface brightness profiles. 

For each dwarf, we have two bands, either $g$ and $r$ or $g$ and $i$. We fit the S\'{e}rsic profiles in the manner of \citet{sbf_calib, carlsten2020a} using \textsc{imfit} \citep{imfit}. The $g$-band image is masked for nearby stars and background galaxies and fit with a S\'{e}rsic profile. This initial masking also includes point sources that are likely NSCs. The $r$ or $i$-band image is then fit, allowing only the amplitude to change while fixing other S\'{e}rsic parameters. 

We subtract this initial single component model from the image and inspect the residuals for possible NSCs. Following \citet{eigenthaler2018} and \citet{rsj2019}, NSCs are determined through a combination of photometry and visual inspection. %Following \citet{rsj2019}, the maximum allowed projected separation from an NSC candidate to the galaxy center is $160$ pc ($\sim5$\arcs at the median LV distance). We note that essentially all of the nuclei candidates are within $<100$ pc of the photocenter. 
We show below that the visual classification of NSCs is essentially a cut in whether there is a point source within a projected radius of $r\lesssim r_e/8$ of the galaxy's photocenter. While it is possible some of the NSC candidates are interloping foreground MW stars or compact background galaxies, we consider this possibility unlikely. \citet{rsj2019} estimate the contamination rate to be $\sim 0.1$\% for the NGVS sample, and we expect a similarly low contamination for the ELVES dwarfs as well. The median density of point sources of the magnitude and color expected for NSCs/GCs (see \ref{sec:gc_preliminaries} below) in the vicinity of the LV dwarf galaxies is $\sim2$ arcmin$^{-2}$. Compared to the median effective radius of the ELVES sample of $0.24$ arcmin, this means the chance of an interloping point source being in the central $r_e/8$ of a dwarf is $\sim0.5$\%, similar to the \citet{rsj2019} estimate. Similarly, we do not expect many NSCs to actually be (non-nuclear) GCs of the dwarf galaxy that are located in the center just by chance projection. A quick estimate indicates that a GC randomly located within $2r_e$ of a dwarf has a roughly 5\% chance of projecting into the inner $r_e/8$ along a random line of sight. Given that dwarfs in the mass range considered here have few, if any, GCs, this is not a large concern.

Once we identify an NSC candidate in the residuals to the single-S\'{e}rsic fit, we refit the galaxy (and nucleus) using a double-S\'{e}rsic profile fit. At the distances of the LV dwarfs, we do not expect the NSCs to be resolved in the ground-based data that we use. However, we find that the double-S\'{e}rsic fit leads to more stable fits than a S\'{e}rsic $+$ point source combination since it is more forgiving of errors in our PSF model (which simply comes from the cutouts of nearby, unsaturated stars). The second S\'{e}rsic profile is limited in effective radius to $r_e < 1$\arcs.

This process is exemplified in Figure \ref{fig:method}. This shows an example of a nucleated dwarf (top left) and the residuals after subtraction of the double-S\'{e}rsic model (top right).

To deal with the fact that sometimes we have $g/r$ and sometimes we have $g/i$ bands, we make use of the conversion 
\beq
g-i = 1.53 (g-r) - 0.032
\label{eq:gi_gr}
\eeq
which we derive from MIST \citep{mist_models} SSP models for a range in age between 3 and 10 Gyr and metallicities in the range $-2<\mathrm{[Fe/H]}<0$. Note that this is calculated specifically for the CFHT filter system.

We derive stellar masses of the dwarf satellites and NSCs from integrated luminosity and color using a color-$M_\star/L$ relation from \citet{into2013}, as described more in Carlsten et al. (submitted). We refer the reader to that paper for more information on how we derive uncertainties in the photometry. The NSC photometric measurements are given in Appendix \ref{app:photo_nsc}.

We estimate the point source completeness of the various imaging datasets we use when analyzing the GC systems, as discussed below. Suffice to say here that in almost all cases the 50\% completeness magnitude is $M_g > -6$ mag at the distance of the hosts surveyed. Considering the magnitudes expected for NSCs \citep[$M_g\gtrsim-7.5$ mag;][]{rsj2019}, we do not expect that we are missing many NSCs due to shallow imaging. We discuss the matter of completeness further in \S\ref{sec:results_nsc}.

\begin{figure}
\includegraphics[width=0.48\textwidth]{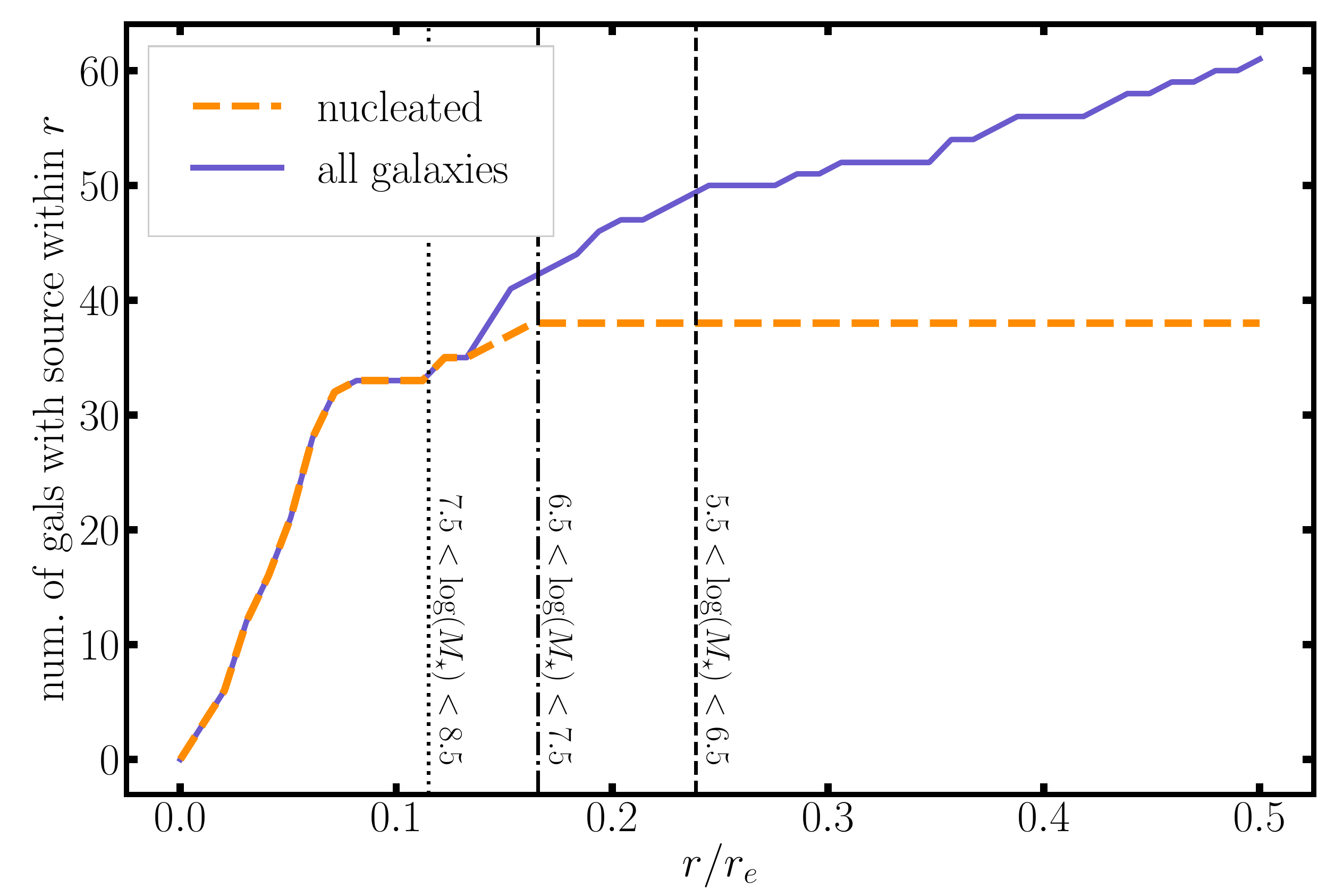}
\caption{Cumulative distribution of galaxies with a detected source within a certain fraction of the galaxy's effective radius. Galaxies classified as nucleated are shown separately from the overall sample of galaxies. The vertical lines show where a physical radius of 100 pc is, considering the average effective radii of galaxies in different mass bins. }
\label{fig:nuc_rads}
\end{figure}

\subsection{NSC Radial Distribution}
\label{sec:nsc_rads}
As stated above in \S\ref{sec:photo}, to be comparable to contemporary analyses of cluster dwarfs, the classification of nucleated vs non-nucleated galaxies is largely based on visual inspection. Below, we show that our nucleation categorization is essentially identical to that of \citet{rsj2019} for the NGVS Virgo sample. 

Figure \ref{fig:nuc_rads} shows the radial distribution (relative to the photometric center of the host galaxy) of nuclei in the LV sample. The cumulative number of galaxies with a source within a certain radius (scaled by the effective radius) is shown for both nucleated galaxies and the whole dwarf sample. The source catalogs used here are the ones input into the GC analysis, as described below in \S\ref{sec:color_cuts}. Any dwarf with a source within $<r_e/8$ is classified as nucleated. Thus, the visual classification of nuclei essentially is a cut on whether there is a point source within the inner $<r_e/8$ region of a dwarf. Since there are no non-nucleated galaxies with point sources within $<r_e/8$, we can quantify our completeness to nuclei using the completeness of the GC catalogs determined below in \S\ref{sec:catalog_completeness}.

\subsection{Catalogs of Candidate GCs}
\label{sec:gc_preliminaries}

With the galaxy profile modelled and subtracted (either a single or double-S\'{e}rsic model), we move to the analysis of the GC systems of these dwarfs. As mentioned above, we use two different methods of estimating the GC system of each galaxy. We know the distance to each dwarf through various methods,  but we have no knowledge of the distance to the various point sources surrounding the galaxies so we have to account for a significant fraction of background (and foreground) contamination. The first method we use to estimate the GC system is a simple, statistical background subtraction. This procedure has been used in several similar previous works \citep[e.g.][]{miller2007, lim2018}. The second method uses a more sophisticated Bayesian model fit to all dwarfs simultaneously. Both methods rely on the same point-source catalogs of eligible GC candidates in the vicinity of each dwarf. We describe the key steps in generating these catalogs in this section. 

\subsubsection{Point Source-Galaxy Separation}

The first step in counting the GC candidates is to procure a catalog of point sources in the surroundings of each dwarf. For point source-background galaxy separation, we measure the magnitude of sources in the image in two different diameter apertures, 6 and 12 pixels. Point sources, like stars and GC candidates, will appear as a distinct horizontal  locus when plotting the difference in the aperture magnitudes versus the total magnitude of sources, which we take from \texttt{SExtractor's} \texttt{MAG\_AUTO} measurement \citep{sextractor}. This is shown in the bottom left panel of Figure \ref{fig:method}. Point sources are those sources which are close to this horizontal locus. To determine the appropriate threshold in the aperture magnitude difference to use, we inject artificial stars into the images using the PSF model. The average recovered difference in the aperture magnitudes gives the horizontal location of the point source locus (red line in Figure \ref{fig:method}). The vertical width of the accepted region, shown by the green points, is determined from the $3\sigma$ spread (as a function of magnitude) in the injected stars. There is a floor in the width of $0.15$ mag. For the Virgo dwarf sample, we found that we needed a less restrictive point source-galaxy separation and so the floor in the width is $0.25$ mag. We do the artificial star injection in both bands but only do the star-galaxy separation in the redder band (either $r$ or $i$) since the PSF is, on average, smaller at longer wavelengths \citep[e.g.][]{scott_psf}.

\subsubsection{Point Source Completeness}
\label{sec:catalog_completeness}

Using the artificial star injection tests, we also derive the completeness of the source catalogs as a function of magnitude. We derive completeness curves for both bands, which are combined into an overall completeness by taking the lower recovery fraction of the two at each magnitude, assuming a fiducial color of $g-r=0.4$ or $g-i=0.75$. This overall completeness curve is shown in the bottom right of Figure \ref{fig:method}. The example dwarf shown in the Figure uses relatively deep CFHT/Megacam data but is the furthest dwarf in the LV sample ($D=11.9$ Mpc). For closer dwarfs, even with the relatively shallow DECaLS imaging, the $90$\% completeness limits are always $\gtrsim 1$ mag fainter than the expected GC luminosity function (GCLF) peak at the distance of the dwarfs \citep[$M_g\sim-7.2$ mag; ][]{jordan2007, villegas2010}.

\subsubsection{Color and Magnitude Cuts}
\label{sec:color_cuts}
We implement broad color cuts on the catalog of point sources that roughly encapsulate the expected range of GCs. We measure colors using the aperture photometry but with aperture correction derived from the artifical star injection process. For dwarfs with $g/r$ imaging, we restrict to sources with $g-r$ in the range $[0.1, 0.9]$ and for dwarfs with $g/i$ imaging, we restrict to sources with $g-i$ in the range $[0.2, 1.1]$. These ranges largely encompass the color spread of GCs in Fornax dwarfs found in \citet{prole2019} and the blue peak of GCs around massive Fornax galaxies found in \citet{dabrusco2016}. Note that since we are focusing just on dwarfs, we expect the GC systems to be fairly metal poor and, hence, blue. We have tried larger ranges in color but find that the results remain largely the same, albeit with lower signal to noise due to higher contamination levels. 

We also cut in luminosity, restricting sources to the range $-9.5<M_g<-5.5$ mag which will encompass the entire expected GCLF. 

These cuts are exemplified in Figure \ref{fig:method}. The green circles in the top right panel show all the point sources while the small red circles denote the sources that satisfy the color and magnitude cuts. 

We can then use these cleaned point-source catalogs to directly determine the overall stacked GCLF or GC radial profile (see below) or to determine an estimate for the number of GCs for each individual dwarf, as we describe in the next two sections.

\subsection{GC Abundance: Simple Background Subtraction}
\label{sec:bkgd_sub}
With the catalog of GC candidates in hand, our first method of inferring GC abundance is to perform a simple background subtraction to estimate the number of GCs around each dwarf. We count the number of sources within a circular annulus of radius $2r_e$. The count of sources within the annulus are then corrected for an estimated background contribution determined from separate $2r_e$ annuli placed around the galaxy at distances $>5r_e$ from the galaxy center. Note that the image cutouts we use are $>20r_e$ on a side so we generally get $>20$ background annuli per galaxy. The distribution of number counts in the background annuli are used to estimate the uncertainty of the background subtracted GC counts. We have checked that using $1r_e$ or $3r_e$ annuli instead do not alter the main conclusions of this work. With a larger annulus, the GC abundance measurements are generally of lower signal to noise. 

The main systematic uncertainty involved in this calculation of $N_{\mathrm{GC}}$ is the unknown radial distribution of GCs for low-mass dwarf galaxies. While it would be possible to count up the number of sources in a much larger aperture ($5 r_e$, for instance) that would capture essentially all of the GCs, larger apertures mean more background sources. Even if we subtract out correspondingly larger estimate of the background contribution, the increased Poisson scatter will cause the measurement to have lower signal to noise. A relatively common approach in the literature is to assume the half-number radius of the GC distribution is $\sim1.5\times$ the galaxy effective radius \citep[e.g.][]{pvd_2017_gcs, lim2018}. Various studies of dwarfs in clusters have found that the radial distribution of GCs in dwarf galaxies is generally less concentrated than the galaxy light profile \citep{amorisco2018, prole2019}. This result has also been found for individual, well-studied dwarfs, including the Fornax dSph \citep{mackey2003}, NGC 147/185 \citep{veljanoski2013}, NGC 6822 \citep{huxor2013, veljanoski2015}, VCC 1087 in Virgo \citep{beasley2006},  VCC 1287 in Virgo \citep{beasley2016}, and DF17 in Coma \citep{peng2016}. 

For the fiducial background subtracted results, we assume that the GC radial distribution is approximated by a Plummer profile with the form:
\beq
\Sigma_{\rm plum}(r;\; r_{\rm GC}) = \frac{1}{\pi}\frac{1}{r_{\rm GC}^2(1+r^2/r_{\rm GC}^2)^2}
\eeq
with half-number radius of $r_{\mathrm{GC}}=1.5\times r_e$. We show this assumption agrees fairly well with the results from the likelihood-based method discussed below, although with that method, we find a slightly smaller value of $r_{\mathrm{GC}}\sim1.2\times r_e$ is more appropriate. In the end, this assumption does not affect the main conclusions since we assume the same distribution for the NGVS comparison sample, and so the cluster-to-LV environment comparison is unchanged\footnote{This assertion does assume that the radial distribution of GCs is the same for dwarfs in clusters as it is for dwarfs in LV-like environments. While increased tidal stripping due to the cluster environment could lead to more extended GC systems \citep{smith2013}, we do not expect the average cluster dwarf to have experienced much stripping \citep[e.g.][]{smith2015}.}. 

Therefore, we correct the background subtracted results for GCs outside of the $2r_e$ aperture simply by integrating the Plummer profile from $2r_e$ to infinity. For the fiducial choice of $r_{\mathrm{GC}}=1.5\times r_e$, this corresponds to increasing the background subtracted $N_{\mathrm{GC}}$ by about $50$\%. Note that a different assumed value of $r_{\mathrm{GC}}$ would correspond to a simple scaling up or down of $N_{\mathrm{GC}}$. At this stage, we also correct for possible luminosity incompleteness in the GC sample by integrating over the expected GCLF multiplied by the completeness function of the data as determined above. We take the GCLF to be a Gaussian with mean of $M_g=-7.2$ mag and dispersion of $0.7$ mag \citep{jordan2007, villegas2010}. We show below that this is quite close to the GCLF properties that we measure for the LV and Virgo samples. As discussed above, the data generally extends to multiple magnitudes below the GCLF so these corrections are small, $\lesssim5$\%.

\subsection{GC Abundance: Likelihood Method}
\label{sec:gc_meth_lik}
We complement the simple background subtraction method with a more sophisticated likelihood-based inference. This method is able to take into account the magnitude and spatial distribution of candidate GCs, allowing for a more precise estimate of GC abundance than the simple background subtraction. Our method is inspired by the methods used by \citet{amorisco2018} and \citet{prole2019}, with the primary difference being that we fit all dwarfs simultaneously. In the mass range studied in this work, dwarfs do not have many (if any) GCs individually. Thus, we fit for underlying, average properties of the GC systems for an ensemble of dwarfs. 

We model the dwarfs as a mixture of systems with no GCs and systems with a non-zero GC population. Thus, we primarily fit for the fraction of dwarfs with GCs and for a relation between average number of GCs (for galaxies with GCs) and dwarf stellar mass. 

We start this procedure with the color restricted catalog of point sources in the vicinity of each dwarf (Section \ref{sec:gc_preliminaries} above). The likelihood that a source (indexed by $i$) of a dwarf galaxy (indexed by $j$) is a GC is

\beq 
\mathcal{L}^{i,j}_{\rm GC} \propto N_{\rm GC}(M_\star^j)\;C^j_\mathrm{mag}\; \Sigma_{\rm plum}(r^i;\; r_{\rm GC})\; \mathcal{N}(m_g^i;\;M_g^0,\;\sigma_g) \\
\eeq

where $r^i$ is the radial distance of the source (in units of $r_e$), $m_g^i$ is the source magnitude, and $\mathcal{N}$ is a Gaussian representing the GCLF with mean at the magnitude $M_g^{0}$ and dispersion $\sigma_g$, which are both also fit for in the procedure. $N_{\rm GC}(M_\star^j)$ is the expected number of GCs for a dwarf with a nonzero GC system based on the dwarf's stellar mass, $M_\star^j$. $C^j_\mathrm{mag}$ accounts for the incompleteness of the source catalogs and is simply

\beq 
C^j_\mathrm{mag} = \int S^j(m) \; \mathcal{N}(m;\;M_g^0,\;\sigma_g) \; dm
\eeq

Here,  $S^j(m)$ is the empirically determined overall completeness function of dwarf $j$ from \S\ref{sec:catalog_completeness}.

The likelihood a source is a contaminant is then

\beq 
\mathcal{L}^{i,j}_{\rm bk} \propto \frac{N^j_{\rm tot} - N_{\rm GC}(M_\star^j)\;C^j_\mathrm{mag}}{A^j} \; \mathcal{F}^j(m_g^i) \\
\eeq

where $N^j_{\rm tot}$ is the total number of sources in the cutout for galaxy $j$ which has area $A^j$, and $\mathcal{F}^j$ is an empirical estimate of the background/foreground source luminosity function constructed from sources beyond $2 r_e$ from the galaxy. 

We do not include color information in the inference. The background sources appear to have a broader distribution in color than the GCs, but the difference is not great. As mentioned earlier, we model the ensemble of dwarfs as a mixture between a sub-population with a nonzero GC system and sub-population without GCs. The overall likelihood for a given dwarf is then a mixture between having a GC system and not:

\beq 
\mathcal{L}^j \propto f^j \prod^i \left[ \mathcal{L}^{i,j}_{\rm GC} + \mathcal{L}^{i,j}_{\rm bk} \right] + (1 -  f^j) \prod^i \mathcal{L}^{i,j}_{\mathrm{bk},\;N_\mathrm{GC}=0}
\label{eq:likelihood}
\eeq

$f^j\equiv f(M_\star^j)$ is the fraction of dwarfs with a nonzero GC system at the stellar mass of the dwarf. The second product term comes from setting $N_\mathrm{GC}=0$ in the equations for $\mathcal{L}^{i,j}_{\rm GC}$ and $\mathcal{L}^{i,j}_{\rm bk}$.

The likelihood for the ensemble of dwarfs is then simply

\beq
\mathcal{L} = \prod^j \mathcal{L}^j
\eeq

The free parameters that are fit for are $N_\mathrm{GC}$, $f$, $r_\mathrm{GC}$, $M_g^0$, and $\sigma_g$.  $N_\mathrm{GC}$ and $f$ are both functions of stellar mass. For $N_\mathrm{GC}$, we parameterize it as a power law

\beq
N_\mathrm{GC}(M_\star) = 1 + T_0 \left( \frac{M_\star}{1\times10^9 M_\odot} \right) ^\alpha
\eeq

thus contributing two free parameters, $T_0$ and $\alpha$. Note there is the $N_\mathrm{GC}=1$ floor in the power law because $N_\mathrm{GC}$ is the average number of GCs for \textit{dwarfs with a nonzero GC system}. For the GC occupation fraction, instead of using a parameterized form (like an error function, for example), we fit for the occupation fraction at five stellar masses and interpolate it for intermediate masses. We fit for the fraction at stellar masses of $\log(M_\star)=[5.5 , 6.25, 7., 7.75, 8.5 ]$, thus contributing five free parameters. We enforce the occupation fraction to be an increasing function with stellar mass through the priors. With $M_g^0$ and $\sigma_g$, this makes 10 total free parameters. The parameters and their priors are listed in Table \ref{tab:likelihood_method}. All priors are taken to be uniform.

\begin{deluxetable}{cc}
\tablecaption{Summary of parameters fit for in the likelihood-based GC ensemble analysis. All priors are taken to be uniform \label{tab:likelihood_method}}
\tablehead{\colhead{Parameter} & \colhead{Range of prior}}
\startdata
$T_0$ & [1, 150] \\
$\alpha$ & [-0.1, 1.4] \\
$M_g^0$ & [-7.5, -6.5] \\
$\sigma_g$ & [0.2, 1.2] \\
$r_\mathrm{GC}$ & [0.25, 2.25] \\
$f_1$ & [0, $f_2$] \\
$f_2$ & [$f_1$, $f_3$] \\
$f_3$ & [$f_2$, $f_4$] \\
$f_4$ & [$f_3$, $f_5$] \\
$f_5$ & [$f_4$, 1] \\
\enddata
\end{deluxetable}

The posterior distributions are sampled via MCMC. The posterior distributions are visually inspected and all appear well-behaved and unimodal. The exceptions are $f_1$ and $f_5$ which, for both the LV and Virgo samples, are pushed against the allowed boundaries of 0 and 1, respectively.

\subsection{Image Simulations}
\label{sec:image_sims}
To test both the simple background subtraction and likelihood-based methods, we perform realistic, end-to-end image simulations of dwarfs with GC systems. Artificial dwarfs are generated in empty fields in the imaging data for three hosts that span the range of data quality and depth of the data used. Data for NGC 4258 and NGC 3379\footnote{In particular, DECaLS data is used for NGC 3379 in this test to represent a worst-case scenario. Most of the dwarfs of NGC 3379 are actually covered by extremely deep HSC data but that is not used in the simulations.} are used to demonstrate a best and worse-case scenario, respectively, for the ELVES sample. The third host included in the simulations is Virgo, to show that the methods also work at the somewhat further distance of Virgo.

We go into more detail of the inputs and recovery performance of the simulations in Appendix \ref{app:simulations}. Suffice to say here that we are able to robustly recover the input GC system parameters, including $N_\mathrm{GC}$ and the occupation fraction, within the estimated uncertainties from the posterior distributions. The simple background subtraction method is able to recover well the average overall trend in $N_\mathrm{GC}$. The simulations are run with $\sim200$ dwarfs, which is similar to the number in the observational datasets, although the simulated dwarfs are given, on average, more GCs than observed. This leads to more precise recovery than expected for the real datasets, but we show that the methods are unbiased in this higher-precision regime. 

Much of the previous work in the field of dwarf galaxy GC systems has been conducted with \emph{HST}. The resolving power of \emph{HST} makes it very useful for photometrically identifying GCs in nearby ($D<10$ Mpc) dwarf galaxies \textit{without the need for a background subtraction.} This is because the GCs are actually resolved as star clusters at these distances. However, through these image simulations and the agreement we get with previous work done with \emph{HST} (see below), we show that studying GCs is not infeasible with ground-based data.

\subsection{NSCs and GCs in the Local Group}
The satellites of the MW and M31 must be treated differently from the other ELVES hosts. Due to the complexities involved in measuring the GC systems, we only use the LG satellites for the nucleation analysis and not for the GC analysis, although we do compare with the LG GC abundances for a check on our method. We take photometry from \citet{mcconnachie2012}\footnote{Except the stellar mass of the disrupting Sgr dSph which we take from \citet{forbes2018_ngc_mhalo}.}. The nucleated LG early-type dwarfs are: Sgr, NGC 147, NGC 205 and M32. NGC 205 and M32 are above our upper mass limit of $10^{8.5}$\msun~and thus are not considered here. The NSC associated with Sgr is M54, and we adopt the stellar mass from \citet{baumgardt2018}. \citet{veljanoski2013} list a GC being at a projected separation of 30 pc ($\sim0.05r_e$) from the center of NGC 147, which would therefore classify as `nucleated' by the criterion used for the LV dwarfs. We take the stellar mass of this GC/NSC from \citet{veljanoski2013}, assuming $M_\star/L_V=1.5$.

%The only two MW early-type satellites with GCs are Sgr and Fornax. The GC abundance for Sgr is taken from \citet{law2010_gcs} and for Fornax is taken from \citet{mackey2003}. The M31 satellites with GCs (in our considered mass range) are NGC 147 and NGC 185\footnote{AndI has a single old cluster \citep{caldwell2017} but at $M_V\sim-3$ mag, it is unlike the GCs considered here. Additionally it would be undetected in almost all of the ELVES data.} the abundances of which are taken from \citet{veljanoski2013}.

\section{Nuclear Star Cluster Results}
\label{sec:results_nsc}

In this section and the next, we present the main results of our analysis of NSCs and GCs in LV dwarfs, starting with the NSC analysis. Throughout, we focus only on dwarfs in the mass range $5.5 < \log(M_\star/{\rm M}_\odot) < 8.5$. The lower limit is roughly the luminosity limit ($M_V=-9$ mag) of the survey, and the upper bound is set so that we avoid many of the highest mass satellites for which our analysis methods start to fail (e.g. they are invariably poorly fit by S\'{e}rsic profiles). There are also very few higher mass early-type dwarfs in MW-sized halos. In this mass range, we characterize 40 nuclei from 177 total LV early-type dwarfs, for an overall nucleation fraction of 0.23.

\begin{figure*}
\includegraphics[width=\textwidth]{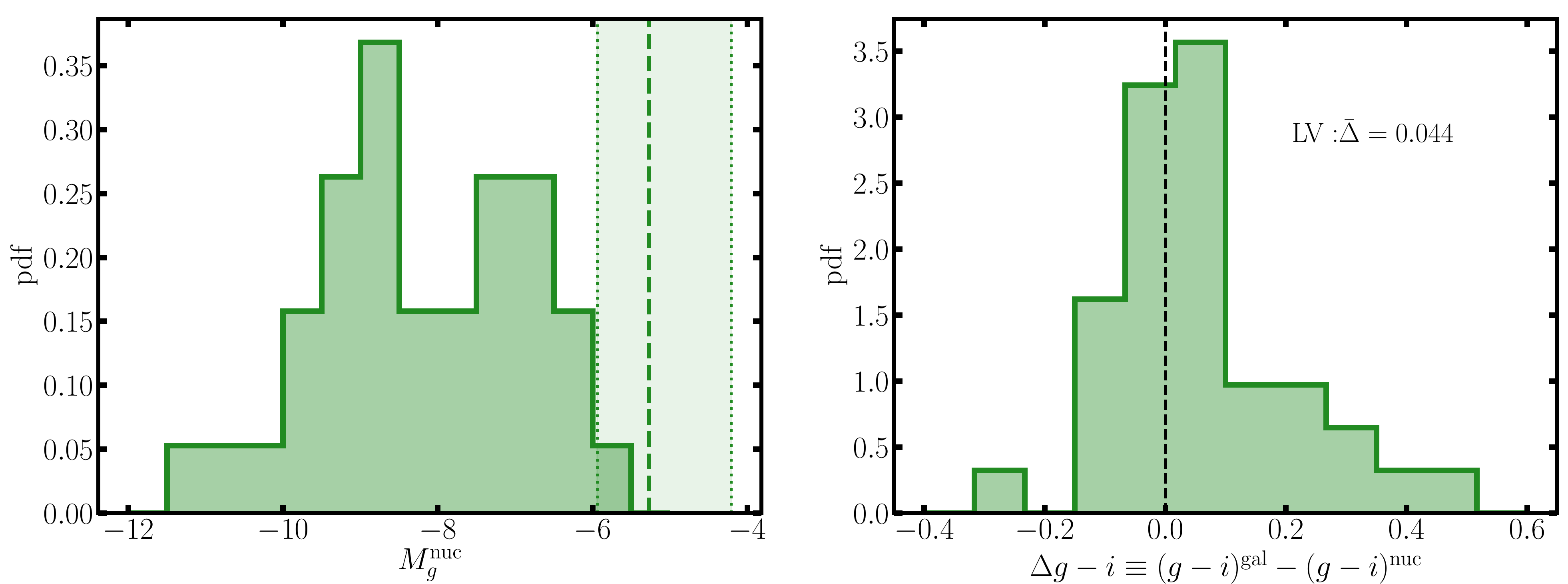}
\caption{NSC photometry for early-type dwarfs in the Local Volume. The left panel gives the absolute $g$ magnitude while the right shows the the color difference between the nucleus and galaxy. In general, nuclei are slightly bluer than their host galaxies. The vertical dashed lines in the left panel show the median 50\% point source completeness limit for the LV dwarfs, with the 16$^{th}$ and $84^{th}$ percentile range shown by the dotted lines.}
\label{fig:nuc_photo}
\end{figure*}

\subsection{NSC Luminosity and Color}
\label{sec:nsc_photo}
Figure \ref{fig:nuc_photo} gives the photometry for the NSCs; NSCs are generally bright with $M_g\sim-9$ mag. The right panel of Figure \ref{fig:nuc_photo} shows the colors of NSCs relative to their host galaxies. By normalizing the color by the host galaxy, we avoid most systematic issues related to the different filter systems used. For dwarfs with $g/r$ photometry, we multiply their colors by 1.53 (cf Equation \ref{eq:gi_gr}) to account for the longer baseline of $g-i$. The nuclei are, on average, slightly bluer than their host galaxies by about $\sim0.05$ mag, in line with the findings of \citet{ordenes-briceno2018_nuc_masses} and \citet{rsj2019}. 

Also shown in the left panel of Figure \ref{fig:nuc_photo} is the distribution of 50\% point source completeness limits for the LV dwarfs from the GC analysis. The data are generally sensitive to NSCs with $M_g>-6$ mag.

\begin{figure*}
\includegraphics[width=\textwidth]{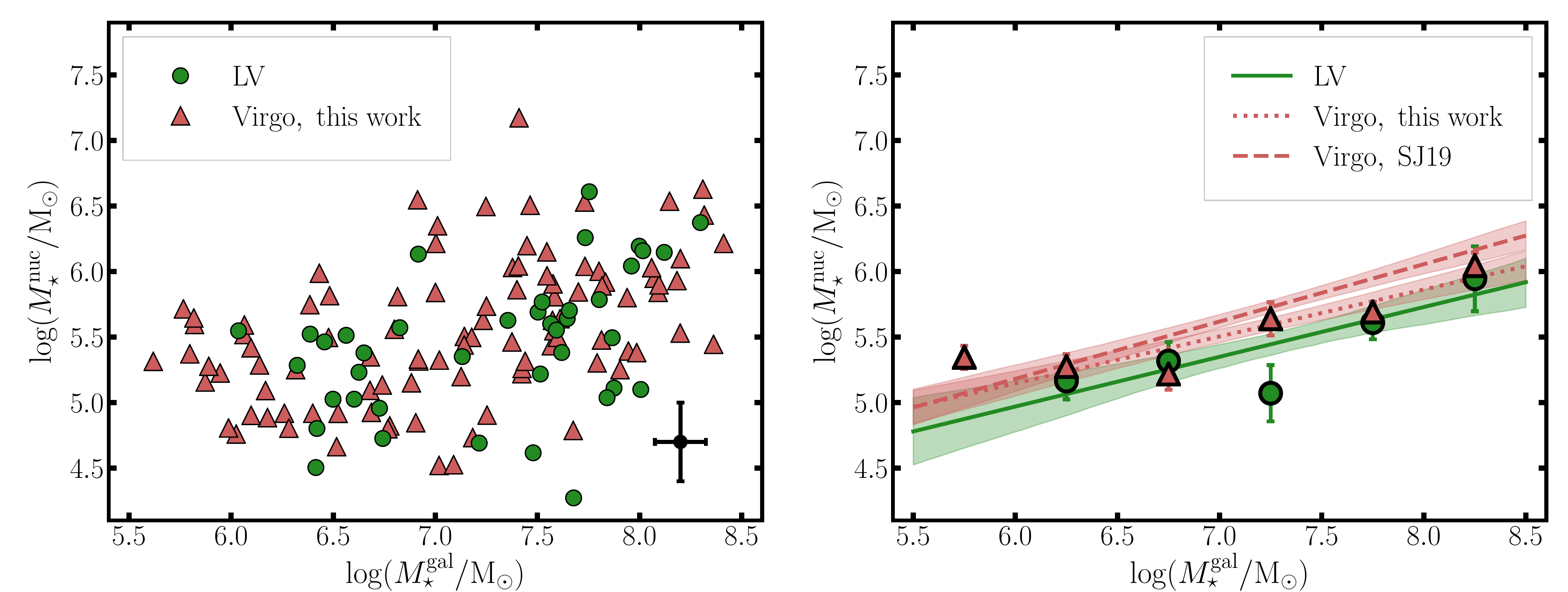}
\caption{NSC stellar mass versus host stellar mass for the LV and Virgo samples. The errorbar in the left panel shows the characteristic errors in stellar mass. The right panel shows power law fits for the two samples, along with a running average in 0.5 dex wide bins of stellar mass.}
\label{fig:nuc_mstar}
\end{figure*}

\subsection{NSC Scaling Relations}
\label{sec:nsc_mstar}
Figure \ref{fig:nuc_mstar} shows the correlation between NSC stellar mass and dwarf-host stellar mass. It compares the 40 LV dwarf nuclei to 98 Virgo dwarf nuclei found in our analysis of the NGVS sample. For this plot, we use our own photometry of the NGVS dwarfs and nuclei to make as direct a comparison as possible. The right panel shows characteristic errorbars for both stellar mass measurements. We assume a constant 0.3 dex uncertainty in the NSC stellar mass, which is the dispersion between the NSC masses we measure for the NGVS sample and those reported in \citet{rsj2019}. 

We fit power laws to the NSC-galaxy stellar mass relation using the \texttt{linmix} algorithm \citep{kelly2007}\footnote{We use the \texttt{python} implementation of \texttt{linmix} written by J. Meyers: \url{https://github.com/jmeyers314/linmix}} of the form:

\beq 
\log(M_\star^{\rm nuc}) = a + b \log(M_\star^{\rm gal}) \\
\eeq

with intrinsic scatter $\sigma$. The best-fitting values and marginalized uncertainties for the two samples are given in Table \ref{tab:nuc_gal_mstar}.

\begin{deluxetable}{cccc}
\tablecaption{Results of fitting the NSC-galaxy stellar mass relation for LV and Virgo dwarfs in the mass range $5.5 < \log(M_\star/{\rm M}_\odot) < 8.5$. $\sigma$ denotes the intrinsic scatter in size at fixed stellar mass.\label{tab:nuc_gal_mstar}}
\tablehead{\colhead{Sample} & \colhead{$a$} & \colhead{$b$} & \colhead{$\sigma$}}
\startdata
LV & $2.68^{+1.05}_{-1.01}$ & $0.38^{+0.14}_{-0.14}$ & $0.41^{+0.08}_{-0.07}$ \\
Virgo & $3.0^{+0.5}_{-0.5}$ & $0.36^{+0.07}_{-0.07}$ & $0.37^{+0.05}_{-0.04}$ \\
\enddata
\end{deluxetable}

Both samples give relations somewhat shallower than the oft-found $M_{\rm nuc} \propto M_{\rm gal}^{0.5}$, although they are consistent within the uncertainties.  We also show a linear fit to the Virgo photometry of \citet{rsj2019} which shows fair agreement with our Virgo results. Our fit to the \citet{rsj2019} photometry gives a power law slope of $b=0.44^{+0.08}_{-0.08}$, which is a little steeper than what our Virgo photometry yields. We note that we use different prescriptions for calculating stellar masses than done in \citet{rsj2019} although that shouldn't cause an offset in this plot. It is possible that our Virgo results disagree from the \citet{rsj2019} results at high galaxy masses because our method of modelling these galaxies as Nucleus$+$S\'{e}rsic is inadequate for many massive dwarfs which often have more complicated brightness profiles. There appears to be some evidence that LV nuclei have lower stellar mass on average at fixed host mass than the Virgo galaxies, although the discrepancy is only at the $\sim1-2\sigma$ level. The intrinsic scatter in the NSC-galaxy stellar mass relation is similar between both environments.

\begin{figure*}
\includegraphics[width=\textwidth]{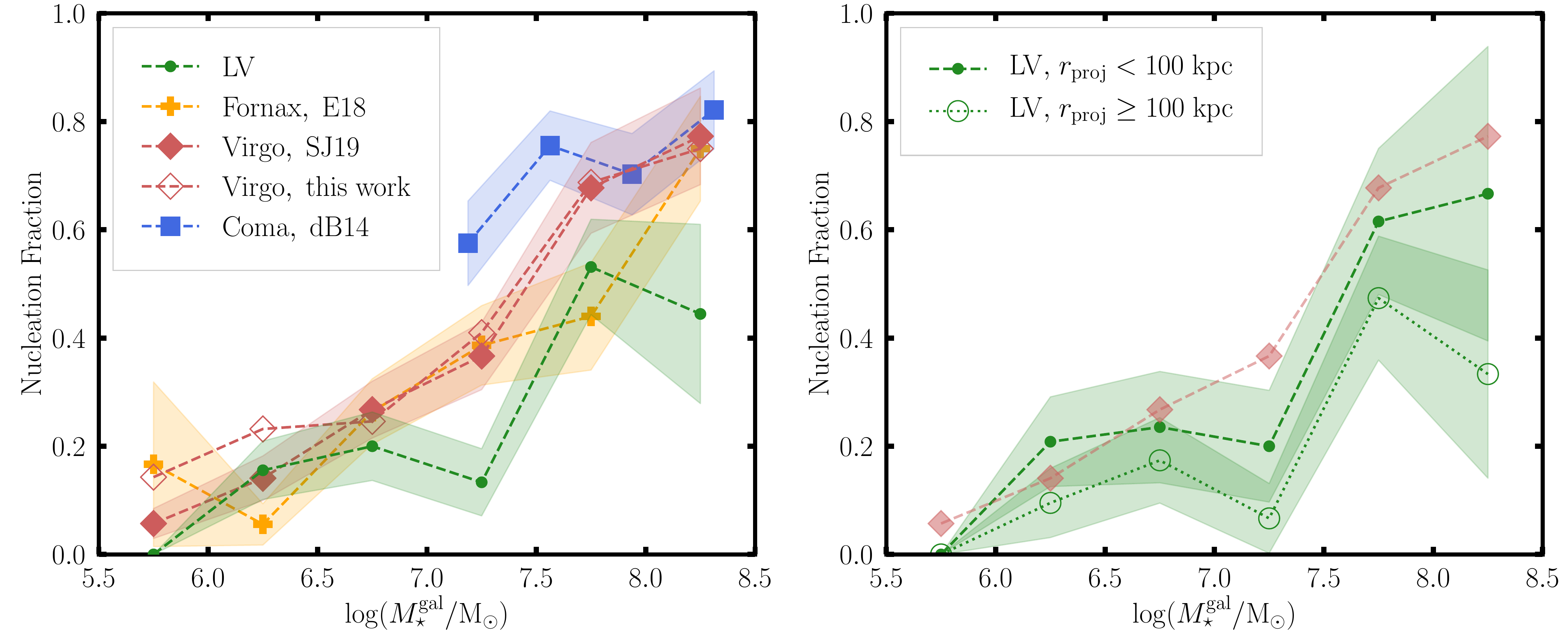}
\caption{\textit{Left}: Nucleation fraction of early-type dwarfs as a function of stellar mass in different environments in the local universe. The Virgo results are from \citet{rsj2019}, although we also show the results from our parallel analysis of the NGVS sample. The Fornax results are from \citet{eigenthaler2018} and Coma from \citet{denbrok2014}. \textit{Right}: The nucleation fraction of LV satellite dwarfs split by their projected separation from their massive host galaxy. The Virgo results are included in this panel for reference. In both panels, there is a clear trend in environment where early-type dwarf satellites in denser environments (either more massive parent halo or closer to their host) are more frequently nucleated at fixed stellar mass.}
\label{fig:nuc_frac}
\end{figure*}

\subsection{NSC Abundance}
\label{sec:nuc_frac}
Figure \ref{fig:nuc_frac} shows the nucleation fraction of dwarfs as a function of stellar mass in the different environments where comparable samples are available. For Virgo, we include both the results directly from \citet{rsj2019} and also from our analysis of the NGVS dwarf sample. They agree quite well, indicating that many systematic concerns in this process, including the visual inspection of candidate nuclei and the determination of stellar masses, have a minor effect. The nucleation fraction from our Virgo analysis is a little higher at the lowest stellar mass bins. This difference is due to 2-3 low-mass dwarfs, which we classify as nucleated but \citet{rsj2019} did not. 

We also show the results for the inner $r\lesssim R_\mathrm{vir}/4$ area of Fornax from the Next Generation Fornax Survey \citet{eigenthaler2018} and for the inner $r\lesssim R_\mathrm{vir}/6$ area of Coma from \citet{denbrok2014}. We describe how we get stellar masses for these dwarf samples in Appendix \ref{app:stellar_masses}.

These results conclusively confirm the result found by \citet{rsj2019} that nucleation fraction of early-type dwarfs has a significant secondary dependence on environment. \citet{rsj2019} just considered early-type dwarfs in the LG and M81 systems while our sample size is nearly an order of magnitude greater. The trend of lower density environments being less conducive to the formation of nuclei extends all the way down to MW-sized halos.

Very recently, \citet{zanatta2021} also confirmed this environmental trend comparing satellites in different mass clusters and lower mass, LV hosts. For the latter, they used satellites partly from the surveys of \citet{carlsten2020a, carlsten2020b}. We have greatly increased the statistics for LV satellites since those works. Additionally, \citet{zanatta2021} include all confirmed and `possible' satellites (candidate satellites where the SBF distances were inconclusive) from those works. From our experience with acquiring deeper follow-up, we suspect roughly half of the latter class are background objects. Since these background objects will likely not have detectable nuclei, their inclusion will lower the average nucleation fraction somewhat. In this work, we only consider fully confirmed satellites and show that, among these objects, the nucleation fraction in LV hosts is still quite a bit less than in the galaxy clusters. We discuss what this finding means for the formation of NSCs below in the Discussion section.

Interestingly, in the right panel of Figure \ref{fig:nuc_frac}, we show that there is an environmental dependence even \textit{within} the LV sample. We split the ELVES sample between dwarfs within 100 projected kpc of their host and those outside. This radius roughly splits the sample into halves. There is a surprisingly consistent trend of lower nucleation fraction for the outer LV dwarfs.

While the inner LV nucleation fractions are still slightly below that of the Virgo sample, the difference is much less. This brings up the intriguing possibility that the difference seen in the left panel and reported by \citet{rsj2019} and \citet{zanatta2021} is actually more of an effect of the cluster samples being very centralized and less due to the parent halo mass in which the dwarfs reside. Unfortunately we do not have the dwarf samples to conclusively show which affect (parent halo mass or location of dwarf within parent halo) is dominant. The full NGVS (that covers the whole virial volume) and final ELVES dwarf samples will be critical in addressing this. We do note that \citet{ordenes-briceno_spatial} presented the Fornax dwarf population over a more extended radial range ($R_\mathrm{vir}/4 < r < R_\mathrm{vir}/2$). The dwarf sample in that radial range had a similar overall nucleation fraction of $106/384=0.28$ as that reported for the inner population \citep[$r < R_\mathrm{vir}/4$;][]{eigenthaler2018} of $75/258=0.29$. We show in Appendix \ref{app:fornax_nuc} that the nucleation fractions at fixed stellar mass are similar between the two radial ranges, hinting that the effect is primarily driven by parent halo mass.

\begin{figure*}
\includegraphics[width=\textwidth]{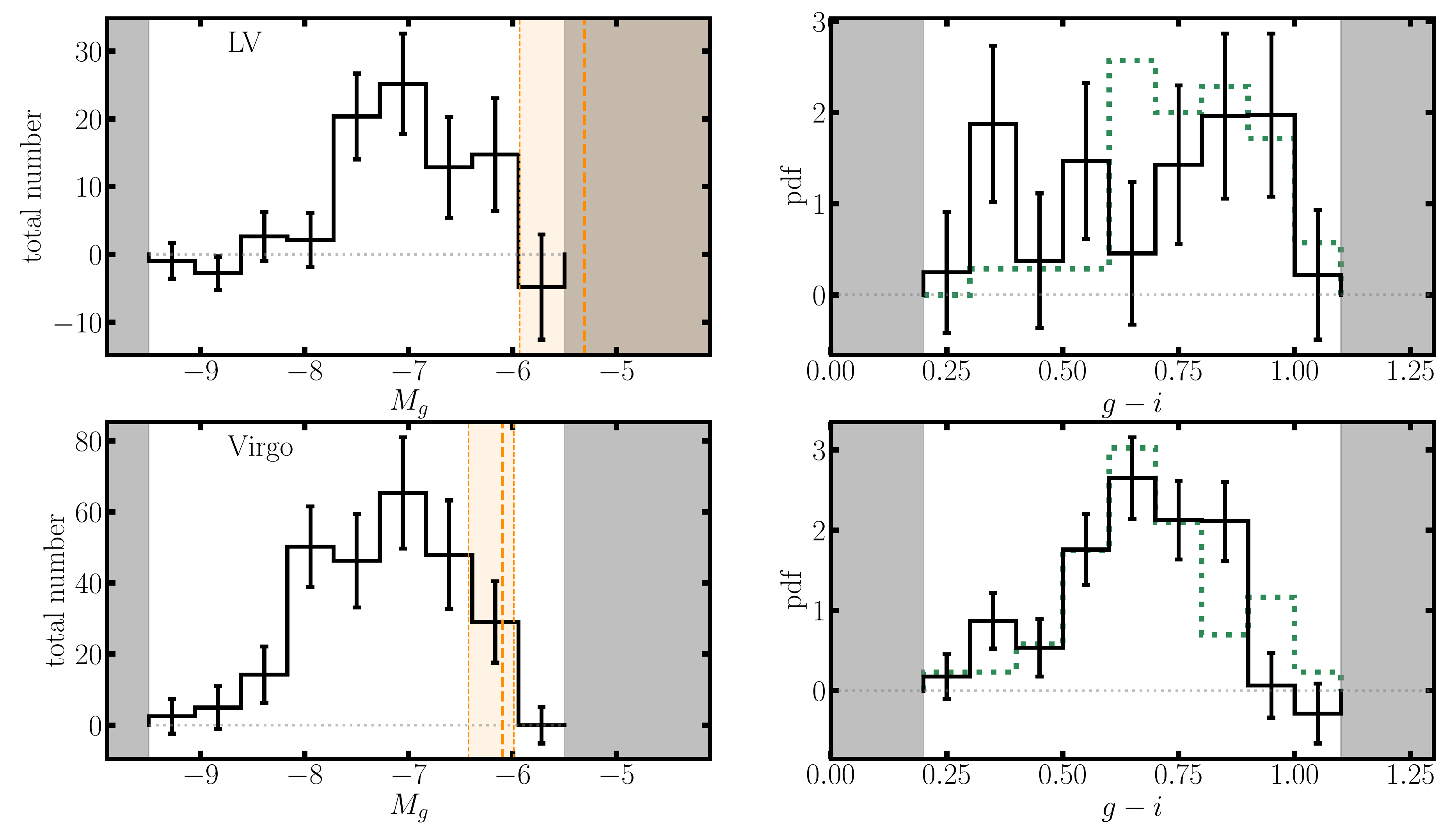}
\caption{Stacked, background subtracted GC luminosity functions and color distributions for GC candidates in the vicinity of the dwarfs. The top shows our results for the LV sample while the bottom shows the analogous results for the Virgo comparison sample. Errorbars show the Poisson uncertainties considering the overall (including background) number of sources in each magnitude bin. The gray areas denote the magnitude and color regions outside of the cuts made on the catalog. The vertical orange line and shaded region denote the 16$^{th}$, 50$^{th}$, and 84$^{th}$ percentile of the 50\% completeness magnitudes for the different dwarfs. In almost all cases, it is more than a magnitude fainter than the GCLF peak. The right panels also show the distributions of nuclei color for each environmental sample as the green, dotted histograms.}
\label{fig:gclf}
\end{figure*}

\section{GC Results}
\label{sec:results_gcs}
In this section, we present the results on the GC analysis of the LV dwarf sample.  We start with a discussion of the observed GCLF, then move to the radial distributions of GCs, and then, finally, present the abundance of GCs as a function of stellar mass with particular focus on a comparison across environments.

\subsection{GC Luminosity Function}
\label{sec:gclf}

Figure \ref{fig:gclf} shows the overall stacked, background-subtracted GCLF and color distributions for the LV and Virgo samples. To make these distributions, we stack the background subtracted histograms for each individual dwarf for sources within a circular $2r_e$ annulus. Note the histograms are not corrected for possible GCs outside of $2r_e$. The errorbars are Poisson uncertainties given the combined total number of sources (before background subtraction) in the $2r_e$ annuli. The GCLF's both show a peak at $M_g\sim-7.2$ mag, which we explore in more detail below. The color distribution for the Virgo dwarfs shows a single peak around $g-i\sim0.7$, similar to the findings of \citet{prole2019} and corresponding to the color range of the metal-poor subpopulation of GCs found around more massive galaxies \citep[e.g.][]{peng2006}. The color distribution for GCs of the LV dwarfs does not show a clear peak, likely due to limited statistics.

\begin{figure}
\includegraphics[width=0.46\textwidth]{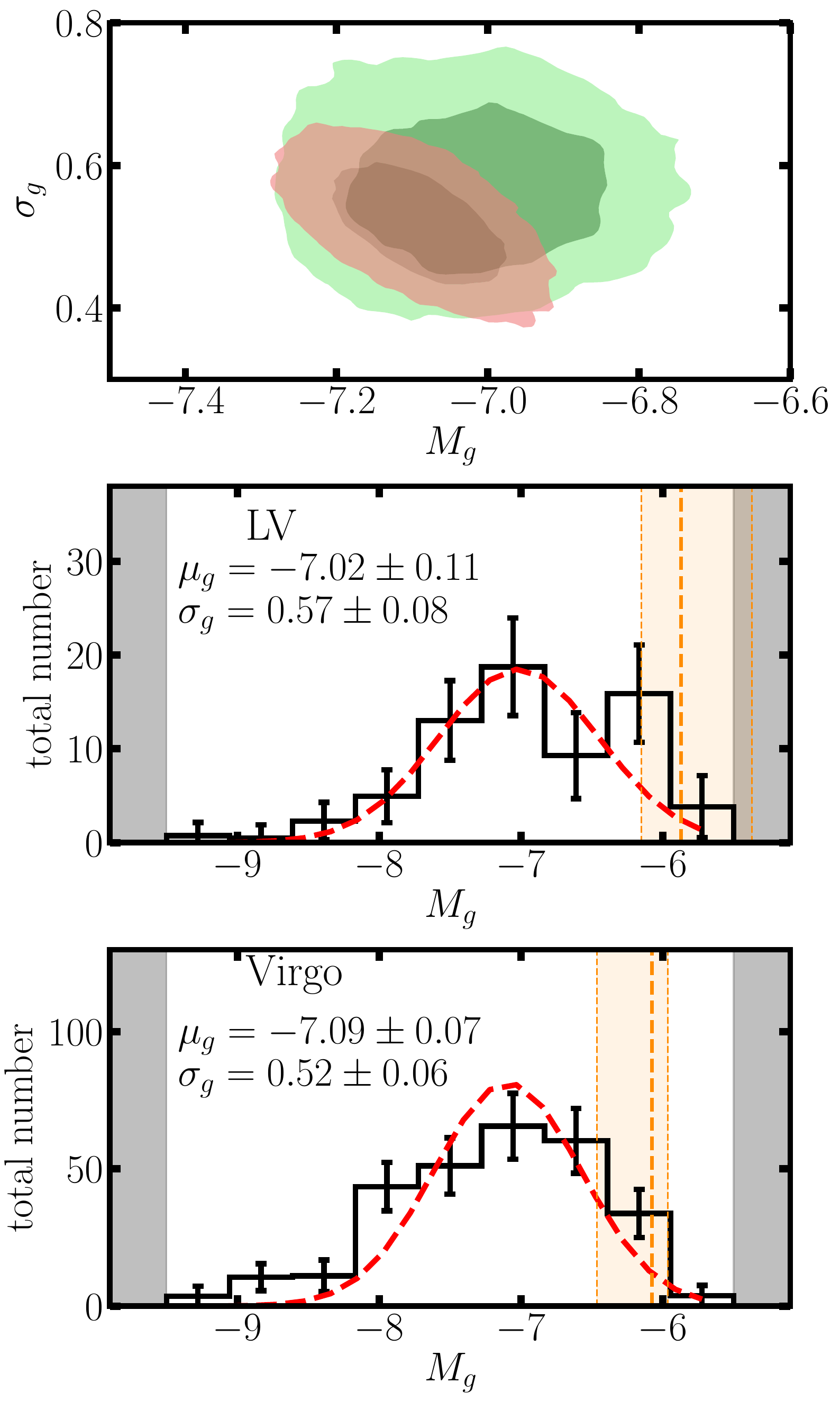}
\caption{\textit{Top}: Contours showing the 1$\sigma$ and 2$\sigma$ confidence regions from the posterior distribution for the GCLF parameters (peak magnitude and dispersion). Green is for the LV sample, red for Virgo. \textit{Middle \& Bottom}: Background subtracted GCLFs restricted to dwarfs with significant GC systems ($N_{\rm GC}>5$). Gaussian GCLFs corresponding to the median parameters from the posterior distributions are shown in dashed red.}
\label{fig:gclf_restricted}
\end{figure}

To get better signal-to-noise on the background subtracted LF, in Figure \ref{fig:gclf_restricted}, we show the background subtracted GCLF but restricted only to dwarfs with $N_{\rm GC}>5$ (as determined by the simple background subtraction method of \S\ref{sec:bkgd_sub}). We show the posteriors of the GCLF parameters from the likelihood-based fit in the top panel and GCLF curves corresponding to the median posterior values in the bottom two panels. The GCLF parameters are similar for both dwarf samples with  $M_g\sim-7.1$ and dispersions of $\sim0.6$ mag. \citet{jordan2007} and \citet{villegas2010} both show that the GCLF dispersion decreases for lower luminosity galaxies, at least for galaxies brighter than $M_B\lesssim-16$ mag. Using Equation 18 of \citet{jordan2007} and assuming $\langle M_B \rangle \sim -12$ mag for our samples, we'd expect $\mu_g\sim0.34$ mag, smaller than the observed $\sim0.6$ mag. This indicates that the dwarf luminosity-GCLF width relation likely flattens out around $M_B\sim-16$ mag, the faintest magnitudes probed by \citet{jordan2007}.

\begin{figure}
\includegraphics[width=0.46\textwidth]{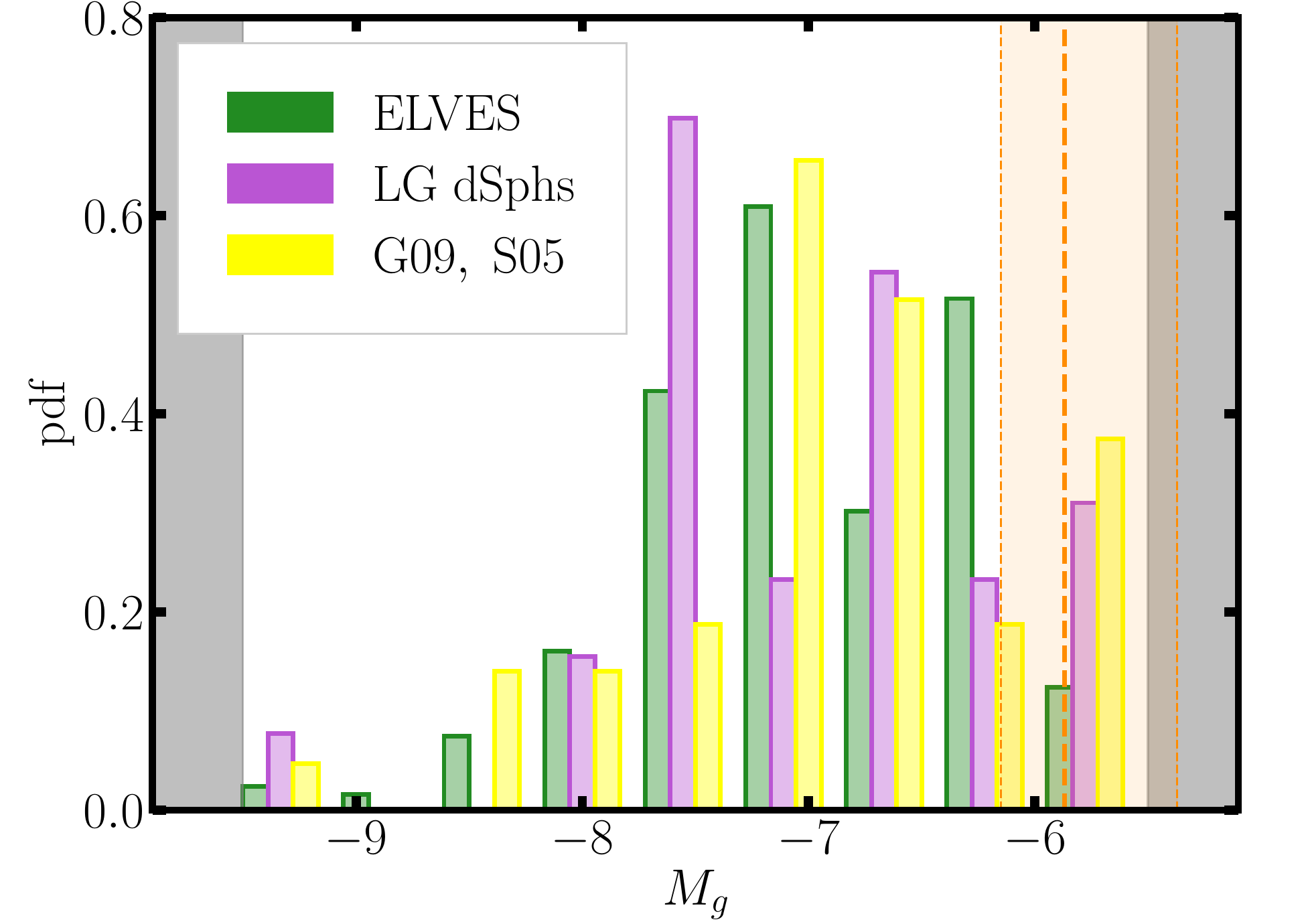}
\caption{The LV GCLF from Figure \ref{fig:gclf_restricted} compared to the GCLF for Local Group early-type satellites and nearby ($D<6$ Mpc) early-type dwarfs from \citet{georgiev2009a} and \citet{sharina2005}. The samples show a pronounced tail towards fainter luminosities. }
\label{fig:gclf_lg}
\end{figure}

The LV dwarf GCLF shown in Figure \ref{fig:gclf_restricted} exhibits a hint of a  non-Gaussian tail extending to faint luminosities. The tail even extends into the magnitude range where some of the data used will be incomplete ($M_g\gtrsim-6$ mag) and is thus a lower limit to the true LF at those magnitudes. To explore this in more detail, in Figure \ref{fig:gclf_lg}, we compare the LV dwarf GCLF from Figure \ref{fig:gclf_restricted} with two other samples of nearby dwarfs. First is the sample of GCs in LG dSphs. We include GCs of NGC 147, NGC 185, NGC 205, Sagittarius, and Fornax. The luminosities of the GCs of M31 dwarfs come from \citet{dacosta1988} and \citet{veljanoski2013}. The list of Sgr GCs comes from \citet{law2010_gcs} with luminosities from \citet{harris1996}. The properties of the Fornax GCs come from \citet{mackey2003}. The second sample of GCs in dwarfs that we include is the combined sample of nearby ($D\lesssim6$ Mpc) dwarfs of \citet{sharina2005} and \citet{georgiev2009a}. Only early-type dwarfs are included. These are all satellites of nearby LV massive hosts. Note, therefore, that many of the \citet{sharina2005} and \citet{georgiev2009a} sample galaxies are included in ELVES, but those works used \emph{HST} to identify GCs (thus obviating the need for a background subtraction) and are still an important comparison to our results. We assume $M_g\approx M_V + 0.2$ for the GCs.

Both of these other samples show remarkably similar GCLF shapes to the ELVES result, including a tail to faint luminosities. Indeed, \citet{sharina2005} comment on this surplus of faint GCs as they observe it extending down to $M\sim-5$ mag, the completeness limit of that work. It is unclear if the dwarfs in the Virgo comparison sample also show this GCLF shape since the completeness limits of the Virgo data are not as faint. An in-depth investigation into this GCLF shape and if it is actually different to the more symmetric and Gaussian GCLFs observed for more massive cluster early-type dwarfs \citep[e.g.][]{miller2007,jordan2007} would be an interesting future venture, particularly after some of the ELVES GCs are confirmed spectroscopically.

\begin{figure*}
\includegraphics[width=\textwidth]{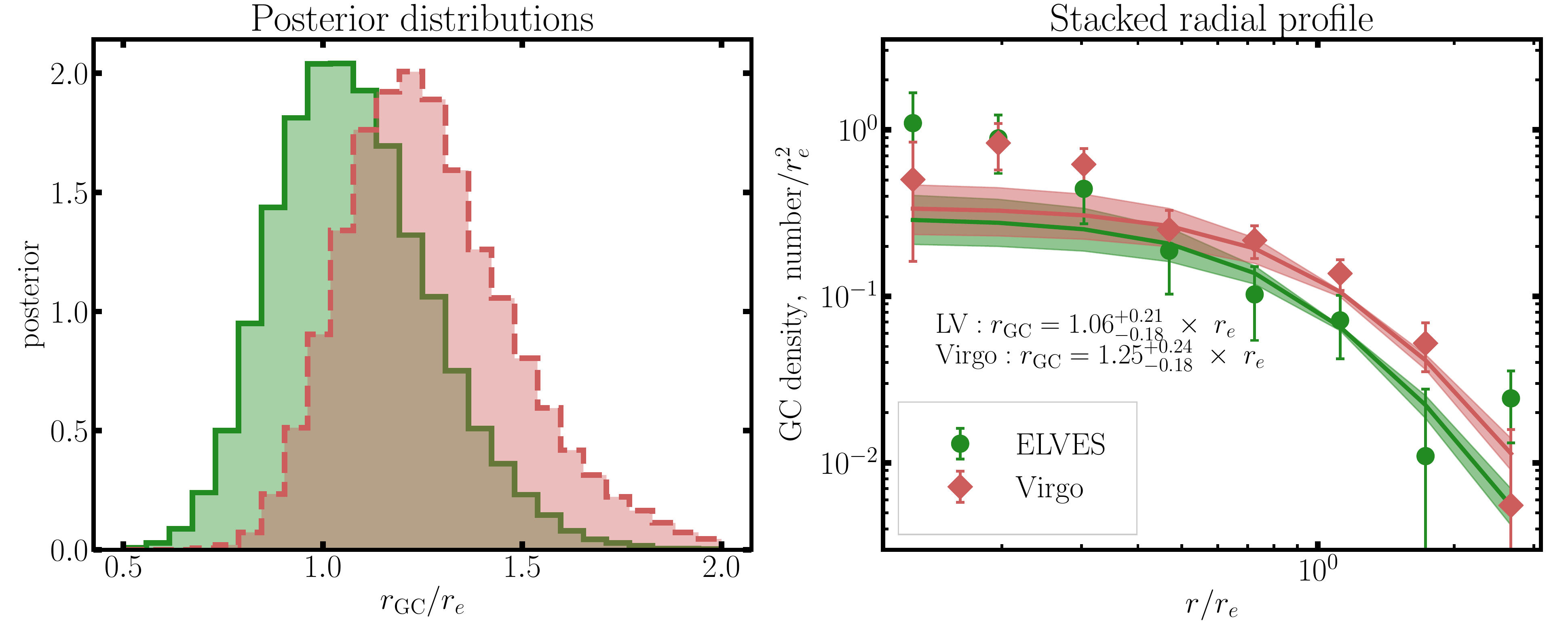}
\caption{\textit{Left}: Posterior distributions from the fit for $r_\mathrm{GC}$, the GC system half-number radius, from the combined likelihood-based analysis. \textit{Right}: Stacked average radial density profiles of the GC systems. The errorbars are Poisson based on the total number of sources (before background-subtraction) in each radial annulus.  The values for $r_\mathrm{GC}$ are from the posterior distributions in the left panel. Plummer profiles with $r_\mathrm{GC}$ drawn from the posterior distributions are also shown.}
\label{fig:rads}
\end{figure*}

\begin{figure}
\includegraphics[width=0.46\textwidth]{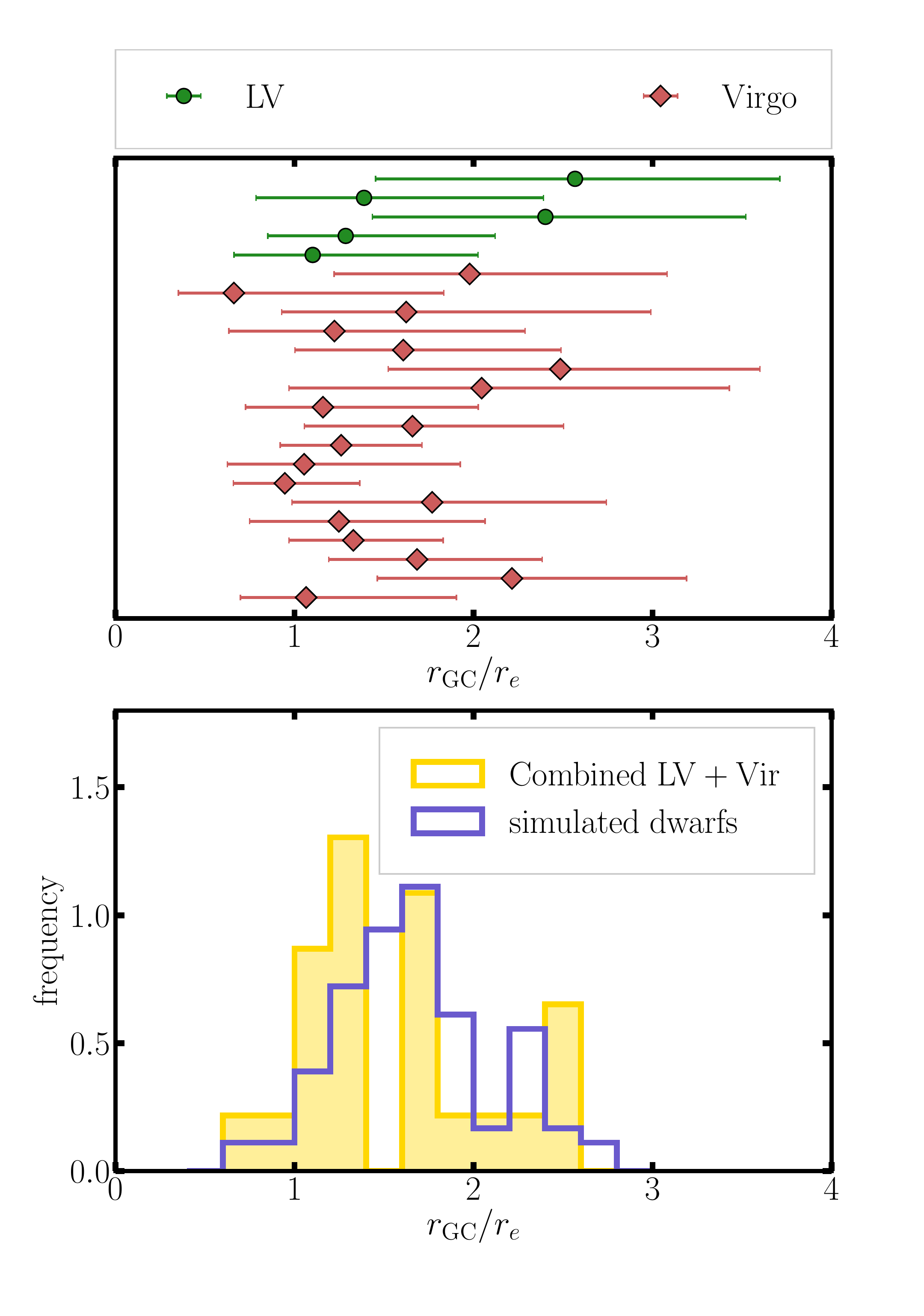}
\caption{\textit{Top}: Constraints on the GC half-number radius, $r_\mathrm{GC}$, from fitting individual dwarfs with rich GC systems (see text for definition). The dwarfs are ordered in terms of increasing stellar mass within each sample. The constraints on $r_\mathrm{GC}$ from most of the individual dwarfs are not tight but almost always $>r_e$. \textit{Bottom}: A histogram showing the constraints on $r_\mathrm{GC}$ from combining the LV and Virgo dwarfs from the top panel. Also shown are the results from fitting for $r_\mathrm{GC}$ using simulated dwarfs with similar GC abundance as the real dwarfs (see \S\ref{sec:image_sims}). The simulations were generated assuming $r_\mathrm{GC}=1.5\times r_e$. }
\label{fig:rad_dists_single}
\end{figure}

\subsection{GC Radial Profiles}
\label{sec:rad_profs}

As discussed in the Methods section, there is significant uncertainty in the radial profiles of the GC systems of dwarf galaxies. The radial profile of dwarf GC systems is a very interesting observable, both from the viewpoint of understanding the formation of the GC system and also the NSCs as they likely formed, at least in part, through GC inspiral. We return to this point in the Discussion section below. Also, the GC radial distribution influences the total GC abundance estimates since we had to assume a certain radial distribution to extrapolate from number counts in an annulus. %In this section, we explore the results for $r_\mathrm{GC}$ from the likelihood-based analysis.

Figure \ref{fig:rads} shows the posterior distributions for $r_\mathrm{GC}$ from the combined fits of all dwarfs in each of the LV and Virgo samples. Unfortunately, neither sample has the statistics to provide tight constraints on the radial distribution, although $r_\mathrm{GC}>r_e$ is mildly preferred for both the LV and Virgo samples. The distribution for the two samples is similar. 

In the right panel of Figure \ref{fig:rads}, we show the average binned radial profile of GCs. For these stacked profiles, we start with the color and magnitude restricted catalogs of candidate GCs from \S\ref{sec:color_cuts}. We then calculate the number density of sources in radial annuli around each dwarf. These curves are background subtracted by calculating the density of background sources outside of $5r_e$ for each dwarf. The radial profiles are scaled by each galaxy's half-light radius and the densities are similarly normalized by $r_e$. To improve the signal-to-noise of the stack, we include only dwarfs in the mass range $6.5 < \log(M_\star/{\rm M}_\odot) < 8.5$.  We also show Plummer profiles with $r_\mathrm{GC}$ values drawn from the posterior distributions.

Several of the dwarfs have reasonably rich GC systems, and it should be feasible to fit for $r_\mathrm{GC}$ for individual dwarfs. To do this, we essentially do the likelihood-based abundance inference but with just a single dwarf. We fix the GCLF parameters to $M_g^0=-7.2$ mag and $\sigma_g=0.7$ mag and assume the dwarf has a non-zero GC system. For this process, we select dwarfs that show significant GC systems from the simple background subtraction results. In particular, we select dwarfs with a background-corrected number of GC candidates greater than 3 in a $1r_e$ annulus and 5 in a $2r_e$ annulus. Either the $1r_e$ or $2r_e$ annulus number counts must also be $>2\sigma$ above the background as estimated by annuli placed in surrounding regions. When corrected for GCs outside of these annuli, the selected dwarfs generally have total $N_\mathrm{GC}\sim7-15$. 

Figure \ref{fig:rad_dists_single} shows the results of these individual fits. The top panel shows the constraints from each individual dwarf, ordered by increasing dwarf stellar mass within each environmental sample. Even among the most GC-rich dwarfs in the sample, the constraint on $r_\mathrm{GC}$ is relatively weak. Still, almost all of the dwarfs show $r_\mathrm{GC}>r_e$. The bottom panel shows the combined histogram of all the $r_\mathrm{GC}$ constraints compared to the $r_\mathrm{GC}$ constraints from individually fitting simulated galaxies (see \S\ref{sec:image_sims}) selected by the same abundance criteria. The simulated galaxies are generated with $r_\mathrm{GC}=1.5\times r_e$. The similarity between the $r_\mathrm{GC}$ results of the real and simulated galaxies indicate that the underlying $r_\mathrm{GC}$ for the real galaxies is $>r_e$, at least for the GC-rich dwarfs.

%Overall, the LV and Virgo dwarf samples do not show a statistically significant difference in $r_{\rm GC}$ between the two environments. Tidal stripping in the severe cluster environment could puff up a GC system, increasing its spatial extent \citep[e.g.][]{smith2013}, but the similarities in $r_{\rm GC}$, at least within the uncertainties, indicate this is not a major effect acting on Virgo dwarfs compared to LV dwarfs.

\begin{figure*}
\includegraphics[width=\textwidth]{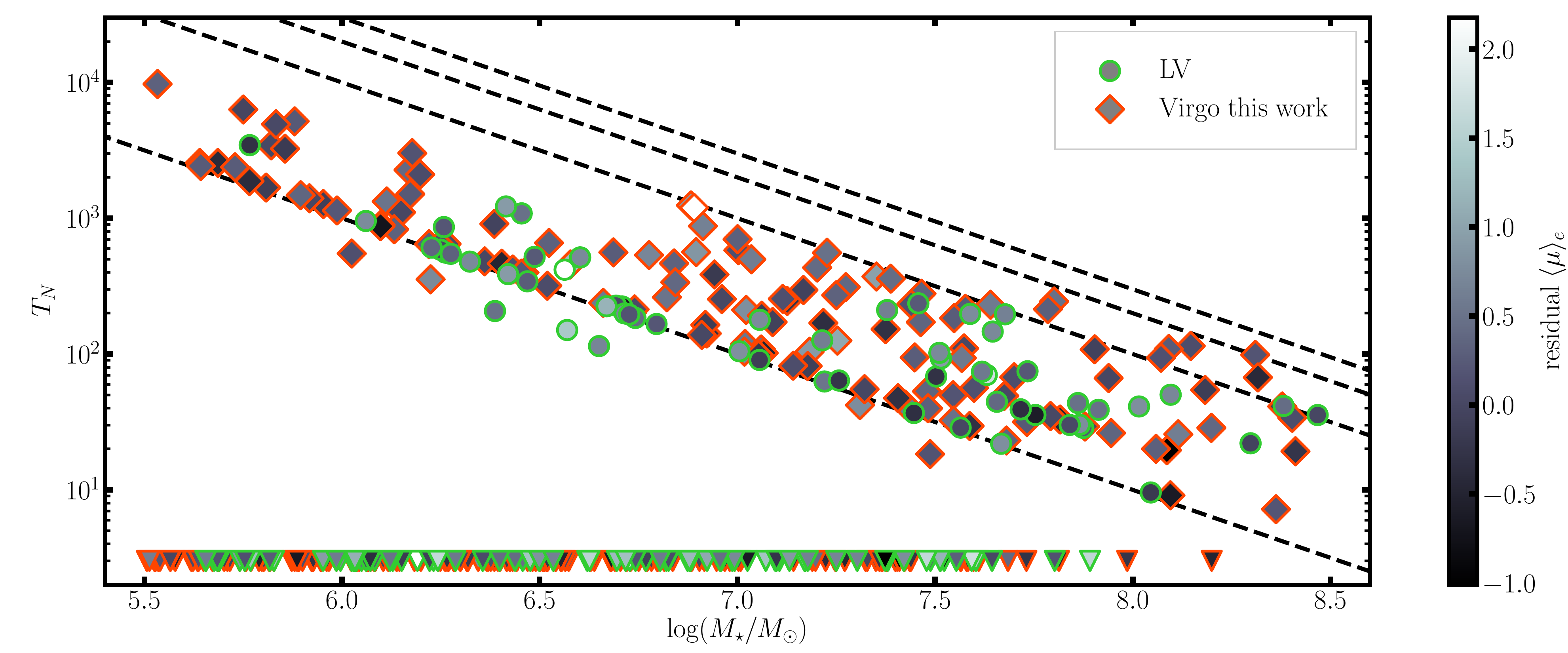}
\caption{GC specific mass frequencies ($T_N\equiv N_\mathrm{GC}\times10^9 M_\odot/M_\star$) for the LV and Virgo samples. Points are colored by their surface brightness, after correcting for the average trend in surface brightness with stellar mass. The triangles indicate dwarfs whose background subtracted GC abundance came out zero or negative. The dashed black lines show constant GC abundances of 1, 10, 20, and 30. The GC abundances are measured within $2r_e$ and have not been corrected for GCs outside of this radius.}
\label{fig:tn_mstar}
\end{figure*}

\subsection{Individual GC Abundances}
\label{sec:gc_abund_ind}
In this section and the next, we explore the GC system abundances, particularly focusing on how it relates to dwarf stellar mass. We focus on the results for individual dwarfs from the background subtraction method in this section, and in the next we present the average trends in GC abundance from the joint likelihood-based inference.

Figure \ref{fig:tn_mstar} shows our main results for the individual, background-subtracted GC abundances. In particular, we show the specific mass frequency $T_N\equiv N_\mathrm{GC}\times10^9 M_\odot/M_\star$ for the LV and Virgo dwarfs as a function of dwarf stellar mass. Note that these values of $T_N$ have \textit{not} been corrected for GCs outside of the $2r_e$ annulus we use to count sources. As mentioned above, if a GC half-number radius of $r_\mathrm{GC}=1.5\times r_e$ is assumed, the total abundance would be $\sim50$\% higher. Also note that many of the GC abundance values are not whole numbers both due to the background subtraction process and the correction for magnitude incompleteness (this last correction is $\lesssim5$\% but will move data points away from the nearest integer). The points are colored by the dwarfs' surface brightness after the general trend of decreasing surface brightness for lower mass dwarfs is removed. There does not appear to be a noticeable trend in surface brightness residual with GC abundance.

In this plot, we see the first indication that dwarfs in Virgo have higher GC abundance than dwarf satellites in lower-mass, MW-sized parent halos in the LV. The Virgo sample has significantly more `rich' GC systems. If we consider a threshold GC abundance of 7 (once corrected for GCs outside of the $2r_e$ annulus, this corresponds to $\gtrsim10$), there are 21 Virgo dwarfs (a fraction of $0.071\pm0.014$) with abundance above this compared to 4 LV dwarfs (a fraction of $0.029\pm0.014$). The GC results for Virgo generally had much larger uncertainties, but this does not appear to be causing this difference since the dwarfs above this threshold in both environments had average S/N$\sim3$ in the abundance measurement. This is a reasonable comparison to make since the stellar mass distributions of Virgo vs LV dwarfs are quite similar (cf. Appendix \ref{app:mass_distributions}).

%The lack of `extreme' GC systems in lower density environments agrees well with results from research in ultra-diffuse galaxies (UDGs). Recent work has indicated that GC systems as rich as some of those found in Coma \citep[e.g.][]{lim2018, pvd_2017_gcs} are generally not found in lower mass clusters like Virgo or Fornax or lower mass groups \citep{prole2019, lim2020, somalwar2020}. We are able to make a particularly strong argument for this since both environmental samples considered here (LV and Virgo) are \textit{volume limited} dwarf samples, thus allowing us to make robust statements on the fractions of all dwarfs in each environment above a certain abundance level. 

\begin{figure*}
\includegraphics[width=\textwidth]{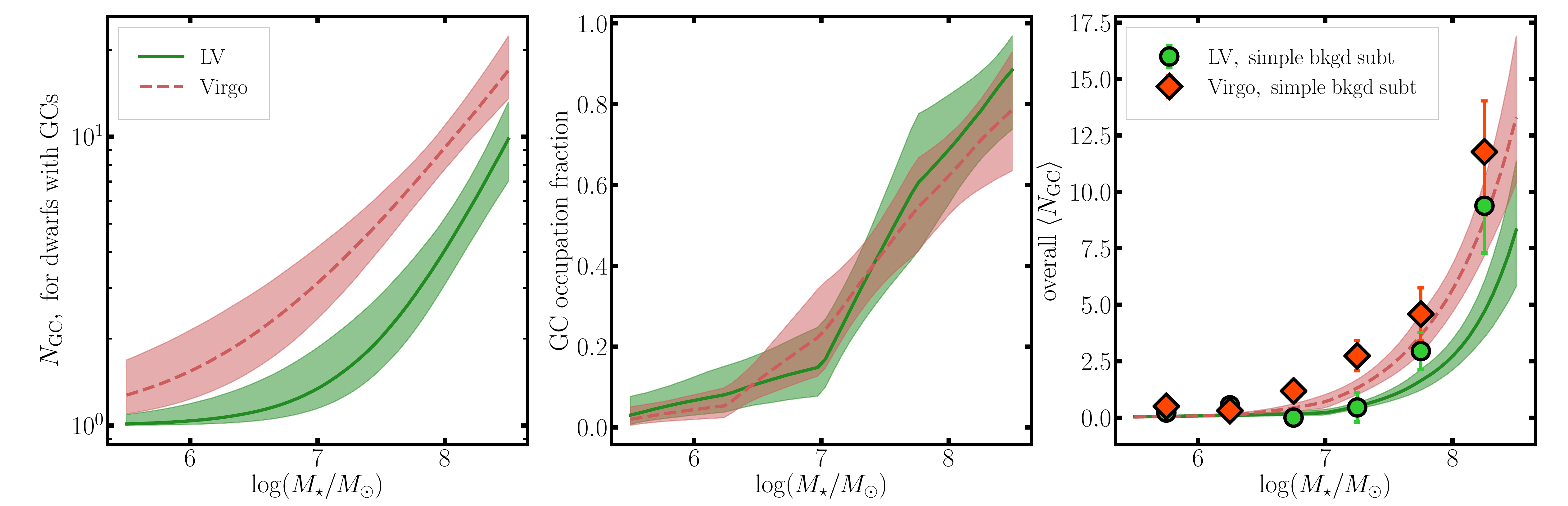}
\caption{The average trends in GC abundance and occupation fraction from the simultaneous fit of all dwarfs in each environmental sample. Each panel shows the median result from the posterior along with $\pm1\sigma$ uncertainties. \textit{Left}: The average number of GCs for dwarfs with GCs is shown as a function of stellar mass. \textit{Middle}: The average occupation fraction of GCs (i.e. fraction of dwarfs with non-zero GC systems) as a function of stellar mass. \textit{Right}: The overall average number of GCs per dwarf, including dwarfs with and without GC systems. Shown in the points are running means from the simple background subtraction analyses, which have been corrected for GCs outside the $2r_e$ annuli used to count GCs.}
\label{fig:ngc_method_compare}
\end{figure*}

\subsection{Average Trends in GC Abundance}
\label{sec:gc_abund_avg}
Figure \ref{fig:ngc_method_compare} shows the main results from the joint fit for all dwarfs in each of the environment samples.  The left panel shows the average number of GCs for dwarfs with non-zero GC systems. This was fit as a power law with a floor of one. The best-fitting power law parameters are given in Table \ref{tab:likelihood_method_results} along with the other best-fitting parameters from the joint fit (cf. \S\ref{sec:gc_meth_lik} for details on what each parameter is). The median posterior values are given along with marginalized uncertainties. The middle panel of Figure \ref{fig:ngc_method_compare} shows the inferred occupation fraction of GCs as a function of dwarf stellar mass. The occupation fraction goes from essentially zero at $M_\star<10^6M_\odot$ to $\sim80$\% at $M_\star\sim10^{8.5}M_\odot$. This is a similar trend to that reported in Figure 6 of \citet{rsj2019} for the NGVS analysis of Virgo dwarfs, although it appears we infer a somewhat smaller occupation fraction for intermediate mass ($M_\star\sim10^7M_\odot$) dwarfs. We compare the GC occupation fraction trends directly to the nucleation fraction trends below in the Discussion.

\begin{deluxetable}{ccc}
\tablecaption{Best-fitting parameters from the joint fit. \label{tab:likelihood_method_results}}
\tablehead{\colhead{Sample} & \colhead{Parameter} & \colhead{Best-fitting value}}
\startdata
LV	  & 	$T_0$ 			& $25.81 ^{+ 17.803 }_{- 11.762}$	\\
	 & 	$\alpha$ 			& $0.945 ^{+ 0.268 }_{- 0.308}$	\\
	 & 	$M_g^0$ 			& $-7.016 ^{+ 0.11 }_{- 0.114}$	\\
	 & 	$\sigma_g$ 			& $0.565 ^{+ 0.079 }_{- 0.072}$	\\
	 & 	$r_\mathrm{GC}$ 			& $1.063 ^{+ 0.21 }_{- 0.177}$	\\
	 & 	$f_1$ 			& $0.03 ^{+ 0.045 }_{- 0.023}$	\\
	 & 	$f_2$ 			& $0.081 ^{+ 0.071 }_{- 0.044}$	\\
	 & 	$f_3$ 			& $0.149 ^{+ 0.101 }_{- 0.072}$	\\
	 & 	$f_4$ 			& $0.602 ^{+ 0.172 }_{- 0.17}$	\\
	 & 	$f_5$ 			& $0.887 ^{+ 0.082 }_{- 0.15}$	\\
\hline
Virgo	 & 	$T_0$ 			& $31.657 ^{+ 17.556 }_{- 10.172}$	\\
	 & 	$\alpha$ 			& $0.59 ^{+ 0.172 }_{- 0.155}$	\\
	 & 	$M_g^0$ 			& $-7.094 ^{+ 0.072 }_{- 0.074}$	\\
	 & 	$\sigma_g$ 			& $0.52 ^{+ 0.055 }_{- 0.055}$	\\
	 & 	$r_\mathrm{GC}$ 			& $1.248 ^{+ 0.236 }_{- 0.183}$	\\
	 & 	$f_1$ 			& $0.02 ^{+ 0.032 }_{- 0.015}$	\\
	 & 	$f_2$ 			& $0.054 ^{+ 0.045 }_{- 0.03}$	\\
	 & 	$f_3$ 			& $0.228 ^{+ 0.122 }_{- 0.098}$	\\
	 & 	$f_4$ 			& $0.541 ^{+ 0.122 }_{- 0.11}$	\\
	 & 	$f_5$ 			& $0.786 ^{+ 0.141 }_{- 0.154}$	\\
\enddata
\end{deluxetable}

The right panel shows the overall average GC abundance as a function of stellar mass, including dwarfs with and without GC systems (i.e. the curves from the left and middle panels multiplied together). We also show the average results from the simple background subtraction process (which have been corrected for GCs outside of the $2r_e$ annulus). The simple background subtracted results are a little higher than the average abundance from the joint fit. This appears to be largely due to assuming $r_\mathrm{GC}=1.5\times r_e$ in correcting for GCs outside of the $2r_e$ annulus while a somewhat smaller $r_\mathrm{GC}$ is likely more realistic.

We find that the Virgo dwarfs have notably higher average GC abundance than the LV dwarfs, particularly in the mass range $10^7 < M_\star < 10^8$ \msun.  Interestingly, this seems to be mostly stemming from the difference in GC abundance for dwarfs \textit{with} GCs and less a difference in average GC occupation fraction.

Since the NGVS dwarf sample is from the innermost $r<R_\mathrm{vir}/5$ region of Virgo, while the LV samples are generally complete to at least $r\gtrsim R_\mathrm{vir}/2$, the same caveat applies to these GC results as for the NSC results discussed in \S\ref{sec:nuc_frac}. With the current dwarf samples, it is not clear whether the difference in GC abundance is primarily an effect of parent halo mass or the dwarfs' current location in the parent halo.

\begin{figure}
\includegraphics[width=0.48\textwidth]{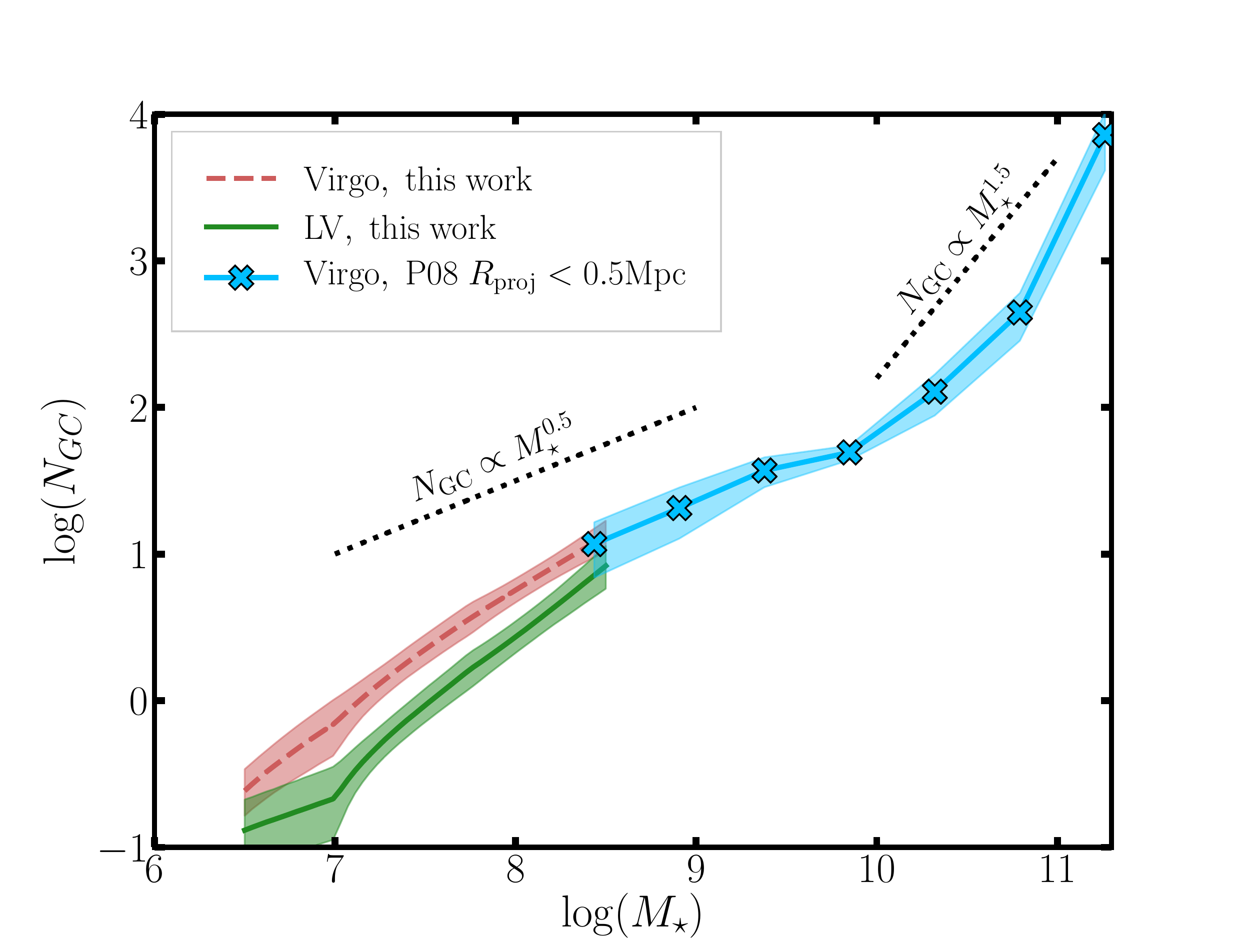}
\caption{We compare the inferred GC abundances for the LV and Virgo samples to the \citet{peng2008} results over a larger dynamic range in galaxy stellar mass.  Our results for both environments show a similar scaling to the \citet{peng2008} results of approximately $N_\mathrm{GC}\propto M^{0.5}_\star$. Above about $M_\star\sim10^{10.5}$, the relation steepens to approximately $N_\mathrm{GC}\propto M^{1.5}_\star$. }
\label{fig:virgo_comp2}
\end{figure}

In Figure \ref{fig:virgo_comp2}, we show the GC abundance results over a larger dynamic range in stellar mass. In addition to the LV and Virgo samples, we include results from \citet{peng2008} which extend to much higher stellar mass. To roughly match the spatial distribution of the NGVS dwarfs within Virgo, the \citet{peng2008} results are  restricted to those within a projected 500 kpc of M87 ($\sim R_\mathrm{vir}/3$). The LV sample is offset below the NGVS sample by roughly $\sim0.3$ dex (about a factor of 2), as was shown in Figure \ref{fig:ngc_method_compare}. The \citet{peng2008} results show a similar average relation as our Virgo results, scaling approximately as $N_\mathrm{GC}\propto M^{0.5}_\star$ below about $M_\star\sim10^{10.5}$. Above that mass, the relation steepens to approximately $N_\mathrm{GC}\propto M^{1.5}_\star$, presumably due to the increasing importance of accreting GCs from cannibalized satellite galaxies. We compare our results in more detail to this and other Virgo dwarf samples in the next section.

\begin{figure*}
\includegraphics[width=\textwidth]{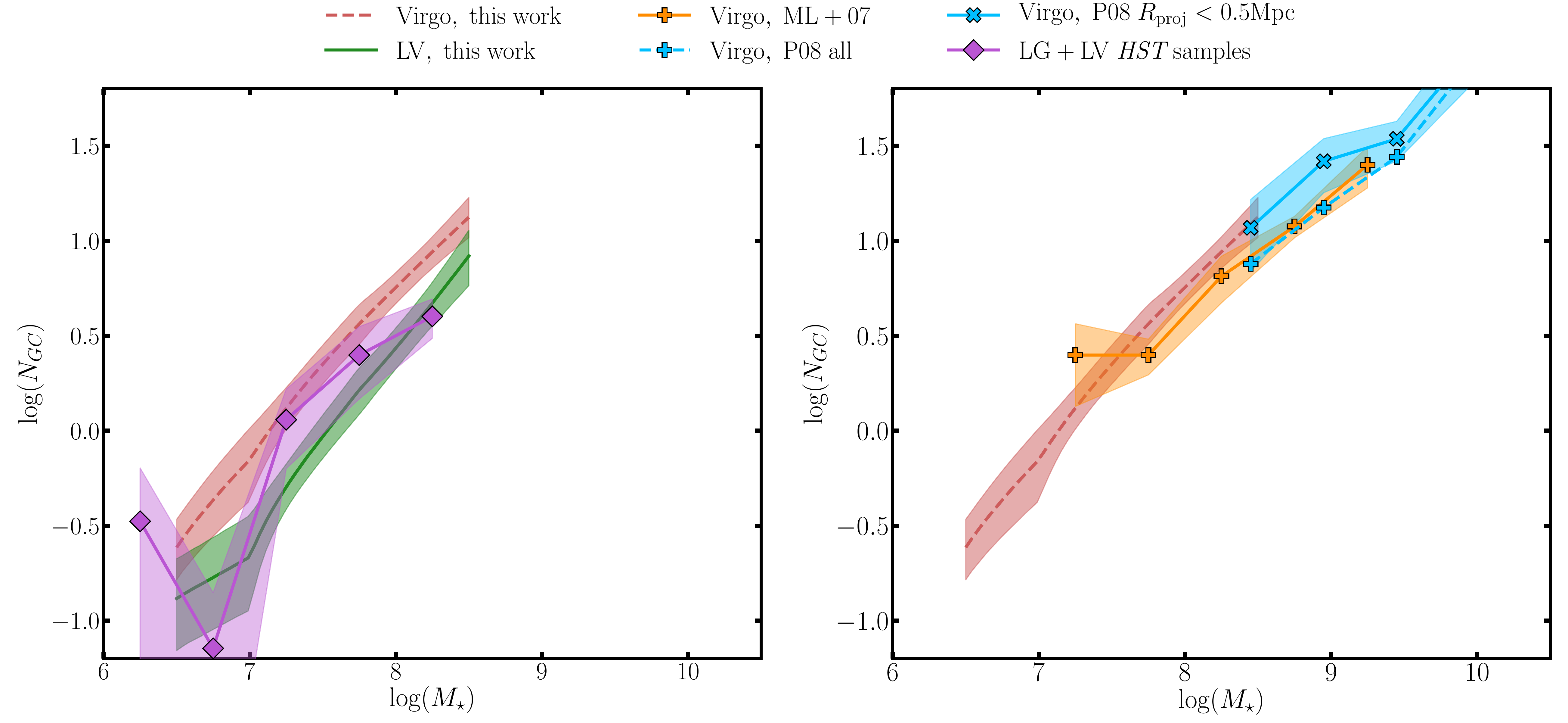}
\caption{\textit{Left}: A comparison between the ELVES sample and other, \emph{HST} based, surveys of LV dwarfs. We also include the LG dSph satellites with these latter surveys. \textit{Right}: The Virgo sample analyzed in this work shown next to two comparable analyses: \citet{miller2007} and \citet{peng2008}.  We find that the sample of \citet{miller2007} shows lower GC abundance at fixed stellar mass than the sample analyzed in this work. To test if this is an environmental effect we show the results of \citet{peng2008} both including all galaxies and also restricting to those within a projected 500 kpc ($\sim R_{\rm vir}/3$) to M87. The inner galaxies show higher GC abundance, appearing to be an extrapolation of the trend found in the current work, indicating the discrepancy with \citet{miller2007} could feasibly be environmental.  The error regions show errors in the mean, not the intrinsic scatter.}
\label{fig:virgo_comp}
\end{figure*}

\subsection{Comparison to Previous Work}
\label{sec:virg_comp}
In this section, we compare our results for the average GC abundance of ELVES and NGVS dwarfs to other comparable analyses of LV and Virgo dwarfs. We find that our inferred GC abundances are similar to what has been found before for dwarfs in these different environments.

In the left panel of Figure \ref{fig:virgo_comp}, we show the LV results compared to the combined results of the LG dSphs and the surveys of \citet{sharina2005} and \citet{georgiev2009a}. As mentioned above, the \citet{sharina2005} and \citet{georgiev2009a} surveys include many of the ELVES satellites but are done with \emph{HST} (obviating the need for a statistical background subtraction\footnote{Recall that these samples are for dwarfs within $D\lesssim6$ Mpc, so GCs are resolved.}) and are important comparison samples for the ELVES results. The references for the LG GC systems were given in \S\ref{sec:gclf}. All GC systems are restricted to GCs with $M_V<-6$ or $M_g<-6$, depending on what is available, to match the approximate magnitude limit of most ELVES dwarfs. For reference, the Fornax dSph has four GCs above this limit while the Sgr dSph has only 2. We describe how we determine stellar masses for these reference samples in Appendix \ref{app:stellar_masses}. The average GC abundance of the LG satellites and \citet{sharina2005} and \citet{georgiev2009a} surveys agrees well with the LV results presented here.

In the right panel of Figure \ref{fig:virgo_comp} we compare the trends of average GC abundance versus dwarf stellar mass for three Virgo samples: our analysis of the NGVS dwarfs, the \emph{HST} survey of \citet{lotz2004} and \citet{miller2007} and the ACS Virgo Cluster Survey sample \citep{cote2004, peng2008}. We detail the sources of the stellar masses for these dwarf samples in Appendix \ref{app:stellar_masses}. For the dwarfs identified as nucleated by \citet{miller2007}, we subtract their reported GC abundance by one since our GC abundances never include the NSC. 

Interestingly, both the \citet{miller2007} and complete \citet{peng2008} (dashed blue) samples show noticeably smaller GC abundance at fixed stellar mass than the NGVS sample (although still consistent with our NGVS results at $\sim2\sigma$). The NGVS sample analyzed in this work is concentrated at the very core of Virgo ($\lesssim R_\mathrm{vir}/5$)  while the \citet{miller2007} sample is distributed all throughout the virial volume of Virgo so this could feasibly be an environmental effect. To test this, we show the \citet{peng2008} sample restricted to those within a projected 500 kpc ($\sim R_{\rm vir}/3$) to M87\footnote{Restricting this sample to $R_\mathrm{proj}< R_\mathrm{vir}/5$ leads to too few statistics.} (solid blue line). As found by \citet{peng2008}, the inner dwarfs show notably higher GC abundance at fixed stellar mass. The centrally concentrated subsample lines up well with the abundance trend of the NGVS sample. This is suggestive that the difference with \citet{miller2007} is indeed environmental. We note, however, that \citet{miller2007} searched for an environmental trend within their sample and found no trend in specific frequency with distance from M87, in seemingly direct conflict with the results of \citet{peng2008}.

\section{Discussion}
\label{sec:disc}
In this paper, we have systematically presented many properties of the NSCs and GC systems of dwarf galaxies in the Local Volume, paying close attention to how these properties are the same as or different from star clusters of dwarfs in other environments, particularly in nearby galaxy clusters. With much improved statistics over what was possible in the past, we have confirmed previous findings indicating that dwarf galaxies are more likely to be nucleated and host richer GC systems, on average, in denser environments. These trends apply even to dwarfs in the halos of MW-sized galaxies. However, other than the abundance of the star clusters, we find that other properties of NSCs and GCs, like luminosity and color, are overall similar across environment. 

In the Introduction, we averred that extending the axis of environment to studies of star clusters in dwarfs would open a new perspective on star cluster formation and how it relates to dwarf galaxy evolution. In this section, we discuss our findings in relation to modern theories of GC and NSC formation. Overall, we find that our observations are well explained, at least qualitatively, by these models. First, we explore the cause of the GC abundance-environment trend and, second, we discuss the NSC properties and their relations to GCs.

\subsection{Effect of Environment on GC Abundance}
\label{sec:disc_gc_abund}

Previous observational studies have generally shown that dwarf galaxies in the denser, inner regions of clusters have higher GC abundances \citep{peng2008, lim2018, liu2019} than dwarfs in the cluster outskirts. In this work, we have confirmed this effect and shown that it extends to even lower density environments, such as the halos of MW-sized galaxies. We will discuss three possible origins of this trend inspired by modern models of GC formation. The first is that dwarfs in denser environments experience increased efficiency of cluster production. The second is the idea that dwarfs in denser environments experience reduced levels of cluster destruction. Finally, the third is that dwarfs in denser environments form fewer field stars, leading to more prominent GC systems.

\subsubsection{Increased GC Production}
The idea of `biased GC formation', first explored in detail by \citet{peng2008}, is that subhalos ending up in a cluster core collapse earlier and form their stars at higher star formation rate (SFR) surface densities compared to subhalos in the cluster outskirts. Higher SFR densities and the associated higher pressures and densities in the ISM are more conducive to the formation of bound stellar clusters, leading to a high cluster formation efficiency \citep{goddard2010, kruijssen2012} and, hence, higher GC specific frequencies. This trend whereby earlier forming galaxies end up hosting larger populations of GCs is found in numerous recent models \citep[e.g.][]{mistani2016, pfeffer2018, carleton2021}. Evidence for this also comes from the elevated [$\alpha$/Fe] ratios of Virgo dwarf galaxies with high specific frequencies \citep{liu2016}, indicating these galaxies formed their stars in a short interval, feasibly accompanied with elevated SFR surface densities and, hence, cluster formation efficiency.

This biased formation scenario implies that dwarf ETGs in the cores of Virgo and Fornax are, generally, not newly quenched dIrrs that recently fell into the cluster from the field. This has been argued based on GC abundance for more massive dwarfs \citep{rsj2012}, and here we show this must apply to lower mass dwarfs with high GC abundance as well. The gas-rich progenitors of these cluster dwarfs would not necessarily be similar to dIrrs that are observed in the field today. However, in a companion paper (Carlsten et al., submitted), we show that the early-type Virgo dwarf population in the mass range we consider in this work is very similar in its structural scaling relations (particularly the mass-size relation) to field dIrrs \citep[this has also been shown by, e.g.,][]{boselli2008_scaling, venhola2019}. Evidently, if the elevated GC abundance in cluster dwarfs is due to an early, intense phase of star-formation with a high cluster formation efficiency, this did not leave a correspondingly-large impact on the structure of the dwarf.

\subsubsection{Decreased GC Destruction}
An alternative way to have higher modern day GC abundance in cluster dwarfs is for these dwarfs to experience lower levels of GC destruction than analogous dwarfs in less dense environments. If GCs are the descendants of star clusters similar to the YMCs currently observed in the local universe, then the initial cluster mass function (ICMF) of GCs will be similar to the mass function of nearby YMCs; namely a power law with slope $\sim-2$ and exponential cutoff at the high mass end \citep[e.g.][]{zhang1999, larsen2009}. To get the observed, peaked GC mass function (GCMF) from this ICMF requires significant levels of disruption and mass loss, particularly among the lowest mass clusters. Various estimates have put the total amount of mass loss of a GC system during its evolution at $\gtrsim90$\% \citep[see][for discussion]{forbes2018}. 

%Early models suggested that two-body relaxation-driven evaporation of GCs was responsible for destroying low-mass clusters, leading to the observed universal GCMF \citep[e.g.][]{fall2001}. However, detailed simulations of cluster evolution indicate that the mass loss due to evaporation depends on the external tidal field \citep{baumgardt2003}. Since the tidal field will clearly be different around different mass hosts and at different galactocentric radii, evaporation has difficulty explaining the essentially universal turn-over mass of the GCMF across different mass hosts and galactocentric radii \citep{vesperini2003}. 

Various dynamical processes have been proposed for the destruction of low mass clusters \citep[e.g.][]{fall2001, baumgardt2003}. The critical constraint is that the turn-over mass of the GCMF is essentially universal across different mass hosts and galactocentric radii \citep{vesperini2003}. In part to explain this, \citet{elmegreen2010} and \citet{kruijssen2015} outline a two-phase model for GC evolution. In the first phase, the newly born GCs are efficiently destroyed via tidal shocks due to GMCs present in the gas-rich environment where the GCs are born. Due to merging at high-redshift and/or quenching of the host galaxy, GCs are moved to the outer regions, away from the GMCs. In this second phase, the GCs slowly evolve via two-body relaxation and evaporation. %Most of the disruption takes place in the first phase, largely setting the GCMF at early times. When the GCs are relocated to the outskirts of the host, they get scrambled in radius, explaining why there are no observed trends in the GCMF with galactocentric radius \citep[see also][for a discussion of the resulting metallicity distribution functions]{lamers2017}.

Based on this model, \citet{kruijssen2015} makes the prediction that dwarfs in denser environments would experience shorter rapid-disruption phases due a combination of increased mergers or earlier quenching through ram-pressure stripping.  

We see evidence for the GC relocation central to this theory within our analysis of the GC radial distributions. While we were unable to place stringent constraints on the GC spatial distribution, Figure \ref{fig:rad_dists_single} shows that our fits prefer the GC distributions to be on average more spatially extended than the stellar body of the dwarfs, as is commonly found in the literature (see \S\ref{sec:bkgd_sub}). Since GCs will form preferentially where the ISM is densest, it seems very unlikely that the GCs formed at the locations we observe them today. As discussed by \citet{kruijssen2015} and \citet{lamers2017}, early mergers are one way to explain the location of GCs. After a merger, the gas will dissipatively sink back to the galactic center, forming new generations of stars while the GCs will be stuck in the galaxy's outskirts. Similarly, \citet{leung2020} argue for a merger origin for the spatially extended current configuration of GCs in the Fornax dSph. 

An intriguing alternative way of getting GCs into the outskirts of their hosts is suggested by the GCs of the LG dwarf NGC 6822 which show a suggestive alignment and spatial distribution \citep[see Fig. 1 of ][]{huxor2013} consistent with the old diffuse stellar halo \citep{battinelli2006} of that dwarf. Many dwarfs are observed to have old/red stellar halos \citep[e.g.][]{kadofong2020}, which are thought to have formed primarily through stellar feedback \citep{stinson2009}. It is unclear if GCs could experience the same gradual orbital expansion due to stellar feedback as stars. In the end, whether it is mergers or stellar feedback that relocates the GCs, the end result of protecting them from disruption in the gas-rich inner regions of the dwarfs would be the same.  Given the importance of GC destruction in setting the universality of the GCLF, it is unclear if this mechanism could reproduce the similarity in the GCLF that we observe between the Virgo and LV dwarfs (cf. Fig \ref{fig:gclf_restricted}). Quantitative modelling would be required to address this.

\subsubsection{Decreased Field Star Formation}
The final explanation for the environmental trend in GC abundance we will discuss is that dwarfs in denser environments experience reduced field star formation efficiencies due to earlier quenching. Closely linked to this is the idea that GC systems constitute a constant mass fraction of the total (i.e. halo) mass of a galaxy \citep[e.g.][]{blakeslee1997, spitler2009, georgiev2010, hudson2014, harris2017}. Thus, at a given halo mass, if the dwarfs in denser environments form fewer stars due to, for example, earlier or more efficient quenching than dwarfs in less dense environments, then they could have more abundant GC systems at fixed \textit{stellar} mass. We note, however, there is significant uncertainty in whether the linearity between GC system mass and halo mass is applicable to the mass range ($M_\mathrm{halo}\lesssim10^{11}$\msun) considered in this work \citep{forbes2018_ngc_mhalo, bastian2020}. 

With that caveat said, essentially this comes down to the question of whether the stellar-to-halo mass relation (SHMR) is different for dwarfs in different environments. One of the best constraints available on the SHMR is simply the observed abundances of satellites of different stellar masses. \citet{grossauer2015} used the NGVS dwarf catalog to infer the SHMR applicable to the Virgo core. In \citet{carlsten2020b}, we showed that the SHMR of \citet{gk_2017} was able to well reproduce the observed abundance of LV satellite systems. Interestingly, the SHMR inferred by \citet{grossauer2015} is quite different than that of \citet{gk_2017}, predicting lower stellar mass at fixed halo mass by nearly an order of magnitude. More work needs to be done to gauge the severity of the difference. Taking it at face value, however, it would indicate that Virgo dwarfs on average do form fewer stars at fixed halo mass and, hence, might explain why they have more GCs at fixed stellar mass. However, we do note that the SHMR is notoriously unconstrained at these mass scales with different observables and simulation projects giving quite disparate constraints.%, and the abundance matching results discussed here are just two discussed options for the SHMR.

One interesting feature from Figure \ref{fig:virgo_comp} is that the $N_\mathrm{GC}-M_\star$ relation has the same slope between the Virgo and LV samples. Thus, if the difference in abundance is due to a constant $M_\mathrm{GC}/M_\mathrm{halo}$ fraction and different SHMRs in the two environments, then the SHMRs must also have the same slope at the low-mass end, with just an offset between them. This would seem to require some fine-tuning, and we note that the SHMRs of \citet{grossauer2015} and \citet{gk_2017} do not have the same slope. It is unclear how significant scatter in the SHMR\footnote{And even different amounts of scatter in dwarfs in Virgo versus the LV environments} could affect this, but a detailed exploration would have to take this into account.

This proposed mechanism would imply that early-type dwarfs in clusters would be, on average, older than in MW-sized halos. The current observations are unclear on this. The age distribution of cluster dEs appears to be quite a mixed bag \citep[e.g.][]{koleva2009}, as does the age distribution of LV satellite dwarfs \citep[e.g.][]{weisz2014, weisz2019}. This is further complicated by the fact that even if cluster dwarfs did on average quench earlier, they might build up stellar mass more rapidly before quenching due to, for example, increased gas availability in a denser environment \citep{joshi2021}.

In closing this section, we note that the higher GC abundance at fixed stellar mass in clusters cannot be due to increased stripping of stars in the cluster environment. GCs have at least as extended spatial distributions as the stars, and the GCs would be stripped first \citep[e.g.][]{smith2015}, leading to lower GC abundance at fixed stellar mass in cluster dwarfs. %Indeed, the fact that we observe cluster dwarfs to have higher GC abundance, on average, means that most cluster dwarfs cannot have experienced significant mass loss through tidal stripping.

\subsubsection{Summary}
We have discussed three possible causes for the increased GC abundance in denser environments: (1) preferential formation of GCs in early collapsing subhalos (increased GC production), (2) shorter duration of an early, rapid-disruption phase in denser environments due to increased merger rates or earlier quenching (decreased GC destruction), and (3) lower efficiency of forming field stars in cluster dwarfs due to earlier quenching (decreased field star formation). Various forms of these ideas have been proposed to explain the original observations of \citet{peng2008}. We confirm the environmental dependence found in that work and show that it extends across environments spanning several orders of magnitude in parent halo mass. Although, due to the different radial extent of the Virgo and LV samples, it is not completely clear whether this dependence is primarily an effect of the parent halo mass or due to the LV dwarfs being, on average, further out in their hosts' virial volumes.

Likely all three of these mechanisms are, to varying degrees, responsible for the increased GC abundance in denser environments. To figure out which mechanism(s) is dominant, quantitative models would be required. %For instance, quantifying the expected increase in SFR surface density or merging rates in cluster environments over LV-like environments could address whether (1) or (2) is more dominant. 
When discussing the increased GC production vs decreased disruption, it is possible that the similarity in GCLF across environments could help elucidate which process dominates. In-depth photometric comparisons between cluster dwarfs and LV dwarfs could reveal interesting differences in average stellar population properties, perhaps illuminating option (3). Additionally, analyses of cluster dwarf samples that are complete to a large fraction of the cluster virial radius would elucidate whether the environmental dependence we observe is primarily an effect of parent halo mass or the dwarfs' location in the parent halo. This could further clarify which of these mechanisms is most important in setting the GC abundance.

\subsection{Connection Between NSCs and GCs}
\label{sec:disc_nscs}
In this section, we discuss the NSC and GC observations presented in this paper in the context of whether a physical connection exists between these two types of clusters. In this work, we showed that the presence of both GCs and NSCs exhibit a significant secondary dependence with host environment, highly suggestive of a connection between GCs and NSCs. Here, we assert that these observations are naturally explained if the inspiral of GCs is the dominant formation pathway of NSCs in dwarfs of this mass range.

As discussed in the introduction, proposed formation mechanisms of NSCs are broadly classed into GC-inspiral and \textit{in-situ} star formation. Support for the GC origin of NSCs comes from the general similarity between NSCs and GCs, including the fact that NSCs almost always have stellar mass above that of a single GC. Both analytic calculations and simulation results have shown that massive GCs starting within a few kpc of the galaxy center can feasibly inspiral within a Hubble time. Additionally, this method has strong support from the fact that many NSCs in dwarf ($M_\star < 10^{9}$ \msun) galaxies have lower metallicity than the surrounding host galaxy \citep{paudel2011, spengler2017, neumayer2020, johnston2020, fahrion2020, fahrion2021}, similar to old, metal-poor GCs which are generally found to be more metal-poor than the host dwarf galaxy \cite[e.g.][]{larsen2012, fensch2019}. Finally, some have argued that the flattened radial distribution of surviving GCs in dEs compared to the stellar distribution is indicative that some of the GCs have decayed into the center and merged \citep{lotz2001}. 

On the other hand, it is possible that NSCs are created through central star-formation. Gas can be diverted to the centers of galaxies through mergers and other ways \citep[e.g.][]{mihos1994, milo2004, bekki2006}. Support for this mechanism comes largely from the presence of young stars in many NSCs which is otherwise hard to explain in the GC-inspiral paradigm \citep{walcher2006, rossa2006, monaco2009, seth2010, paudel2011, carson2015, spengler2017, johnston2020}. The significantly flattened NSCs observed in galaxies of relatively high stellar mass \citep{seth2006, spengler2017} also points to \textit{in-situ} star formation.

As stated in the introduction, there is a growing consensus that the GC-inspiral mechanism dominates at low ($M_\star<10^9$\msun) masses while \textit{in-situ} star formation dominates at higher masses \citep{turner2012, rsj2019, neumayer2020, fahrion2021}. This is particularly true if we separately consider how NSCs \textit{form} versus how they \textit{grow} \citep{neumayer2020}. NSCs in dwarf galaxies almost certainly started as GC-like clusters. This would explain the absence of NSCs of lower mass than typical GCs and their similarities in size and density to GCs. As we show in Figure \ref{fig:nuc_photo}, we are complete to nuclei down to $M_g\sim-5$ mag for many of the LV galaxies, yet the observed NSC luminosity function drops off steeply fainter than $M_g\sim-7$ mag. 

After forming, the NSCs could have grown either through merging of other GCs or through \textit{in-situ} star formation. The NSCs with lower metallicity than the dwarf hosting them likely just grew via GC merging while NSCs with young stellar populations must have had some \textit{in-situ} star formation. As a dwarf is being quenched via ram pressure stripping either in the cluster environment or in the halo of a MW-like host, the central regions would be the last to quench, as evinced by the blue cores of many dEs \citep[e.g.][]{urich2017,hamraz2019}. Thus, it is not surprising that many NSCs have a subpopulation of young stars. This idea is exemplified elegantly in the stellar populations of M54, the Sagittarius dSph's NSC \citep{siegel2007, alfaro-cuello2019}. A young metal-rich (age $\sim2$ Gyr, [Fe/H]$\sim0$) population is superimposed on an old metal-poor population (age $\sim12$ Gyr, [Fe/H]$\sim-1.5$). The old stars likely originated as one or more globular clusters while the young population was believably formed when Sgr was stripped of its gas as it was accreted by the MW.

\begin{figure}
\includegraphics[width=0.46\textwidth]{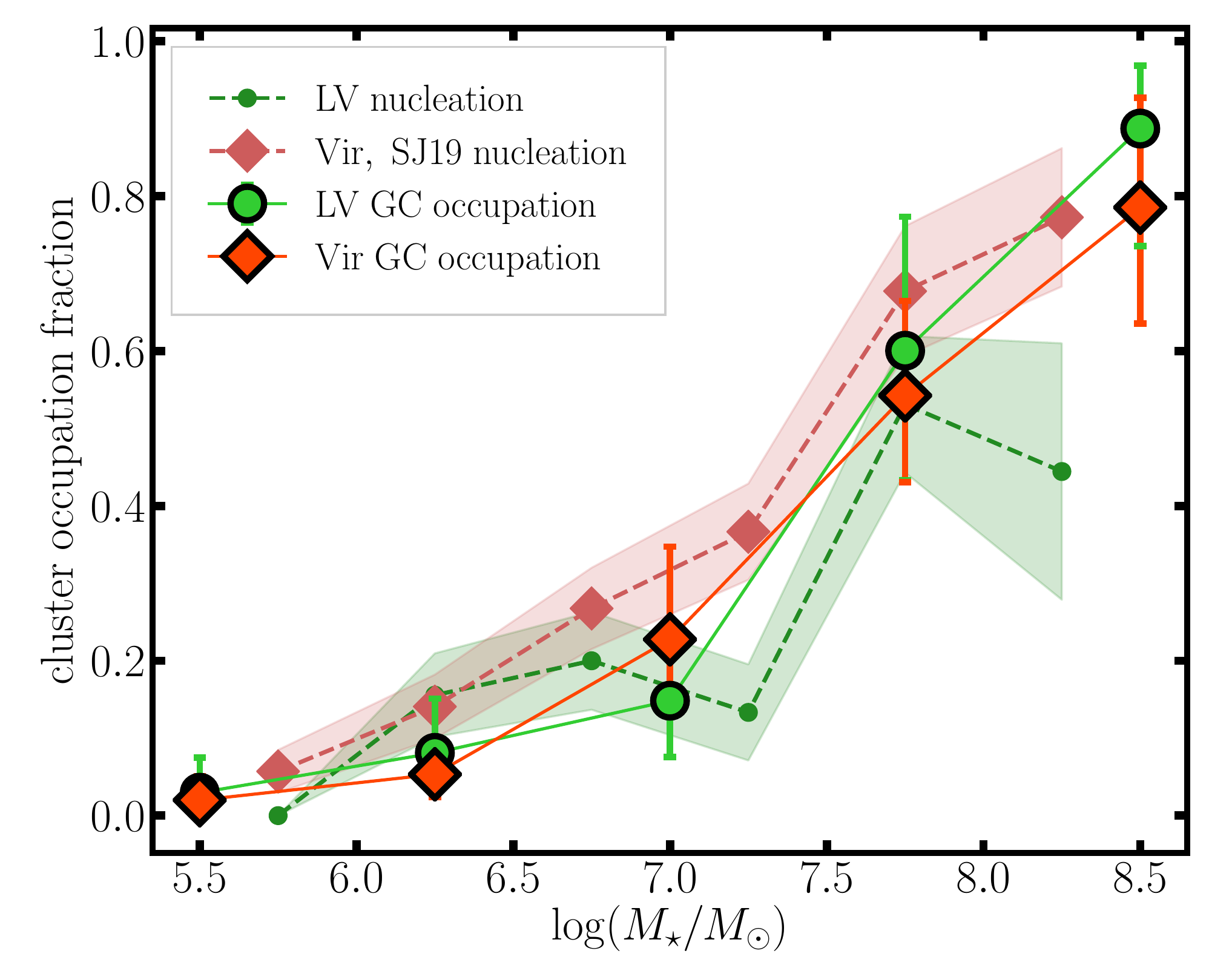}
\caption{The occupation fraction of GCs and NSCs as a function of dwarf stellar mass. The NSC occupation fraction is the nucleation fraction from Figure \ref{fig:nuc_frac} while the GC occupation fraction comes from the joint fit shown in Figure \ref{fig:ngc_method_compare}. There are clear similarities between the occupation rates of the different types of star cluster. The difference in GC occupation between the two environments is less significant than the difference in NSC occupation. }
\label{fig:gc_nuc_occ}
\end{figure}

The results of the current study fit well into this picture. We confirm that there is a strong correspondence between GCs and NSCs in low mass early-type dwarfs. We show one example of this in Figure \ref{fig:gc_nuc_occ} which directly compares the NSC occupation fraction (i.e. nucleation fraction) from Figure \ref{fig:nuc_frac} with the GC occupation fraction from Figure \ref{fig:ngc_method_compare}. Broadly speaking, there are clear similarities in the occupation fraction between the two types of clusters with both types of clusters going from occupation fractions of essentially zero at $M_\star<10^6$ \msun~ to $\sim60-80$\%  at $M_\star=10^{8.5}$ \msun. This is similar to the result found by \citet{rsj2019} for their analysis of the NGVS sample. The errorbars on the inferred GC occupation fraction are quite large so it is hard to compare the GC curves to the NSC curves in detail, but it does appear that the GC occupation fractions for both environments are less discrepant than the nucleation fractions, suggestive that there is more to forming a nucleus than simply the presence of GCs. We note, however, that \citet{rsj2019} found the NGVS GC occupation fraction to more closely track the trend in nucleation fraction. It is unclear if this is due to that team using more than just two bands in analyzing the NGVS sample or a difference in methodology.

Additionally, we confirm previously found trends that both the NSC frequency and GC abundance show a significant environmental dependence, indicating a strong connection between the two types of clusters. In the context of a picture where NSCs form primarily through GC inspiral, the decreased NSC occupation fraction at fixed stellar mass in LV hosts is because these dwarfs have fewer GCs which can act as the seed for NSC formation. In Figure \ref{fig:nuc_mstar}, we show evidence that NSCs in LV-like environments are lower mass at fixed host stellar mass. This is consistent with NSC growth being due to GC merging since the LV dwarfs would have fewer GCs to fuel this growth.

The current study also highlights some interesting open questions in the GC-NSC connection. Our results suggest that the GC radial profiles are more extended than the stellar distributions, in line with numerous previous findings. As we argue above, it seems very unlikely that the GCs formed in their current locations. Candidate mechanisms for the relocation of GCs outward include mergers at high redshift and stellar feedback. Clearly, to understand the formation of NSCs through GC-inspiral and quantitatively test that model, we must know the initial galactocentric radii of GCs and understand how they got there. For instance, it is unclear if NSCs formed before or after GCs were relocated outwards. Similarly, did the NSCs in early-type dwarfs form before or after the galaxy was quenched \citep[e.g.][]{guillard2016}? Understanding the radial distribution of GCs would likely also be a prerequisite to understanding why there is roughly equipartition in mass between NSCs and the surviving GC system for nucleated dwarfs in Virgo \citep{cote2006, rsj2019}. Getting IFU spectra that allows us to measure and compare the stellar populations of the NSC, GC system, and host galaxy \citep[e.g.][]{fahrion2019, johnston2020} will be critical in elucidating the GC-NSC connection.

Finally, we note that an interesting avenue to explore in the GC-NSC connection is the difference of GC-inspiral based NSC formation in different dark matter profiles \citep[e.g.][]{leaman2020}. The fact that the Fornax dSph \textit{does not} have a NSC even though it has several GCs has led to a contentious debate about whether it can be hosted in a cuspy profile \citep[e.g.][]{goerdt2006, meadows2020, shao2020}. Exploring this with the large samples of dwarfs presented here and elsewhere would be an interesting future direction.

\section{Conclusions}
\label{sec:concl}

In this work, we have systematically presented the NSC and GC properties of early-type dwarfs in the mass range $10^{5.5} < M_\star < 10^{8.5}$\msun~ in low-density environments. In particular, we consider a sample of satellite galaxies of roughly MW-mass hosts in the Local Volume from the ongoing ELVES Survey. The goal of the survey is to catalog all satellites down to $M_V\sim-9$ and surface brightness of $\mu_{0,V}\lesssim 26.5$ mag arcsec$^{-2}$ for all massive hosts ($M_K<-22.4$ mag) in the Local Volume. While previous surveys have characterized the GCs and NSCs of early-type dwarfs in clusters or field late-type dwarfs, the sample we consider here is by far the largest sample to-date of early-type dwarf satellites in lower-mass parent halos. Throughout this paper, we analyze data from the Next Gen Virgo Survey (NGVS) in the same way as for the LV sample, providing a direct comparison between two nearly volume limited samples of early-type dwarfs in drastically different environments: the halos of MW-mass galaxies versus the core of the Virgo cluster.

To briefly describe our methodology, we search for candidate NSCs in our galaxies through a combination of visual inspection and photometry. Candidate GCs are identified as point sources in the vicinity of the dwarf. We deal with contamination from background galaxies and foreground MW stars among the GC candidates in two main ways: (1) a simple background subtraction and (2) a likelihood-based joint fit for the average GC system properties across all dwarfs. Overall we find the two methods to be in agreement.

The main results regarding the nuclei are as follows:

\begin{itemize}
    \item We find the nucleation fraction of dwarf satellites of LV hosts is significantly lower than that of dwarfs in clusters (Figure \ref{fig:nuc_frac}), confirming that there is a noticeable secondary dependence of nucleation on environment, beyond the primary dependence on stellar mass.
    \item The NSCs are located within a projected radius of $r\lesssim r_e/8$ of the galaxy center, corresponding to $\lesssim 75$ pc for the average dwarf in the sample.
    \item The NSC luminosities show a steep dropoff fainter than $M_g\gtrsim-7$ mag (Figure \ref{fig:nuc_photo}) even though we are complete to point sources down to $M_g\sim-5$ mag in many of the LV dwarfs.
    \item The LV and Virgo samples of NSCs follow a similar relation with dwarf stellar mass of approximately $M_\mathrm{NSC}\propto M_\mathrm{gal}^{0.4}$ (Figure \ref{fig:nuc_mstar}). The LV NSCs appear to be lower mass at fixed host mass, at the $\sim1-2\sigma$ level.
\end{itemize}

The main GC results are as follows:

\begin{itemize}
    \item Fitting a Gaussian to the GCLF, we find a peak of $M_g^0\sim-7.1$ mag and dispersion $\sigma_g\sim0.6$ mag with no discernible difference between the LV and Virgo samples (Figure \ref{fig:gclf_restricted}). The GCLF exhibits a noticeable, non-Gaussian tail to fainter luminosities, similar to that found for LG dSphs and in previous GC studies of LV dwarfs with \emph{HST}. 
    \item We do not have the statistics to robustly constrain the radial distribution of GCs in either environmental sample, but GC systems more extended than the galaxy light profiles are favored in both samples, particularly among the GC-rich dwarfs, in agreement with numerous previous findings (Figure \ref{fig:rads}).
    \item We find that Virgo has a significantly higher fraction of dwarfs with rich GC systems. The fraction of dwarfs with $N_\mathrm{GC}>10$ is twice as high in Virgo as in the LV sample (Figure \ref{fig:tn_mstar}).
    \item From the joint fit, we find that Virgo dwarfs have, on average, higher GC abundance than the LV dwarfs. This appears to be primarily due to a higher average GC abundance for dwarfs with GCs, and less due to higher GC occupation fraction (Figure \ref{fig:ngc_method_compare}).
    \item Our results for the Virgo sample agree well with previous results for higher mass dwarfs in the Virgo core (Figure \ref{fig:virgo_comp}). Our inferred abundances are higher than previous work considering dwarf samples from throughout the Virgo cluster, confirming that the local environment of a dwarf in the cluster impacts its GC abundance  \citep[e.g.][]{peng2008}.
    \item The GC occupation fraction as a function of dwarf stellar mass broadly follows the nucleation fraction, but the difference between environmental samples appears to be less than that for the nucleation rates (Figure \ref{fig:gc_nuc_occ}). 
\end{itemize}

The main result of this paper is that we confirm that environment plays a strong role in the star cluster properties (primarily their abundance) of low-mass galaxies. These results strengthen the connection between NSCs and GCs in low-mass dwarfs. It is very likely that the nuclei of dwarfs in the mass range studied here form from a GC that spiralled into the galaxy center with later grow either through further GC inspiral or \textit{in-situ} star formation. We identify three possible explanations for the greater cluster abundance in dense environments often discussed in the literature:

\begin{enumerate}
    \item \textit{Increased GC Production}: GCs are thought to form in the regions of a galaxy's ISM with the highest pressures and densities. Thus, galaxies with high star formation rate densities are expected to form higher fractions of their stars in bound clusters. Dwarfs in the cores of galaxy clusters will form in earlier collapsing subhalos, on average, experiencing higher star formation rate densities, possibly explaining their higher star cluster abundance \citep[e.g.][]{peng2008, mistani2016}.
    \item \textit{Decreased GC Destruction}: GCs are known to experience significant tidal processing, with many lower mass clusters being disrupted before the present day. Of particular importance to GC destruction is tidal shocks from giant molecular clouds. Dwarfs in denser environments, like cluster cores, would likely experience earlier quenching or higher rates of merging, effectively `saving' the GCs from destruction \citep[e.g.][]{kruijssen2015}.
    \item \textit{Decreased Field Star Formation}: Due to dwarfs in cluster cores likely experiencing earlier quenching, it is possible that the cluster dwarfs have assembled less stellar mass overall. This would suggest that dwarfs in denser environments at a certain stellar mass would be hosted by more massive subhalos, on average, which could lead to higher GC abundances.
\end{enumerate}

It is likely that multiple, or even all, of these effects are at work. Exploring which physical process dominates, if any, through quantitative models will be important to explore in future work.

On the observational front, it will be interesting to further explore these effects with dwarf samples from galaxy clusters that are complete to the cluster virial radius. Due to the fact that the LV dwarfs are, on average, further out from their host ($r\sim R_\mathrm{vir}/2$) than the cluster dwarfs ($r\lesssim R_\mathrm{vir}/4$), it is not clear whether the environmental effects we observe are primarily due to the parent halo mass or more due to the dwarf satellite's current location in the parent halo. Addressing this with dwarf samples from clusters that extend to the cluster virial radius or with an analysis that quantifies the local environment of a dwarf (e.g. a metric that measures the local tidal field) could greatly help to illuminate the important physical processes involved in star cluster formation in dwarf galaxies.

\section*{Acknowledgements}
% -- RLB 

Support for this work was provided by NASA through Hubble Fellowship grant \#51386.01 awarded to R.L.B. by the Space Telescope Science Institute, which is operated by the Association of  Universities for Research in Astronomy, Inc., for NASA, under contract NAS 5-26555. J.P.G. is supported by an NSF Astronomy and Astrophysics Postdoctoral Fellowship under award AST-1801921. J.E.G. is partially supported by the National Science Foundation grant AST-1713828. S.G.C acknowledges support by the National Science Foundation Graduate Research Fellowship Program under Grant No. \#DGE-1656466.

Based on observations obtained with MegaPrime/MegaCam, a joint project of CFHT and CEA/IRFU, at the Canada-France-Hawaii Telescope (CFHT) which is operated by the National Research Council (NRC) of Canada, the Institut National des Science de l'Univers of the Centre National de la Recherche Scientifique (CNRS) of France, and the University of Hawaii. This research was made possible through the use of the AAVSO Photometric All-Sky Survey (APASS) \citep{apass}, funded by the Robert Martin Ayers Sciences Fund and NSF AST-1412587

\software{ \texttt{astropy} \citep{astropy} \texttt{sep} \citep{sep} \texttt{imfit} \citep{imfit} \texttt{emcee} \citep{emcee} \texttt{SExtractor} \citep{sextractor}}

\bibliographystyle{aasjournal}
\bibliography{calib}

\appendix

\section{Dwarf Photometry and GC Results}
\label{app:photo_gcs}
In Table \ref{tab:gc_res}, we list the early-type LV satellites analyzed in this work, presenting their photometry and GC results. All dwarfs are listed together along with which host they are a satellite of, their host's distance, the dwarf's stellar mass, GC abundance results from the simple background subtraction analysis, filter combination ($g/r$ or $g/i$), and their source of data (CFHT/MegaCam, DECaLS, etc.). Other photometry results, such as sizes and surface brightness, can be found in other papers in this series. Note that other papers in this series might use photometry from different sources (e.g. prioritizing DECaLS over deeper data for the sake of homogeneity). While we find overall good agreement in the photometry from different sources, values will not be exactly the same.

As mentioned in the main text, we do not do the GC analysis for satellites of the MW, M31, NGC5236, or CenA. These are still listed in the table as they are part of the nuclei analysis, but do not have entries for the GC abundances. Additionally there are two other LV dwarfs for which we do not list GC abundance results. These are heavily tidally disrupted systems with very large on-sky sizes, which made the GC analysis infeasible. 

The listed GC abundances have all been corrected for GCs outside of the $2r_e$ annulus by assuming the GC half-number radius is $r_\mathrm{GC}=1.5\times r_e$. This correction amounts to dividing by a factor of 0.646. The results have also been corrected for incompleteness to faint GCs that are below the data detection limits. These corrections are generally on the order of $\sim5$\%.

\startlongtable
\begin{deluxetable}{ccccccc}
\tablecaption{Dwarf Photometry and GC Results\label{tab:gc_res}}
\tablehead{
\colhead{Name} & \colhead{Host} & \colhead{$D_\mathrm{host}$} & \colhead{$\log(M_\star^\mathrm{gal}/M_\odot)$} & \colhead{$N_\mathrm{GC}$}  & \colhead{Filters}    & \colhead{Data Source}   \\ 
\colhead{} & \colhead{} & \colhead{(Mpc)}  & \colhead{}  & \colhead{} & \colhead{} & \colhead{} }
\startdata
UGC8882  &  NGC5457  &  6.5  &  7.75  &  $3.1^{+0.0}_{-1.6}$  &  gi  &  CFHT  \\
 M101-DF1  &  NGC5457  &  6.5  &  6.26  &  $1.6^{+0.0}_{-0.0}$  &  gi  &  CFHT  \\
 M101-DF3  &  NGC5457  &  6.5  &  5.77  &  $3.1^{+1.5}_{-3.1}$  &  gi  &  CFHT  \\
 M101-dwA  &  NGC5457  &  6.5  &  6.05  &  $0.0^{+0.0}_{-1.6}$  &  gi  &  CFHT  \\
 dw1350p5441  &  NGC5457  &  6.5  &  6.74  &  $1.6^{+1.6}_{-1.6}$  &  gr  &  DECaLS  \\
 dw1255p4035  &  NGC4736  &  4.2  &  6.19  &  $-1.6^{+1.6}_{-1.6}$  &  gr  &  DECaLS  \\
 dw1251p4137  &  NGC4736  &  4.2  &  5.95  &  $0.0^{+0.0}_{-1.6}$  &  gr  &  DECaLS  \\
 dw1234p2531  &  NGC4565  &  11.9  &  7.65  &  $10.0^{+3.3}_{-3.3}$  &  gr  &  CFHT  \\
 dw1239p3251  &  NGC4631  &  7.4  &  5.77  &  $0.0^{+0.0}_{-1.6}$  &  gr  &  CFHT  \\
 dw1240p3216  &  NGC4631  &  7.4  &  6.25  &  $1.6^{+0.0}_{-0.0}$  &  gr  &  CFHT  \\
 dw1240p3247  &  NGC4631  &  7.4  &  7.53  &  --  &  gr  &  CFHT  \\
 dw1242p3158  &  NGC4631  &  7.4  &  6.15  &  $0.0^{+0.0}_{-0.0}$  &  gr  &  CFHT  \\
 dw1243p3228  &  NGC4631  &  7.4  &  7.06  &  $1.6^{+1.6}_{-0.1}$  &  gr  &  CFHT  \\
 dw1248p3158  &  NGC4631  &  7.4  &  6.88  &  $0.0^{+0.0}_{-2.0}$  &  gr  &  DECaLS  \\
 NGC4258-DF6  &  NGC4258  &  7.2  &  6.36  &  $-1.6^{+1.6}_{-1.6}$  &  gr  &  CFHT  \\
 KDG101  &  NGC4258  &  7.2  &  7.73  &  $6.3^{+3.1}_{-1.6}$  &  gr  &  CFHT  \\
 dw1220p4649  &  NGC4258  &  7.2  &  6.23  &  $1.6^{+0.0}_{-1.6}$  &  gr  &  CFHT  \\
 dw1223p4739  &  NGC4258  &  7.2  &  6.46  &  $0.0^{+0.0}_{-1.6}$  &  gr  &  CFHT  \\
 dw0932p2127  &  NGC2903  &  9.0  &  7.87  &  $3.3^{+1.7}_{-1.7}$  &  gr  &  CFHT  \\
 dw0933p2030  &  NGC2903  &  9.0  &  6.99  &  $0.0^{+0.0}_{-1.9}$  &  gr  &  DECaLS  \\
 dw0932p1952  &  NGC2903  &  9.0  &  6.44  &  $0.0^{+0.0}_{-0.0}$  &  gr  &  DECaLS  \\
 dw1300p1843  &  NGC4826  &  5.3  &  6.62  &  $0.0^{+1.6}_{-1.6}$  &  gr  &  DECaLS  \\
 dw1302p2159  &  NGC4826  &  5.3  &  7.1  &  $0.0^{+0.0}_{-1.6}$  &  gr  &  DECaLS  \\
 dw1252p2215  &  NGC4826  &  5.3  &  5.68  &  $0.0^{+0.0}_{-1.6}$  &  gr  &  DECaLS  \\
 dw1258p2329  &  NGC4826  &  5.3  &  5.83  &  $0.0^{+1.6}_{-1.6}$  &  gr  &  DECaLS  \\
 dw0221p4221  &  NGC891  &  9.12  &  7.11  &  $0.0^{+1.6}_{-1.6}$  &  gr  &  CFHT  \\
 dw0222p4242  &  NGC891  &  9.12  &  6.96  &  $-6.5^{+3.3}_{-4.8}$  &  gr  &  CFHT  \\
 dw1232p0015  &  NGC4517  &  8.34  &  7.03  &  $0.0^{+0.0}_{-1.6}$  &  gi  &  HSC  \\
 dw1238m0035  &  NGC4517  &  8.34  &  6.13  &  $0.0^{+0.0}_{-0.0}$  &  gi  &  HSC  \\
 dw1238p0028  &  NGC4517  &  8.34  &  6.0  &  $0.0^{+0.0}_{-1.6}$  &  gi  &  HSC  \\
 dw1238m0105  &  NGC4517  &  8.34  &  5.65  &  $0.0^{+0.0}_{-0.0}$  &  gi  &  HSC  \\
 dw1907m6342  &  NGC6744  &  8.95  &  6.71  &  $1.8^{+1.8}_{-3.3}$  &  gi  &  DECam  \\
 dw1906m6357  &  NGC6744  &  8.95  &  7.46  &  $10.5^{+3.5}_{-3.5}$  &  gi  &  DECam  \\
 dw1913m6154  &  NGC6744  &  8.95  &  7.5  &  $3.4^{+3.4}_{-3.4}$  &  gi  &  DECam  \\
 dw1905m6316  &  NGC6744  &  8.95  &  6.69  &  $1.7^{+1.7}_{-3.5}$  &  gi  &  DECam  \\
 dw1310p4153  &  NGC5055  &  8.87  &  6.48  &  $0.0^{+1.6}_{-1.6}$  &  gr  &  DECaLS  \\
 dw1312p4158  &  NGC5055  &  8.87  &  6.23  &  $0.0^{+1.6}_{-1.6}$  &  gr  &  DECaLS  \\
 dw1312p4147  &  NGC5055  &  8.87  &  8.1  &  $9.7^{+3.2}_{-4.8}$  &  gr  &  DECaLS  \\
 dw1313p4202  &  NGC5055  &  8.87  &  6.87  &  $-1.7^{+1.7}_{-1.7}$  &  gr  &  DECaLS  \\
 dw1315p4123  &  NGC5055  &  8.87  &  5.74  &  $0.0^{+0.0}_{-1.7}$  &  gr  &  DECaLS  \\
 dw1315p4130  &  NGC5055  &  8.87  &  6.92  &  $-1.8^{+2.5}_{-1.8}$  &  gr  &  DECaLS  \\
 dw1321p4226  &  NGC5055  &  8.87  &  6.06  &  $0.0^{+0.0}_{-0.0}$  &  gr  &  DECaLS  \\
 Sgr  &  MW  &  0.0  &  8.12  &  $2.0^{+0.0}_{-0.0}$  &    &    \\
 Fnx  &  MW  &  0.0  &  7.5  &  $4.0^{+0.0}_{-0.0}$  &    &    \\
 Leo1  &  MW  &  0.0  &  6.78  &  $0.0^{+0.0}_{-0.0}$  &    &    \\
 Scl  &  MW  &  0.0  &  6.36  &  $0.0^{+0.0}_{-0.0}$  &    &    \\
 Leo2  &  MW  &  0.0  &  5.89  &  $0.0^{+0.0}_{-0.0}$  &    &    \\
 Umi  &  MW  &  0.0  &  5.58  &  $0.0^{+0.0}_{-0.0}$  &    &    \\
 Car  &  MW  &  0.0  &  5.76  &  $0.0^{+0.0}_{-0.0}$  &    &    \\
 NGC147  &  M31  &  0.78  &  8.01  &  $7.0^{+0.0}_{-0.0}$  &    &    \\
 NGC185  &  M31  &  0.78  &  8.05  &  $4.0^{+0.0}_{-0.0}$  &    &    \\
 AndI  &  M31  &  0.78  &  6.81  &  $0.0^{+0.0}_{-0.0}$  &    &    \\
 AndIII  &  M31  &  0.78  &  6.08  &  $0.0^{+0.0}_{-0.0}$  &    &    \\
 AndV  &  M31  &  0.78  &  5.81  &  $0.0^{+0.0}_{-0.0}$  &    &    \\
 AndXXXII  &  M31  &  0.78  &  6.98  &  $0.0^{+0.0}_{-0.0}$  &    &    \\
 AndXV  &  M31  &  0.78  &  5.74  &  $0.0^{+0.0}_{-0.0}$  &    &    \\
 AndII  &  M31  &  0.78  &  7.12  &  $0.0^{+0.0}_{-0.0}$  &    &    \\
 AndVII  &  M31  &  0.78  &  7.39  &  $0.0^{+0.0}_{-0.0}$  &    &    \\
 AndXXXI  &  M31  &  0.78  &  6.74  &  $0.0^{+0.0}_{-0.0}$  &    &    \\
 AndVI  &  M31  &  0.78  &  6.64  &  $0.0^{+0.0}_{-0.0}$  &    &    \\
 dw0133p1543  &  NGC628  &  9.77  &  6.22  &  $0.0^{+0.0}_{-0.0}$  &  gr  &  DECaLS  \\
 dw0134p1544  &  NGC628  &  9.77  &  7.07  &  $0.0^{+1.9}_{-3.7}$  &  gr  &  DECaLS  \\
 dw0136p1628  &  NGC628  &  9.77  &  7.1  &  $0.0^{+0.0}_{-1.6}$  &  gr  &  DECaLS  \\
 dw0137p1537  &  NGC628  &  9.77  &  7.17  &  $0.0^{+0.0}_{-1.7}$  &  gr  &  DECaLS  \\
 dw0137p1607  &  NGC628  &  9.77  &  6.24  &  $0.0^{+0.0}_{-1.6}$  &  gr  &  DECaLS  \\
 KDG61  &  M81  &  3.61  &  7.51  &  $-0.8^{+2.3}_{-2.3}$  &  gr  &  DECaLS  \\
 BK5N  &  M81  &  3.61  &  6.65  &  $0.8^{+0.8}_{-2.4}$  &  gr  &  DECaLS  \\
 IKN  &  M81  &  3.61  &  7.67  &  $1.6^{+4.2}_{-1.6}$  &  gr  &  DECaLS  \\
 Fm1  &  M81  &  3.61  &  7.17  &  $-2.3^{+0.8}_{-2.3}$  &  gr  &  DECaLS  \\
 KDG64  &  M81  &  3.61  &  7.6  &  $-3.1^{+1.6}_{-3.1}$  &  gr  &  DECaLS  \\
 F8D1  &  M81  &  3.61  &  6.73  &  $0.0^{+1.6}_{-3.2}$  &  gr  &  DECaLS  \\
 KK77  &  M81  &  3.61  &  7.63  &  $4.6^{+1.0}_{-4.1}$  &  gr  &  DECaLS  \\
 D1006p67  &  M81  &  3.61  &  5.69  &  $0.0^{+0.0}_{-1.5}$  &  gr  &  DECaLS  \\
 KDG63  &  M81  &  3.61  &  7.51  &  $4.7^{+0.2}_{-3.1}$  &  gr  &  DECaLS  \\
 DDO78  &  M81  &  3.61  &  7.89  &  $-4.8^{+4.8}_{-3.9}$  &  gr  &  DECaLS  \\
 HS117  &  M81  &  3.61  &  6.82  &  $0.0^{+1.6}_{-1.6}$  &  gr  &  DECaLS  \\
 D0944p71  &  M81  &  3.61  &  7.18  &  $-1.6^{+1.6}_{-1.6}$  &  gr  &  DECaLS  \\
 BK6N  &  M81  &  3.61  &  6.7  &  $-3.1^{+3.1}_{-1.6}$  &  gr  &  DECaLS  \\
 D0934p70  &  M81  &  3.61  &  6.03  &  $-1.6^{+1.6}_{-1.6}$  &  gr  &  DECaLS  \\
 dw1237m1125  &  M104  &  9.55  &  6.77  &  $0.0^{+0.0}_{-1.7}$  &  gi  &  CFHT  \\
 dw1239m1159  &  M104  &  9.55  &  6.08  &  $-1.6^{+1.6}_{-1.6}$  &  gi  &  CFHT  \\
 dw1239m1143  &  M104  &  9.55  &  7.56  &  $1.6^{+1.6}_{-1.8}$  &  gi  &  CFHT  \\
 dw1239m1113  &  M104  &  9.55  &  6.88  &  $0.0^{+0.0}_{-3.2}$  &  gi  &  CFHT  \\
 dw1239m1120  &  M104  &  9.55  &  6.07  &  $0.0^{+0.0}_{-1.6}$  &  gi  &  CFHT  \\
 dw1239m1144  &  M104  &  9.55  &  7.06  &  $3.1^{+3.1}_{-7.9}$  &  gi  &  CFHT  \\
 dw1240m1118  &  M104  &  9.55  &  7.8  &  $-0.8^{+2.4}_{-2.4}$  &  gi  &  CFHT  \\
 dw1240m1140  &  M104  &  9.55  &  6.92  &  --  &  gi  &  CFHT  \\
 dw1241m1131  &  M104  &  9.55  &  6.04  &  $0.0^{+0.0}_{-1.7}$  &  gi  &  CFHT  \\
 dw1241m1153  &  M104  &  9.55  &  6.71  &  $1.6^{+1.0}_{-1.6}$  &  gi  &  CFHT  \\
 dw1241m1155  &  M104  &  9.55  &  7.13  &  $-3.1^{+3.1}_{-1.6}$  &  gi  &  CFHT  \\
 dw0233p3852  &  NGC1023  &  10.4  &  6.49  &  $2.5^{+4.1}_{-2.5}$  &  gi  &  CFHT  \\
 dw0237p3836  &  NGC1023  &  10.4  &  6.74  &  $0.0^{+1.6}_{-1.6}$  &  gi  &  CFHT  \\
 dw0239p3926  &  NGC1023  &  10.4  &  6.72  &  $-0.8^{+2.4}_{-2.5}$  &  gi  &  CFHT  \\
 dw0239p3903  &  NGC1023  &  10.4  &  5.68  &  $0.0^{+0.0}_{-1.6}$  &  gi  &  CFHT  \\
 dw0239p3902  &  NGC1023  &  10.4  &  6.02  &  $0.0^{+0.0}_{-1.6}$  &  gi  &  CFHT  \\
 CenA-mm-dw1  &  NGC5128  &  3.66  &  7.36  &  --  &  gr  &  DECam  \\
 CenA-mm-dw2  &  NGC5128  &  3.66  &  6.35  &  --  &  gr  &  DECam  \\
 CenA-mm-dw4  &  NGC5128  &  3.66  &  6.19  &  --  &  gr  &  DECam  \\
 CenA-mm-dw8  &  NGC5128  &  3.66  &  6.58  &  --  &  gr  &  DECam  \\
 CenA-mm-dw11  &  NGC5128  &  3.66  &  5.87  &  --  &  gr  &  DECam  \\
 dw1323-40B  &  NGC5128  &  3.66  &  6.02  &  --  &  gr  &  DECam  \\
 dw1323-40  &  NGC5128  &  3.66  &  6.23  &  --  &  gr  &  DECam  \\
 dw1342-43  &  NGC5128  &  3.66  &  6.21  &  --  &  gr  &  DECam  \\
 ESO269-066  &  NGC5128  &  3.66  &  8.0  &  --  &  gr  &  DECam  \\
 KK189  &  NGC5128  &  3.66  &  6.82  &  --  &  gr  &  DECam  \\
 KK197  &  NGC5128  &  3.66  &  7.96  &  --  &  gr  &  DECam  \\
 KK213  &  NGC5128  &  3.66  &  6.51  &  --  &  gr  &  DECam  \\
 KKS55  &  NGC5128  &  3.66  &  7.14  &  --  &  gr  &  DECam  \\
 dw1335-29  &  NGC5236  &  4.7  &  5.88  &  --  &  gr  &  DECam  \\
 dw1340-30  &  NGC5236  &  4.7  &  6.22  &  --  &  gi  &  DECam  \\
 KK208  &  NGC5236  &  4.7  &  7.5  &  --  &  gr  &  DECam  \\
 KK218  &  NGC5236  &  4.7  &  6.76  &  --  &  gr  &  DECam  \\
 dw1122p1326  &  NGC3627  &  10.5  &  8.47  &  $16.2^{+1.6}_{-3.2}$  &  gr  &  DECaLS  \\
 dw1122p1258  &  NGC3627  &  10.5  &  7.52  &  $0.0^{+1.6}_{-1.6}$  &  gi  &  HSC  \\
 dw1119p1419  &  NGC3627  &  10.5  &  7.59  &  $11.9^{+3.4}_{-4.4}$  &  gr  &  DECaLS  \\
 dw1119p1404  &  NGC3627  &  10.5  &  7.68  &  $14.4^{+1.6}_{-1.7}$  &  gi  &  HSC  \\
 dw1118p1233  &  NGC3627  &  10.5  &  7.49  &  $-1.6^{+1.6}_{-1.6}$  &  gr  &  DECaLS  \\
 dw1120p1332  &  NGC3627  &  10.5  &  7.38  &  $-4.7^{+6.6}_{-14.1}$  &  gi  &  HSC  \\
 dw1118p1348  &  NGC3627  &  10.5  &  6.4  &  $0.0^{+1.6}_{-3.2}$  &  gi  &  HSC  \\
 dw1121p1326  &  NGC3627  &  10.5  &  6.67  &  $1.6^{+1.6}_{-3.2}$  &  gi  &  HSC  \\
 dw1050p1316  &  NGC3379  &  10.7  &  8.38  &  $15.5^{+3.4}_{-5.0}$  &  gr  &  DECaLS  \\
 dw1048p1408  &  NGC3379  &  10.7  &  8.3  &  $6.8^{+1.7}_{-3.4}$  &  gr  &  DECaLS  \\
 dw1052p1102  &  NGC3379  &  10.7  &  8.05  &  $1.6^{+1.6}_{-3.3}$  &  gr  &  DECaLS  \\
 dw1051p1250  &  NGC3379  &  10.7  &  7.86  &  $4.9^{+3.3}_{-3.3}$  &  gr  &  DECaLS  \\
 dw1044p1356  &  NGC3379  &  10.7  &  7.72  &  $3.2^{+3.2}_{-4.7}$  &  gi  &  HSC  \\
 dw1051p1320  &  NGC3379  &  10.7  &  7.87  &  $3.4^{+5.1}_{-3.4}$  &  gr  &  DECaLS  \\
 dw1046p1219  &  NGC3379  &  10.7  &  7.66  &  $3.1^{+1.6}_{-3.1}$  &  gi  &  HSC  \\
 dw1052p1501  &  NGC3379  &  10.7  &  7.64  &  $0.0^{+0.0}_{-1.7}$  &  gr  &  DECaLS  \\
 dw1050p1221  &  NGC3379  &  10.7  &  7.21  &  $3.2^{+1.6}_{-4.7}$  &  gi  &  HSC  \\
 dw1042p1208  &  NGC3379  &  10.7  &  7.59  &  $0.0^{+3.4}_{-3.4}$  &  gr  &  DECaLS  \\
 dw1046p1401  &  NGC3379  &  10.7  &  7.51  &  $5.1^{+1.7}_{-3.4}$  &  gr  &  DECaLS  \\
 dw1046p1259  &  NGC3379  &  10.7  &  7.22  &  $1.6^{+3.2}_{-1.6}$  &  gi  &  HSC  \\
 dw1051p1406  &  NGC3379  &  10.7  &  7.26  &  $1.8^{+0.0}_{-1.8}$  &  gr  &  DECaLS  \\
 dw1055p1220  &  NGC3379  &  10.7  &  7.38  &  $7.8^{+2.0}_{-3.9}$  &  gr  &  DECaLS  \\
 dw1046p1145  &  NGC3379  &  10.7  &  6.83  &  $-1.6^{+1.6}_{-1.6}$  &  gi  &  HSC  \\
 dw1047p1248  &  NGC3379  &  10.7  &  6.45  &  $4.8^{+1.6}_{-3.2}$  &  gi  &  HSC  \\
 dw1046p1247  &  NGC3379  &  10.7  &  6.99  &  $0.0^{+1.6}_{-1.6}$  &  gi  &  HSC  \\
 dw1046p1257  &  NGC3379  &  10.7  &  6.8  &  $1.6^{+1.6}_{-1.6}$  &  gi  &  HSC  \\
 dw1048p1259  &  NGC3379  &  10.7  &  6.03  &  $0.0^{+1.6}_{-3.3}$  &  gi  &  HSC  \\
 dw1047p1257  &  NGC3379  &  10.7  &  6.8  &  $-1.6^{+1.6}_{-1.6}$  &  gi  &  HSC  \\
 dw1044p1351  &  NGC3379  &  10.7  &  6.47  &  $1.6^{+0.0}_{-3.1}$  &  gi  &  HSC  \\
 dw1048p1303  &  NGC3379  &  10.7  &  6.42  &  $1.6^{+1.6}_{-1.6}$  &  gi  &  HSC  \\
 dw1048p1158  &  NGC3379  &  10.7  &  6.5  &  $0.0^{+0.0}_{-3.1}$  &  gi  &  HSC  \\
 dw1047p1258  &  NGC3379  &  10.7  &  6.39  &  $0.8^{+0.8}_{-0.8}$  &  gi  &  HSC  \\
 dw1049p1233  &  NGC3379  &  10.7  &  6.6  &  $3.2^{+1.6}_{-1.7}$  &  gi  &  HSC  \\
 dw1049p1247  &  NGC3379  &  10.7  &  6.26  &  $2.4^{+0.8}_{-2.4}$  &  gi  &  HSC  \\
 dw1047p1202  &  NGC3379  &  10.7  &  6.32  &  $1.6^{+1.6}_{-0.0}$  &  gi  &  HSC  \\
 dw1043p1415  &  NGC3379  &  10.7  &  6.27  &  $1.6^{+1.6}_{-1.6}$  &  gi  &  HSC  \\
 dw1044p1351B  &  NGC3379  &  10.7  &  6.29  &  $0.0^{+1.6}_{-1.6}$  &  gi  &  HSC  \\
 dw1043p1410  &  NGC3379  &  10.7  &  6.03  &  $-1.6^{+1.6}_{-0.0}$  &  gi  &  HSC  \\
 dw1050p1236  &  NGC3379  &  10.7  &  6.03  &  $0.0^{+1.6}_{-1.6}$  &  gi  &  HSC  \\
 dw1048p1154  &  NGC3379  &  10.7  &  6.56  &  $2.4^{+3.9}_{-7.1}$  &  gi  &  HSC  \\
 dw1044p1359  &  NGC3379  &  10.7  &  5.99  &  $0.0^{+0.0}_{-1.6}$  &  gi  &  HSC  \\
 dw1042p1359  &  NGC3379  &  10.7  &  5.82  &  $0.0^{+0.0}_{-1.6}$  &  gi  &  HSC  \\
 dw1050p1213  &  NGC3379  &  10.7  &  5.75  &  $0.0^{+0.0}_{-1.6}$  &  gi  &  HSC  \\
 dw1005m0744  &  NGC3115  &  10.2  &  8.02  &  $6.6^{+4.9}_{-4.6}$  &  gr  &  DECaLS  \\
 dw1000m0821  &  NGC3115  &  10.2  &  7.91  &  $4.9^{+1.6}_{-3.3}$  &  gr  &  DECaLS  \\
 dw1000m0741  &  NGC3115  &  10.2  &  7.3  &  $-1.7^{+1.7}_{-3.4}$  &  gr  &  DECaLS  \\
 dw1007m0715  &  NGC3115  &  10.2  &  7.25  &  $-1.6^{+1.6}_{-3.2}$  &  gr  &  DECaLS  \\
 dw1006m0730  &  NGC3115  &  10.2  &  6.96  &  $0.0^{+0.0}_{-1.7}$  &  gr  &  DECaLS  \\
 dw1004m0657  &  NGC3115  &  10.2  &  6.63  &  $0.0^{+1.6}_{-3.2}$  &  gr  &  DECaLS  \\
 dw1004m0737  &  NGC3115  &  10.2  &  6.57  &  $0.9^{+2.6}_{-4.3}$  &  gr  &  DECaLS  \\
 dw1006m0732  &  NGC3115  &  10.2  &  6.53  &  $-1.7^{+1.7}_{-0.0}$  &  gr  &  DECaLS  \\
 dw1006m0730-n2  &  NGC3115  &  10.2  &  6.41  &  $4.9^{+1.6}_{-1.6}$  &  gr  &  DECaLS  \\
 dw1007m0830  &  NGC3115  &  10.2  &  6.12  &  $0.0^{+0.0}_{-1.8}$  &  gr  &  DECaLS  \\
 dw1000m0831  &  NGC3115  &  10.2  &  6.06  &  $1.7^{+0.0}_{-1.7}$  &  gr  &  DECaLS  \\
 dw1002m0642  &  NGC3115  &  10.2  &  5.66  &  $0.0^{+0.0}_{-1.6}$  &  gr  &  DECaLS  \\
 dw1104p0003  &  NGC3521  &  11.2  &  7.55  &  $-4.0^{+2.4}_{-5.6}$  &  gr  &  CFHT  \\
 dw1110p0037  &  NGC3521  &  11.2  &  7.62  &  $4.8^{+1.6}_{-5.0}$  &  gr  &  HSC  \\
 dw0855p3333  &  NGC2683  &  9.4  &  7.51  &  $-1.7^{+3.3}_{-5.0}$  &  gr  &  DECaLS  \\
 dw0323m4040  &  NGC1291  &  9.08  &  7.84  &  $3.2^{+1.6}_{-3.2}$  &  gr  &  DECaLS  \\
 dw0317m4142  &  NGC1291  &  9.08  &  7.42  &  $-1.6^{+1.6}_{-3.2}$  &  gr  &  DECaLS  \\
 dw0317m4058  &  NGC1291  &  9.08  &  7.48  &  $-3.2^{+3.2}_{-3.2}$  &  gr  &  DECaLS  \\
 dw0318m4101  &  NGC1291  &  9.08  &  6.73  &  $1.6^{+0.0}_{-1.6}$  &  gr  &  DECaLS  \\
 dw0507m3744  &  NGC1808  &  9.29  &  7.37  &  $0.0^{+0.0}_{-1.6}$  &  gr  &  DECaLS  \\
 dw0508m3808  &  NGC1808  &  9.29  &  7.45  &  $1.6^{+0.0}_{-1.6}$  &  gr  &  DECaLS  \\
 dw0507m3739  &  NGC1808  &  9.29  &  7.01  &  $1.6^{+1.6}_{-3.3}$  &  gr  &  DECaLS  \\
 \enddata
\tablecomments{The main GC abundance results in this work}
\end{deluxetable}

\begin{figure}
\includegraphics[width=0.5\textwidth]{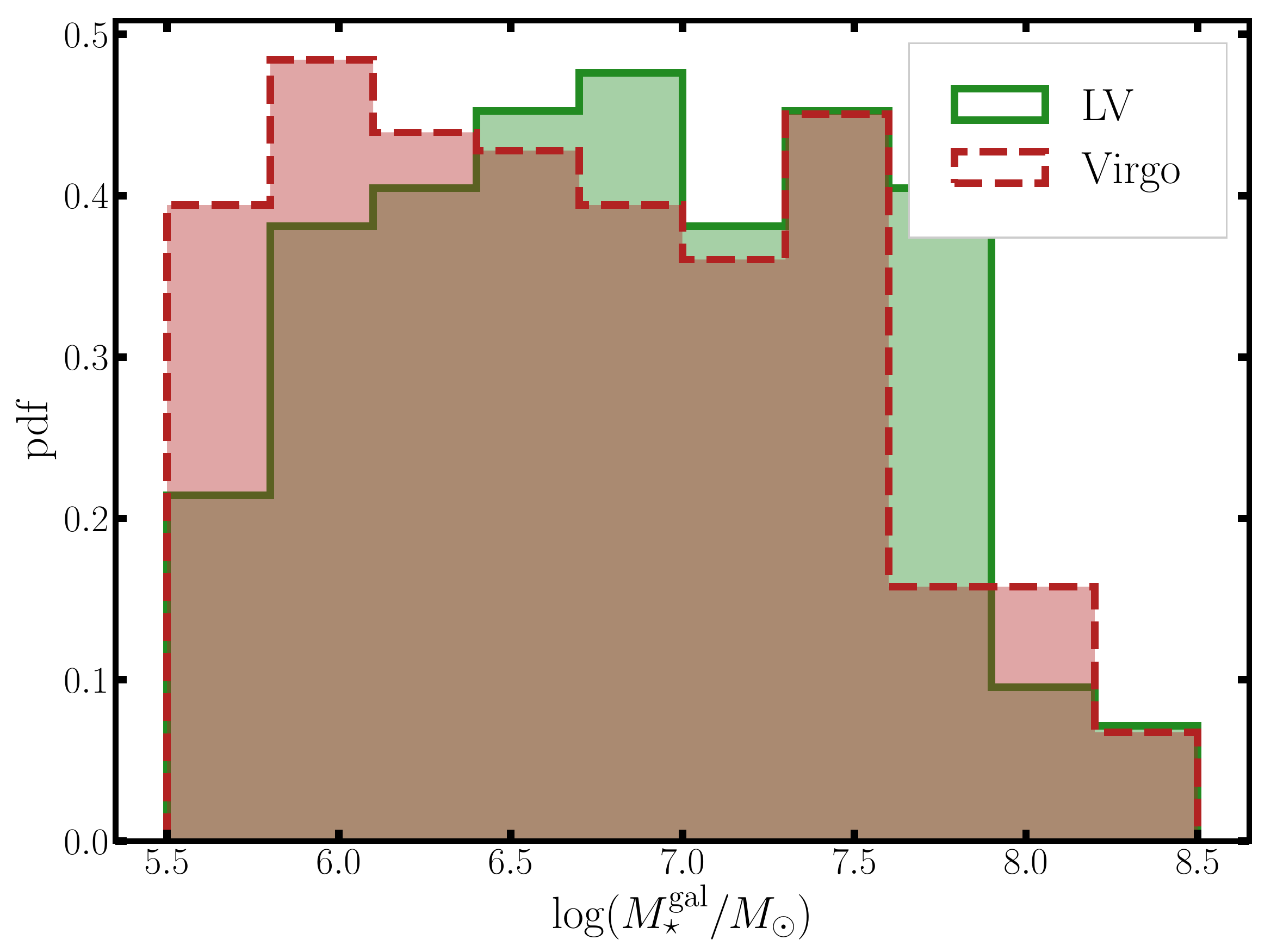}
\caption{The mass distributions for the early-type dwarfs used in the GC analysis in both the LV and Virgo samples. The relative distribution of dwarf masses in the included range of $5.5<\log(M_\star/M_\odot) < 8.5$ is quite similar between the two environmental samples. }
\label{fig:mass_dists}
\end{figure}

\section{ELVES and Virgo Dwarf Sample Mass Distributions}
\label{app:mass_distributions}
In Figure \ref{fig:mass_dists}, we compare the stellar mass distributions between the LV and Virgo samples used in the GC analysis. They are quite comparable across the analyzed mass range. While we would expect that the overall dwarf populations in these two environments to have similar mass distribution shapes \cite[i.e. the luminosity functions likely have similar power law slopes, see][]{carlsten2020b}, the \textit{early-type} sub-sample will differ due to the dearth of massive early-type dwarfs in the LV environments. It appears that this is largely balanced out by the fact that we are only including LV dwarf satellites which have measured distances, which is more likely for the more massive dwarfs. Thus the LV sample has a boost in massive dwarfs compared to lower-mass dwarfs. Along this vein, the turn-down at very low masses of the mass distribution is not incompleteness in the detection catalogs but due to the fact that many low-mass dwarf candidate satellites don't have conclusive distance confirmation yet. Analyses of satellites luminosity functions will clearly need to account for this, but leaving these candidate satellites out for the analysis done here will not affect the conclusions of this work.

\section{NSC Photometry}
\label{app:photo_nsc}
In Table \ref{tab:nuc_phot}, we list the NSCs found in the LV satellites. We list the host dwarf's name, NSC $M_g$, color, and stellar mass.

\begin{deluxetable}{cccc}
\tablecaption{Dwarf Nuclei Photometry\label{tab:nuc_phot}}
\tablehead{
\colhead{Name} & \colhead{$\log(M_\star^\mathrm{NSC}/M_\odot)$} & \colhead{$M_g^\mathrm{NSC}$}  & \colhead{$(g-i)^\mathrm{NSC}$}   \\ 
\colhead{} & \colhead{} & \colhead{(mag)}  & \colhead{(mag)}  }
\startdata
UGC8882  &  6.61  &  -11.34  &  0.7  \\
 dw1234p2531  &  5.64  &  -9.2  &  0.63  \\
 KDG101  &  6.26  &  -10.6  &  0.68  \\
 dw1223p4739  &  5.46  &  -8.6  &  0.68  \\
 dw0932p2127  &  5.11  &  -7.08  &  0.91  \\
 dw1913m6154  &  5.69  &  -10.23  &  0.34  \\
 Sgr  &  6.15  &  --  &  --  \\
 NGC147  &  5.1  &  --  &  --  \\
 KDG61  &  5.22  &  -7.16  &  0.97  \\
 BK5N  &  5.38  &  -6.3  &  1.41  \\
 dw1239m1143  &  5.6  &  -8.55  &  0.79  \\
 dw1240m1118  &  5.79  &  -9.46  &  0.65  \\
 dw1240m1140  &  6.13  &  -9.31  &  0.97  \\
 dw1241m1131  &  5.55  &  -7.95  &  0.93  \\
 dw1241m1155  &  5.35  &  -8.26  &  0.69  \\
 dw0237p3836  &  4.73  &  -6.88  &  0.63  \\
 CenA-mm-dw1  &  5.63  &  -8.58  &  0.83  \\
 ESO269-066  &  6.19  &  -9.67  &  0.94  \\
 KK197  &  6.04  &  -9.48  &  0.88  \\
 dw1122p1258  &  5.77  &  -9.8  &  0.53  \\
 dw1119p1404  &  4.27  &  -6.26  &  0.47  \\
 dw1048p1408  &  6.37  &  -9.74  &  1.08  \\
 dw1051p1320  &  5.49  &  -8.53  &  0.74  \\
 dw1046p1219  &  5.7  &  -8.44  &  0.9  \\
 dw1050p1221  &  4.69  &  -6.6  &  0.69  \\
 dw1042p1208  &  5.55  &  -7.84  &  1.03  \\
 dw1046p1145  &  5.57  &  -8.54  &  0.77  \\
 dw1048p1303  &  4.8  &  -6.94  &  0.67  \\
 dw1048p1158  &  5.03  &  -7.2  &  0.76  \\
 dw1047p1258  &  5.52  &  -9.0  &  0.59  \\
 dw1049p1233  &  5.03  &  -7.43  &  0.69  \\
 dw1047p1202  &  5.28  &  -8.19  &  0.66  \\
 dw1048p1154  &  5.51  &  -8.62  &  0.7  \\
 dw1005m0744  &  6.16  &  -9.29  &  1.05  \\
 dw1004m0657  &  5.23  &  -7.26  &  0.95  \\
 dw1006m0730-n2  &  4.5  &  -5.95  &  0.77  \\
 dw1110p0037  &  5.38  &  -7.96  &  0.84  \\
 dw0323m4040  &  5.04  &  -6.67  &  0.98  \\
 dw0317m4058  &  4.62  &  -6.27  &  0.76  \\
 dw0318m4101  &  4.96  &  -6.9  &  0.83  \\
 \enddata
\tablecomments{The photometry for the nuclei of the LV dwarfs considered in the current work}
\end{deluxetable}

\begin{figure*}
\includegraphics[width=\textwidth]{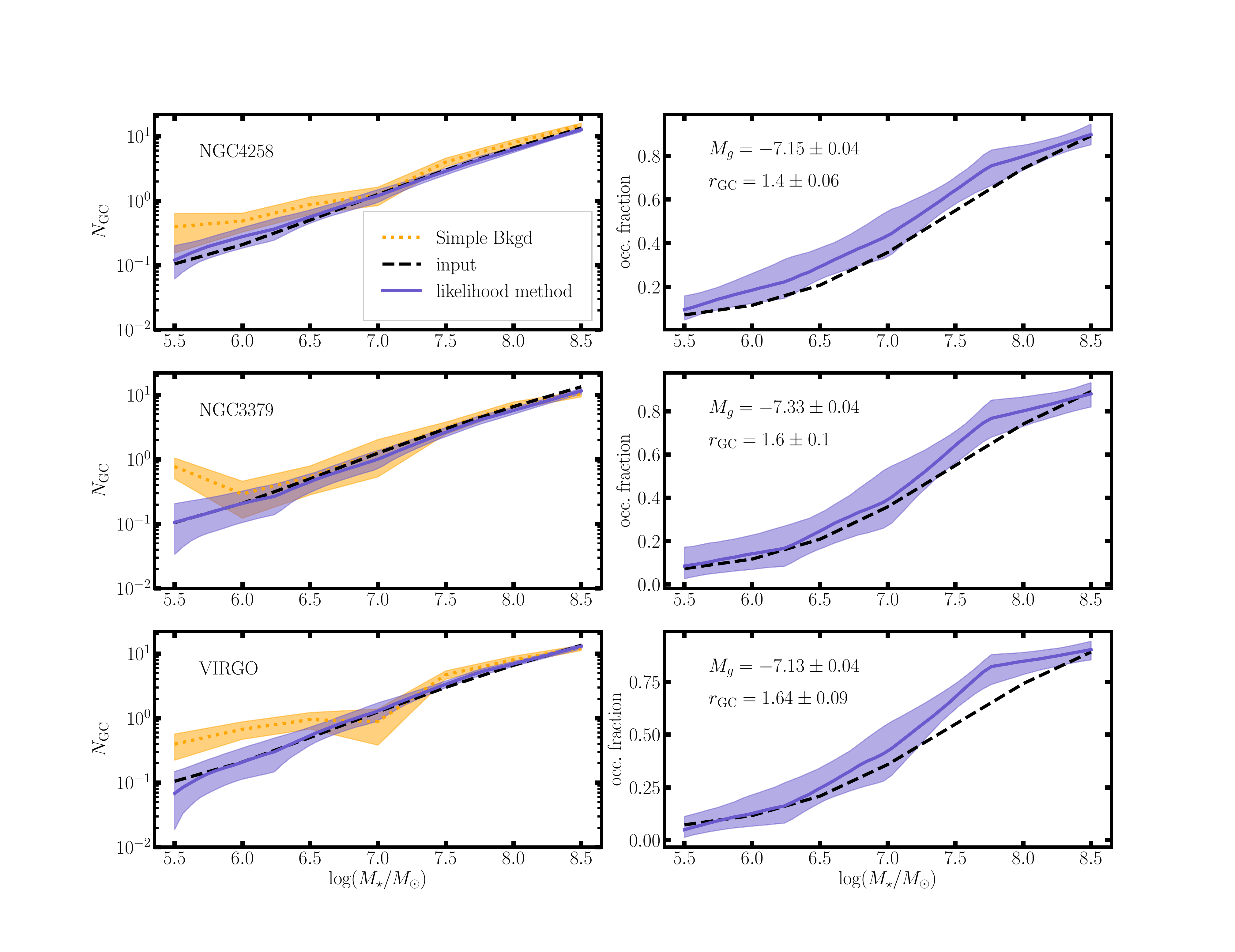}
\caption{The results from realistic end-to-end simulations of artificial dwarfs with GC populations used to test our GC analysis methods. Each row corresponds to a different host whose data quality typifies the range of data used in this work. The left column shows the overall average GC abundance as a function of stellar mass. Both the average result from the simple background subtraction analysis and the result from the likelihood-based inference are shown compared to the input relation, given by the dashed black line. Both methods recover the input within the estimated uncertainties. The right column shows the occupation fraction of GCs (i.e. fraction of dwarfs with non-zero GC systems) inferred from the likelihood-based analysis. Again, we are able to recover the input relation well. The numbers in the corner show the inferred GCLF peak luminosity and the GC half-number radius. For these quantities, the input values were $M_g=-7.2$ mag and $r_\mathrm{GC}=1.5$ (in units of $r_e$). }
\label{fig:sims_results}
\end{figure*}

\section{GC Analysis Tests Using Image Simulations}
\label{app:simulations}
In this section, we describe in more detail the image simulations we use to verify our GC analysis procedures, both the simple background subtraction and the joint likelihood-based inference. We chose three hosts which span the range in data quality used throughout this paper and simulate satellites at the distance of these hosts in empty fields adjacent to the real satellites. In particular we simulate dwarfs using CFHT/MegaCam data for NGC 4258, DECaLS data for NGC 3379, and CFHT/MegaCam data for Virgo. Due to the depth of the data and relatively nearby distance of NGC 4258 ($D=7.4$ Mpc), this represents a `best-case' for the LV sample while the shallower DECaLS data for NGC 3379, which is one of the most distant hosts ($D=10.7$ Mpc), represents a `worst-case'. Due to the importance of the Virgo analysis as a comparison to the LV sample, we include it in the simulations as well.

We simulate dwarfs at 7 stellar masses: $\log(M_\star/M_\odot)=5.5, 6.0, 6.5, 7.0, 7.5, 8.0, 8.5$. Thirty artifical dwarfs are simulated at each stellar mass for each host. The dwarfs are generated using S\'{e}rsic profiles with $n=0.9$ and ellipticity uniformly drawn from the range [0,0.5]. The sizes are drawn from roughly the same mass size distribution as that observed for the real LV dwarf satellites. Dwarfs are simulated with $g-i=0.7$ or $g-r=0.5$, depending on the filter combination used for that host.

The fraction of dwarfs at each stellar mass that are given a non-zero GC system is taken from a smooth function that starts at $\sim0$ at $M_\star=10^{5.5} M_\odot$ and reaches 1 at $M_\star\sim10^{9} M_\odot$. The dwarfs assigned a GC system are given a number of GCs according to $N_\mathrm{GC}=1+25(M_\star/10^9M_\odot)^{0.5}$, which is roughly the same scaling observed for the real dwarfs, albeit with a little higher abundance. The GC luminosities are drawn from a Gaussian with mean $M_g=-7.2$ mag and dispersion $\sigma_g=0.7$ mag. The GCs are assigned colors from a Gaussian with mean of 0.75 mag and scale of 0.15 mag if $g/i$ is used and mean of 0.4 mag and scale of 0.1 mag if $g/r$ is used. The GCs are spatially located around the dwarf according to a Plummer profile with half-number radius $r_\mathrm{GC}=1.5\times r_e$. 

At this point, the artificial dwarfs are processed through the same pipeline as the real galaxies, including modelling of the galaxy light profile and counting of the candidate GCs. Figure \ref{fig:sims_results} shows the results of the simulations, comparing both the inferred average GC abundance and occupation fractions to the input relations. We can see that both GC counting methods recover the input average GC abundance within the errors. The likelihood-based method also recovers the GC occupation fraction, GCLF peak, and GC half-number radius well, although it seems to infer slightly higher occupation fractions than are input. Since this offset is about the same in the Virgo and LV simulations, it will not affect the main conclusions in this work about the effect of environment.

\section{Stellar Masses of Reference Dwarf Samples}
\label{app:stellar_masses}
 In this section, we describe how we calculate stellar masses for the reference environmental samples of dwarfs that we compare with in the NSC and GC analyses. 
 
 \textbf{\citet{eigenthaler2018} NGFS Fornax Sample:} \citet{eigenthaler2018} report stellar masses for NGFS dwarfs, however, in Carlsten et al. (submitted) we find a significant, unexplained bias between our photometry of NGFS dwarfs using DECaLS with that reported in \citet{eigenthaler2018}. Thus, for the nucleation fraction comparison, we use stellar masses for Fornax dwarfs that we measure from DECaLS images (see Carlsten et al., submitted, for details). These stellar masses use the same color-$M/L$ relation \citep[from][]{into2013} as used for the LV dwarfs so assumptions about the IMF, etc. are the same.

 \textbf{\citet{ordenes-briceno_spatial} NGFS Fornax Sample:} For the more extensive NGFS dwarf sample (going out to $R_\mathrm{vir}/2$ as opposed to $R_\mathrm{vir}/4$) of \citet{ordenes-briceno_spatial}, we do not perform our own photometry like in the preceding paragraph. Instead, we use the results described above for the inner NGFS sample to fit a line between $M_i$ reported by the NGFS team to the $M_\star$ we derive from our own photometry. We then use this relation to convert the  $M_i$'s given by \citet{ordenes-briceno_spatial} into stellar masses.

 \textbf{Coma Sample:} For the stellar masses of the Coma sample \citet{denbrok2014},  we  use  a  relation  between $g-i$ color and $M_i$ derived from the Virgo dwarfs to assign average colors to each Coma dwarf based on its reported $M_I$ luminosity. Then we use the color-$M/L$ relation from \citet{into2013} to determine a stellar mass.
 
 \textbf{\citet{miller2007} Virgo Sample:} We derive stellar masses for the \citet{miller2007} dwarfs from S\'{e}rsic photometry using DECaLS images in the same way as for the LV early-type dwarfs \citep[i.e. the color-$M/L$ relations are taken from][]{into2013}. 

\textbf{\citet{peng2008} Virgo Sample:} \citet{peng2008} list stellar masses for their Virgo galaxy sample, and we use those. Those stellar masses are calculated assuming a \citet{chabrier2003} IMF while the other stellar masses in this work assume a \citet{kroupa1998} IMF as that is what \citet{into2013} use. This could lead to a $\sim0.2$ dex bias in stellar mass, but that would not change the interpretation of the \citet{peng2008} comparison.

\textbf{LG and \emph{HST} LV Samples:} We get $V$-band luminosities for the LG dwarfs from \citet{mcconnachie2012} and for the \emph{HST} LV sample from \citet{georgiev2009a} and \citet{sharina2005}. To convert $M_V$'s into stellar masses, we use a relation between $M_\star/L_V$ and $M_V$ that we derive for the ELVES dwarfs using the color-M/L relations from \citet{into2013}.

\begin{figure}
\includegraphics[width=0.5\textwidth]{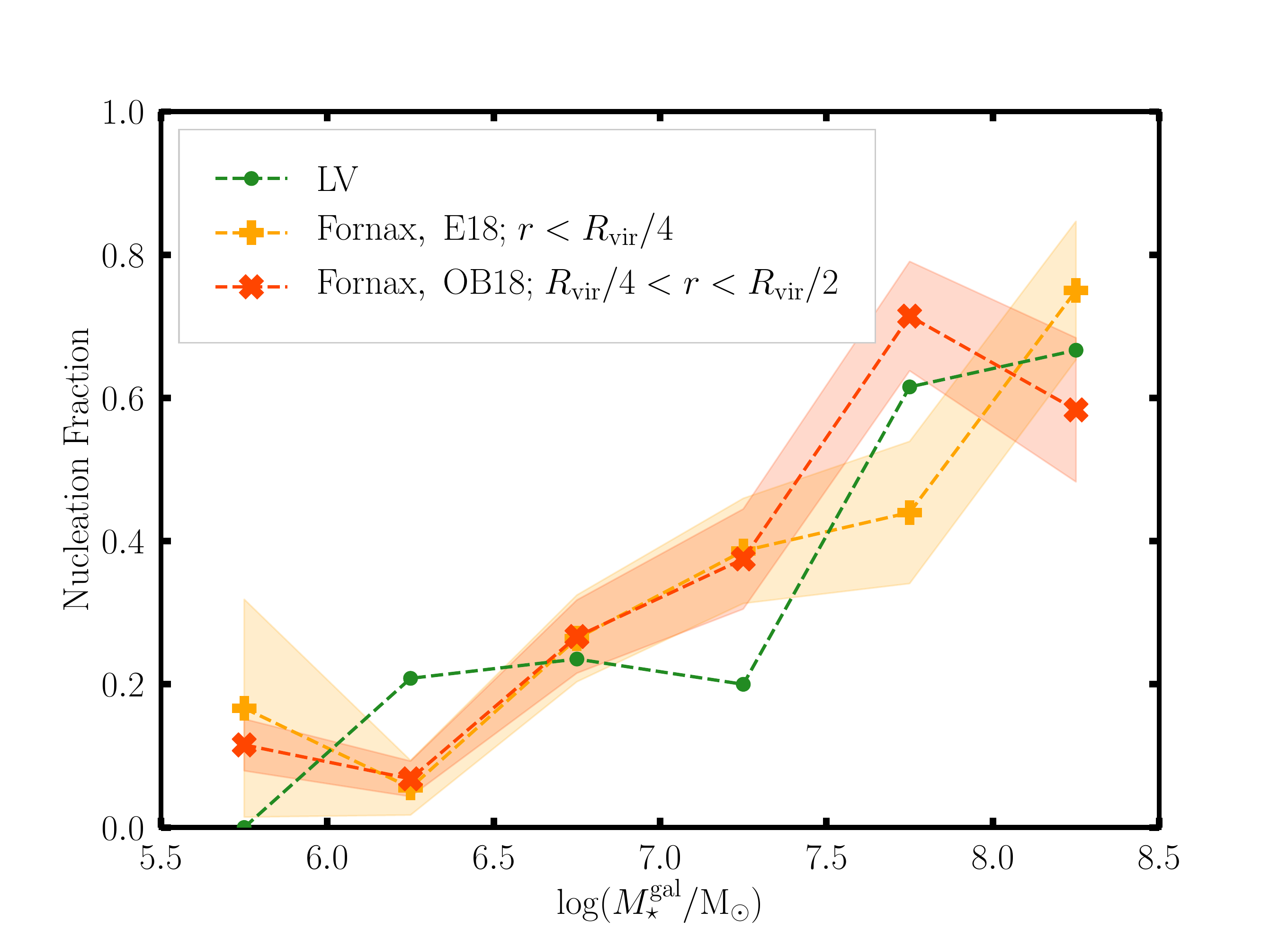}
\caption{The nucleation fraction as a function of dwarf stellar mass for the two NGFS dwarf samples. That of \citet{eigenthaler2018} includes the innermost ($r<R_\mathrm{vir}/4$) dwarfs while that of \citet{ordenes-briceno_spatial} includes dwarfs in the radial range ($R_\mathrm{vir}/2<r<R_\mathrm{vir}/4$). There is no significant difference between the two radial ranges.}
\label{fig:fornax_nuc}
\end{figure}

\section{Fornax Cluster Nucleation Fraction}
\label{app:fornax_nuc}
In Figure \ref{fig:fornax_nuc}, we show the nucleation fraction of the two NGFS samples of \citet{eigenthaler2018} and \citet{ordenes-briceno_spatial}. \citet{eigenthaler2018} includes the innermost ($r<R_\mathrm{vir}/4$) dwarfs while \citet{ordenes-briceno_spatial} includes dwarfs in the radial range $R_\mathrm{vir}/2<r<R_\mathrm{vir}/4$. There does not seem to be any noticeable difference in nucleation fractions, indicating that the environmental trends we see in, e.g., Figure \ref{fig:nuc_frac}, are due primarily to parent halo mass, and less to dwarf location in the parent halo. However, this is difficult to reconcile with the results of \citet{lisker2007} and \citet{ordenes-briceno_spatial} which conclude that nucleated dwarfs are more spatially centrally concentrated in Virgo and Fornax than non-nucleated dwarfs. Dwarf samples from the entire virial regions of Fornax and Virgo are likely needed to fully understand this.

\end{document}